\documentclass{emulateapj}
\slugcomment{{\sc Published in ApJ 07/2013} }

\newcommand{\summation}[3]{\sum^{#2}_{#1} #3\, }

\newcommand{\Msun}[0]{M_\odot}

\newcommand{\hinv}[0]{\mbox{ }h_{\rm 100}^{-1}}
\newcommand{\etal}[0]{et al.}

\newcommand{\simLA}[0]{{\mathcal{L}_0}}
\newcommand{\simLB}[0]{{\mathcal{L}_1}}
\newcommand{\simLC}[0]{{\mathcal{L}_2}}
\newcommand{\simSA}[0]{{\mathcal{S}_0}}
\newcommand{\simSB}[0]{{\mathcal{S}_1}}
\newcommand{\simSC}[0]{{\mathcal{S}_2}}

\newcommand{\simSi}[0]{{\mathcal{S}_{\rm i}}}
\newcommand{\simLi}[0]{{\mathcal{L}_{\rm i}}}

\newcommand{\gcc}[0]{\mbox{ g}\mbox{ cm}^{-3}}
\newcommand{\fig}[1]{Figure~\ref{#1}}		
\newcommand{\sect}[1]{Section\ref{#1}}			
\newcommand{\tabl}[1]{Table~\ref{#1}}		

\begin{document}

\title{Hybrid Cosmological Simulations with Stream Velocities}
\author{Mark L. A. Richardson\altaffilmark{1}, Evan Scannapieco\altaffilmark{1}, and Robert J. Thacker\altaffilmark{2}}
\altaffiltext{1}{School of Earth and Space Exploration, Arizona State University, Tempe, AZ 85287, USA}
\altaffiltext{2}{Department of Astronomy and Physics, Saint Mary's University, Halifax, B3H 3C3, Canada}

\setcounter{footnote}{2}

\begin{abstract}
In the early universe, substantial relative ``stream" velocities between the gas and dark matter arise due to radiation pressure and persist after recombination.  To assess the impact of these velocities on high-redshift structure formation, we carry out a suite of high-resolution adaptive mesh refinement (AMR) cosmological simulations, which use smoothed particle hydrodynamic datasets as initial conditions, converted using a new tool developed for this work. These simulations resolve structures with masses as small as a few $100 \Msun$, and we focus on the $10^6 \Msun$  ``mini-halos'' in which the first stars formed. At $z \approx 17,$ the presence of stream velocities  has only a minor effect on the number density of halos below $10^6 \Msun$, but it greatly  suppresses gas accretion onto all halos and the dark matter structures around them. Stream velocities lead to significantly lower halo gas fractions, especially for $\approx 10^5 \Msun$ objects, an effect that is likely to depend on the orientation of a halo's accretion lanes.  This reduction in gas density leads to colder, more compact radial profiles, and it substantially delays the redshift of collapse of the largest halos, leading to delayed star formation and possibly delayed reionization.  These many differences suggest that future simulations of early cosmological structure  formation should include stream velocities to properly predict  gas evolution, star-formation, and the epoch of reionization.\end{abstract}
\maketitle

\section{Introduction}

A cold dark matter dominated model universe including a cosmological constant term ($\Lambda$CDM) 
provides multiple predictions that are in excellent agreement with observations 
(e.g., Spergel \etal\ 2007; Larson \etal\ 2011).  In this theory,  the evolution of density perturbations at very high 
redshifts can be well understood  by working to first order, where the amplitude of the perturbations remains small 
(e.g., Peebles 1974; Ciardi \& Ferrara 2005). At this epoch, dark matter perturbations are able to grow by 
gravitational collapse, but gas must wait until after recombination \citep{Ciardi05}, when photons decouple from the 
baryons, removing the radiation pressure  and allowing gravity to become dominant.   Soon after, these 
perturbations grow more overdense than unity, and detailed modeling must proceed using large, multi-resolution 
simulations. 

Recently, \citet{Tseliakhovich10} showed that this evolution will depend on stream velocities that are present 
between the dark matter and gas. This effect is due to quadratic terms in the cosmic perturbation theory that account 
for the relative velocity between the dark matter and gas due to the impact of radiation pressure before 
recombination. At $z=z_{\rm dec}\simeq1020$, the redshift of decoupling, this stream velocity is coherent over large 
scales (a few comoving Mpc), with a typical value of $30\mbox{ km s}^{-1}$. \citet{Tseliakhovich10} determined that the inclusion of this 
streaming term could reduce the growth of structure at scales below 17 comoving kpc, with suppression in the 
matter power spectrum peaking at about 5 comoving kpc. They also found that the inclusion of the streaming term 
would reduce the halo number density, with the largest suppression occurring at
$\approx 2 \times 10^6\Msun$.

To better understand the effects of stream velocity on structure formation, multiple recent works have probed 
different scales with different techniques. \citet{Maio11} performed smoothed particle hydrodynamic 
(SPH) cosmological simulations with GADGET-2 (Springel \etal\ 2001; Springel 2005) and examined the effect of varying the 
magnitude of the stream velocity on star formation.   Their simulations had  fixed dark matter and gas particles 
resolutions of $800\Msun\hinv$ and $160\Msun\hinv$, respectively, and they found that star formation was 
delayed by a few tens of Myr,  dark matter halos were reduced in gas content by up to 50\%, and there was a
minimal effect on the dark matter halo statistics. \citet{Greif11} performed multiple cosmological hybrid 
simulations that combined grid and particle-based techniques, using the AREPO code \citep{Springel10}. Their 
simulations included nested levels of resolution, with the highest level having dark matter and gas particle 
masses of 3.53 and 0.72$\Msun$, respectively. They found there was a delayed collapse in minihalos when a 
stream velocity was included, requiring an increase in the virial mass by a factor of 3 before reaching a critical 
density, which, in turn, delayed the onset of Population III  star formation. \citet{Naoz12} and \citet{Naoz13} 
performed multiple cosmological GADGET-2 simulations of various resolution that included multiple stream 
velocities. Their fixed-resolution simulations had particle masses ranging from a few $\times 10^1 - 10^2\Msun$, 
and they looked at large-scale statistics, with a particular focus on larger mass halos ($> 10^4 \Msun$). In 
\citet{Naoz12}, they found that the introduction of a stream velocity suppressed the clumping of baryons and the halo mass function in 
mass and redshift, and that for very large stream velocities, many halos were 
devoid of gas. They also considered the effect of adding a physical offset between the gas and dark matter 
corresponding to advection that would occur between $z_{\rm dec}$ and $z_{\rm init}$, the redshift at which they start 
their simulation. They saw minor effects from including this spatial offset, although we argue that such an offset 
would cause more specific changes on individual halos since the majority of this displacement is accrued at large redshifts. It is 
likely these effects may be missed when looking at the 
statistics of a large volume. In \citet{Naoz13}, they altered the halo gas fraction fits of \citet{Gnedin00} to 
accommodate the effects of stream velocity. Finally, OÕLeary \& McQuinn (2012) performed a suite of uniform resolution SPH simulations (with masses of 10 Ð 20 Mo), uniform grid simulations, and adaptive mesh refinement (AMR) simulations whose initial conditions were created via a transfer function that self-consistently accounts for stream velocity. They looked at the effect of box size, resolution, and initial redshift on the formation of the first structures in the universe with stream velocity included. They found that the box size must be at least $1$ Mpc h$^{-1}$ to properly resolve the halo mass function up to masses where halos can cool via molecular hydrogen and form stars. They also found that the best initial redshift was around 200 to 400, since increasing the initial redshift at these early times adds little extra computational effort, yet is an effective way to reduce noise in the power spectrum. The authors then looked at the specifics of gas accretion onto halos, finding that by adding a stream velocity a significant fraction of the gas is moved downwind of halos, with the effect most notable with larger stream velocities. The authors also note that the orientation of filaments with respect to the stream direction can influence the effect of stream velocities on gas accretion.

In this work, we adopt a new simulation technique and perform multiple cosmological hybrid simulations combining the 
particle-based code GADGET-2 and the grid-based code FLASH3.2 \citep{Fryxell00} with nested resolution levels 
to probe the smallest structures. We investigate the effect of multiple stream velocity magnitudes, and probe this
effect on a range of halo mass scales both by looking at individual halos in detail, and probing halo statistics. We 
will also demonstrate how our new simulation method compares with existing SPH and hybrid methods.

Our hybrid approach taps into the strengths of these two simulation methods.
SPH, which uses a collection of Lagrangian particles, is economical in time and memory, can be implemented fairly 
easily, and naturally provides high resolution in the denser areas in which galaxies form.  However,  mixing, shocks, 
and shear layers are difficult to accurately model with this method as they require excellent capturing of density 
fronts, which typically requires implementation of an ad hoc artificial viscosity (e.g., Thacker \etal\ 
2000; O'Shea \etal\ 2005; Agertz \etal\ 2007; Sijacki \etal\ 2011). Also, mixing, which requires significant resolution, 
often occurs in low density regions where the SPH method inherently has less resolution. AMR, on the other hand, uses an Eulerian grid of varying sized cells, which change in resolution 
according to criteria specified by the user. An AMR method requires significant memory overhead, but can add resolution 
as needed to low density regions and can also circumvent using an artificial viscosity, making it much better suited for low-density dynamics, mixing, shocks, and shear 
layers, and can achieve the multi-scale resolution requirements provided the memory and computational hardware 
is available (e.g., Fryxell \etal\ 2000; O'Shea \etal\ 2005; Agertz \etal\ 2007).

To tap into the strengths of both these methods, we have developed a tool to take SPH simulations from  codes such 
as HYDRA \citep{Thacker06A} and GADGET-2 and map them to AMR datasets that can then be evolved further with 
the FLASH code \citep{Fryxell00}. Combined, these two schemes will efficiently produce the initial 
conditions of structure formation, while properly modeling the dynamics of hierarchical merging. Note that this is one 
of many scenarios where converting between these methods would greatly benefit the models. Such hybrid 
approaches are becoming more popular and are finding new applications, particularly simulations of cosmological 
structure formation (e.g., Agertz \etal\ 2007; Sijacki \etal\ 2011).

For this work, we first perform two sets of SPH simulations, including stream 
velocities between the gas and dark matter. We then implement our tool at an intermediate redshift to 
map the SPH datasets into high-precision initial conditions for an AMR simulation. We determine the effect of stream velocity 
on the gas fraction and number density of halos, 
and on the evolution of structure at various densities, and how these effects vary between methodologies.

The structure of this paper is as follows. In \sect{codes} we describe the combined $N$-body and SPH 
simulations performed with GADGET-2, our method of mapping the particle simulations to an AMR scheme, and
the simulation carried out with the FLASH code. In \sect{Results} we discuss our results, beginning 
with the three largest halos and 
the effect of stream velocities on their structure and gas fraction. We then focus on radial density profiles 
of the largest halo, and finally discuss statistics describing all of the virialized halos in our volume. We summarize our work and conclusions are given in \sect{Conc}. Throughout this paper we use 
($\Omega_{\Lambda}, \Omega_{\rm M}, \Omega_{\rm b}, n, \sigma_{\rm 8}, \hinv$) = (0.734, 0.266, 0.0449, 0.963, 0.801, 0.71) 
\citep{Larson11}.

\section{Numerical Methods}\label{codes}
\begin{deluxetable*}{lcccccc}
\tabletypesize{\scriptsize}
\tablewidth{0pc}
\tablecaption{Simulations Summary
\label{tab_sims}}
\startdata
General Parameters & SPH $z_{\rm init}$ & AMR $z_{\rm init}$ & $z_{\rm f}$ & SPH $m_{\rm DM} $\tablenotemark{a} &  SPH $m_{\rm gas}$\tablenotemark{a} & AMR $\Delta x$  \\ \hline
& 199 & 39 & 17.18 & 3.01  & 0.611 &  3.09 pc \\ \hline \hline
Simulation: & $\simLA$ & $\simLB$ & $\simLC$ & $\simSA$ & $\simSB$ & $\simSC$  \\ \hline
$v_{\rm s}$\tablenotemark{b} & 0 & 5.88 & 11.76 & 0 & 5.88 & 11.76 \\
$M_{\rm group}$\tablenotemark{c} & 12.83 & 12.83 & 12.83 & 2.45 & 2.45 & 2.45
\enddata
\tablenotetext{a}{$\Msun h^{-1}$. \ $^{\rm b}$km s$^{-1}$. \ $^{\rm c}10^6 \Msun$. }
\end{deluxetable*}
\subsection{Particle Simulations}\label{part}
We first performed one low-resolution cosmological simulation using GADGET-2 where our box was 0.75 comoving Mpc$\hinv$ on a side.
This low-resolution simulation had 256$^3$ dark matter particles (each with mass $1.54 \times 10^{3}\Msun\hinv$ ) and 256$^3$ gas
particles (each with mass $3.13 \times 10^2\Msun\hinv$). To set up the initial conditions we us the transfer function from CAMB (Lewis et al. 2000 and references therein), assuming
an initial spectral slope of $n=0.963$ \citep{Larson11}. CAMB uses a line-of-sight implementation of the linearized equations of the covarient approach to cosmic microwave background
(CMB) anisotropies. This results in different transfer functions for the dark matter and baryon components, with the two weighted together
to yield the total transfer function:
\begin{equation}
\delta_{\rm Tot} = f_{\rm DM}\delta_{\rm DM} + f_{\rm B}\delta_{\rm B},
\end{equation}
where $\delta_{\rm DM}$, $\delta_{\rm B}$, and $\delta_{\rm Tot}$ are the resulting linear overdensity for the dark matter, baryon, and combined components, respectively, and $f_{\rm DM}$ and $f_{\rm B}$ are the mean cosmic dark matter and baryon fractions, respectively.
 This simulation began at $z=199$ and was evolved to $z=17.18$. We
did not include star formation or feedback, but did include atomic and molecular cooling. We calculated the different ionization states of hydrogen and 
helium from the density and temperature following \citet{Katz96}. Although at such high redshifts collisional ionization equilibrium is perhaps not appropriate, it is irrelevant
since the material remains neutral. We assumed a primordial number density fraction of molecular hydrogen
of  H$_2$/H $= 1.1\times10^{-6}$ following \citet{Palla00} and  a primordial deuterium  number density fraction of D/H $= 2.7\times 10^{-5}$ following \citet{Steigman09}.
Deuterated hydrogen was set to be HD/H$_2 = 6\times10^{-4}$ by combining \citet{Palla00} and \citet{Steigman09} with their consideration 
of the photon-baryon ratio. Given these abundances, we employed the molecular cooling rates of \citet{Gray10} and cooling rates for Compton scattering against CMB photons 
as given in \citet{Barkana01}.

We then used the friends-of-friends algorithm \citep{Davis85} to determine groups that corresponded to an overdensity of 180, 
expected from a spherical top-hat collapse. We took two isolated groups (i.e.\ not within a few virial radii of another group of equal or larger 
mass) with masses of 2.45$\times 10^6 \Msun$ and 12.83$\times 10^6 \Msun$, and performed two simulations statistically identical to the 
low-resolution simulation, but with additional resolution centered on each of these two groups. We focus on these masses as they are near the critical 
halo mass needed to form stars via molecular cooling \citep{Yoshida03}, and also in the range where we expect the most significant
suppression in halo abundance \citep{Tseliakhovich10}. We added three spherically-nested resolution levels 
resulting in particle masses of $3.01\Msun\hinv$ and $0.611\Msun\hinv$ for dark matter and baryons, respectively, at the highest
resolution. At each level of resolution the modes of the initial spectrum are truncated to ensure there is no aliasing of high $k$ modes into the low $k$ values (as in Thacker \& Couchman 2000). 

We then created two additional instances of each simulation in which we added a stream velocity  to the baryons with respect
to the dark matter and a spatial offset corresponding to the displacement accrued between $z_{\rm dec}$, the redshift of decoupling, and $z=199$:
\begin{equation} \label{voffset}
\Delta x_{\rm s}(a_{\rm f}) = \int_0 ^{t_{\rm f}} \frac{v_{\rm s} (a)}{a} dt,
\end{equation}
where $\Delta x_{\rm s}(a_{\rm f})$ is the comoving displacement of a parcel of gas due to the stream velocity, $v_{\rm s}$, by a final time, $t_{\rm f}$ 
and corresponding final scale factor, $a_{\rm f}$ \citep{Naoz12}. 
The added stream velocities were $v_{\sigma}\hat{x}$ and $2v_{\sigma}\hat{x}$, where $v_\sigma$ is the typical stream velocity found in \citet{Tseliakhovich10}, roughly 
$30\mbox{ }{\rm km}\mbox{ }{\rm s}^{-1}$ at $z_{\rm dec}$. These correspond to $\Delta x_{\rm s}(z=199)$ of 18.5 kpc $h^{-1}$, and 37 kpc $h^{-1}$, respectively. We enforced the stream velocity over the whole box, as \citet{Tseliakhovich10} showed that
such streams should be coherent over a few Mpc. The resulting six simulations, summarized in \tabl{tab_sims}, are referred to as $\simLA$, $\simLB$, $\simLC$, $\simSA$, $\simSB$, 
and $\simSC$, corresponding to the fiducial run, the run with a stream velocity of $v_\sigma$ and $2v_{\sigma}$ centered on the larger 
group and the fiducial run, the run with a stream velocity of $v_\sigma$ and $2v_{\sigma}$ centered on the smaller group, respectively. We  refer
to $\simLA$, $\simLB$, and $\simLC$ as a group as $\simLi$, and $\simSA$, $\simSB$, and $\simSC$ as a group as $\simSi$. 
These high-resolution initial conditions were evolved to $z=17.18$, before the nonlinear scale reaches an appreciable fraction of the small simulation volume. At $z=39$ we 
mapped (see \sect{mapping}) the GADGET-2 snapshots to the 
AMR code FLASH3.2, corresponding to a scale factor slightly less than half than the final scale factor, which is 
before significant collapse of structures. 

\subsection{Mapping SPH to AMR}\label{mapping}

To map the gas particle field to a grid configuration, we first relate the particle
variables, i.e.\ position, $\textbf{x}_{\rm i}=(x_{\rm i},y_{\rm i},z_{\rm i})$, velocity,
$\textbf{v}_{\rm i}=(\dot{x}_{\rm i},\dot{y}_{\rm i},\dot{z}_{\rm i})$, mass, $m_{\rm i}$, and internal specific
energy, $e_{\rm int,i}$, for any given particle, \emph{i}, to any point in space. We map to the gas center-of-mass frame, 
which translates the stream velocity to the dark matter particles.
Each gas particle also has a smoothing length, $h_{\rm i}$, and obeys the smoothing kernel $w(r,h_{\rm i})$
given by \citep{Springel01}

\begin{equation} \label{kernel}
  w(r,h_{\rm i}) = \frac{8}{\pi h_{\rm i}^3} \left\{ \begin{array}{ll}
  1 - 6\left( \frac{r}{h_{\rm i}} \right)^2 + 6\left(\frac{r}{h_{\rm i}} \right)^3 & \mbox{if $0\le r \le \frac{h_{\rm i}}{2}$;}\\
  2\left[1-\left(\frac{r}{h_{\rm i}}\right) \right]^3 & \mbox{if $\frac{h_{\rm i}}{2} \le r \le h_{\rm i}$;}\\
  0 & \mbox{if $h_{\rm i} \le r$.} 
   \end{array} \right. 
\end{equation}

The smoothing kernel has several important properties: it has both continuous first and second derivatives,
and its volume integral over all space is unity. 
Thus the density at any point, $\textbf{r}$, a distance
$r_{\rm i}=|\textbf{r}-\textbf{x}_{\rm i}|$ from particle $i$ caused by this particle is
simply
\begin{equation}\label{rhor}
\rho (\textbf{r})_{\rm i} = m_{\rm i} w(r_{\rm i},h_{\rm i}).
\end{equation}
The total density is then just the sum over all gas
particles' density contribution to that point.

Since the conserved quantities are mass, energy (internal and kinetic), and momentum, while 
the particles track the internal specific energy and the specific momentum,
 the velocity and internal specific energy at any point must be the sum of the
contributing particle's velocity and internal specific energy, weighted by the
particle's density, and normalized by the total density at that
point.  

We are now in a position to discretize space such that it is
represented by a three-dimensional grid of cells. Each cell can be 
thought of as a point at
its center whose physical values are equal everywhere in the
cell. This is different from the SPH particles in two main ways:
first, as viewed from the grid, every point in space is
characterized by only one particle and its corresponding cell center; and
second, the density is also uniform in a given cell, compared with a
density that drops off from an SPH particle according to the kernel,
$w(r,h_{\rm i})$.

We evaluate the $(\rho, v, e=e_{\rm int}+v^2/2)$ fields at each cell center and assign these values
to the cell. However, this method will not  conserve mass, thus for each particle
we determine a mass correction factor, equal to the ratio of the mass mapped to the particle's 
actual mass and multiply this value to the particle's contribution to the density in each cell. 
In the situation where a particle's smoothing length does
not reach a single cell center, then its quantities are added to the cell in which it is fully contained.
The final quantities in a cell \emph{j} are then:
\begin{eqnarray}
\rho_{\rm j} & = & \summation{i=1}{N}{f_{\rm i}\rho({\textbf{r}_{\rm j}})_{\rm i}}  \equiv  \summation{i=1}{N}{\rho_{\rm ij}} \\
\textbf{v}_{\rm j} & = & \frac{1}{\rho_{\rm j}}\summation{i=1}{N}{\textbf{v}_{\rm i} \rho_{\rm ij}} \\
e_{\rm j} & = & \frac{1}{\rho_{\rm j}}\summation{i=1}{N}{(e_{\rm int,i} + \frac{v_{\rm i}^2}{2}) \rho_{\rm ij}} \\
e_{\rm int,j} & = & e_{\rm j} - \frac{v_{\rm j}^2}{2},
\end{eqnarray} 
where subscript \emph{i} implies it is the quantity associated with particle \emph{i},
subscript \emph{j} implies it is the quantity associated with cell \emph{j},
and subscript \emph{ij} implies it is the contribution from particle \emph{i} 
in cell \emph{j}. The mass correction factor is given by $f_{\rm i}$, and cell $j$ has center position $\textbf{r}_{\rm j}$. \emph{N} is the number of particles, and recall that a cell farther 
than $h_{\rm i}$ from particle \emph{i} will not have any contribution from \emph{i}
mapped into it.

In preserving the total energy, the resulting internal energy will be slightly increased by the mapping process. The reason for this is that  the total kinetic energy of the AMR grid will always be less than or equal to the kinetic energy of the original SPH simulation, since mapped momentum from one particle can reduce the mapped momentum from another. Thus to preserve total energy, we must add the missing energy to the internal specific energy. This is acceptable as we should interpret this scenario as having phase velocities in the fluid that are not mapped over.

The angular momentum is not preserved in general. Consider the case where a particle does not influence any cells but the one in which it lies. The result to conserve mass, momentum and internal energy is just to move the particle to the center of the cell and add it to the other values that cell already has. This clearly changes the angular momentum of that particle with respect to all points in space. However, for sufficient resolution this effect should be small, and when a particle is located at the center of a cell, its angular momentum is conserved. When a particle influences many cells but is not at a cell center, then angular momentum is changed very slightly.

\subsection{AMR Simulations}\label{grid}

The mapped $z=39$ GADGET-2 snapshots were treated as initial conditions for six FLASH3.2 simulations. 
FLASH is an AMR code with blocks of $N_{\rm B,X} \times N_{\rm B,Y} \times N_{\rm B,Z}$ cells divided between processors. Each subsequent level in refinement increases the
spatial resolution by a factor of two in each dimension. The relative difference in resolution between a block and any of its neighbors can be at most a factor of two. For our simulations we used
$N_{\rm B,X} = N_{\rm B,Y} = N_{\rm B,Z}=16$, which maximized the ratio of memory and time spent on the active cells versus the inactive guard cells, while maintaining an efficient
division of the simulation volume. Although the SPH simulations had periodic boundary conditions, and used comoving coordinates,
we performed the AMR simulations with physical coordinates and outflow
boundary conditions, preferring a fixed maximum physical resolution over a fixed maximum comoving resolution. Our simulation volume has a side-length of 25.35 physical kpc. We evolved the AMR simulations to $z=17.18$, in agreement with the SPH simulations. 

The grid was composed of a high-resolution region, extending 25\% beyond the high-resolution region of the SPH volume to allow for Hubble expansion, and 
beyond this was a low-resolution region where we enforced derefinement, since here we only needed to accurately model the tidal field. The high-resolution region, centered on the group of interest, 
allowed for refinement following the procedure in \citet{Turk12}. 
The first refinement criteria depended on the minimum Jeans length in a 
block, where the Jeans' length is given by:
\begin{equation}
\lambda_{\rm J}  = \sqrt{\frac{15 (\gamma -1) e_{\rm int}}{4\pi G \rho_{\rm m}}},
\end{equation}
where $\gamma$ is the ratio of specific heats of the gas, and $\rho_{\rm m}$ is the total mass density.
We ensure that we resolved the Jeans' length by a factor $n_{\rm j}$, which we call the Jeans parameter. \citet{Truelove97} demonstrated that this must be at least 4 to stop
artificial fragmentation, and for this work we use a value of 16. 
Second, we refined the coarsest blocks in this region if its maximum 
density was four times denser than cosmological mean. For each additional level of refinement we increased this overdensity threshold 
by a factor of $8\times2^{-0.3}$. Derefinement was allowed if 1/48 the minimum Jeans' length was resolved and the density was below the lower 
refinement level's density threshold. 

To ensure no material from the low-resolution region
fell into the high-resolution region and merges with our group,  
a mass scalar was used with an enforced value of unity in the low-resolution region and initially zero in the high-resolution region. This scalar moved with the
fluid flow, and we found that it did not fall into the high resolution region.

During the AMR simulations we used the same cooling source terms as in the SPH runs, although we also accounted for possible chemistry following \citet{Gray10}. We argue that
up until $z=39$ little chemistry has taken place, thus not accounting for chemistry in GADGET up to this point should be acceptable. 

FLASH also includes particles that we used to represent dark matter, which we simply move directly from GADGET to FLASH.
Dark matter mass is mapped to the grid after each hydro step using the cloud-in-cell method \citep{Birdsall97}. The 
gravitational potential is then calculated based on the total mass density, and accelerations are interpolated back on to the particles such that there is no self-force. When
a dark matter particle is advected beyond the simulation volume it is removed from the simulation.

We set the maximum resolution to 1/8192 the size of the box, corresponding to 10 levels of refinement.
 This was chosen to ensure a single dark matter particle was not mapped to a sufficiently small cell as to overwhelm
the total density, which would lead to unrealistic collapse of gas. 
This maximum resolution, combined with the Truelove condition that the Jeans' parameter, $n_{j,T}$, be at least 4, sets a maximum density 
for gas with some specific internal energy, above which unrealistic
fragmentation would occur. This maximum baryon density is then given by
\begin{eqnarray*}
\rho_{\rm max,b} &=& \frac{15\Omega_{\rm b} (\gamma -1) e_{\rm int}}{4\pi G \Omega_{\rm m} n_{\rm j,T}^2 x_{\rm min}^2} \\
&=&   8.2\times 10^{-22} \left(\frac{e_{\rm int}}{10^{11}\mbox{ erg g}^{-1}}
\right) \left(\frac{\Omega_{\rm b}}{\Omega_{\rm m}}\right)\gcc ,
\end{eqnarray*}
where $x_{\rm min}$ is the minimum zone size which for our simulations is 3 pc.

\section{Results}\label{Results}

\subsection{Delay of Structure and Gas Fraction}\label{threeHalos}
\begin{figure*}[t!] 
\centering
\includegraphics[scale=0.21]{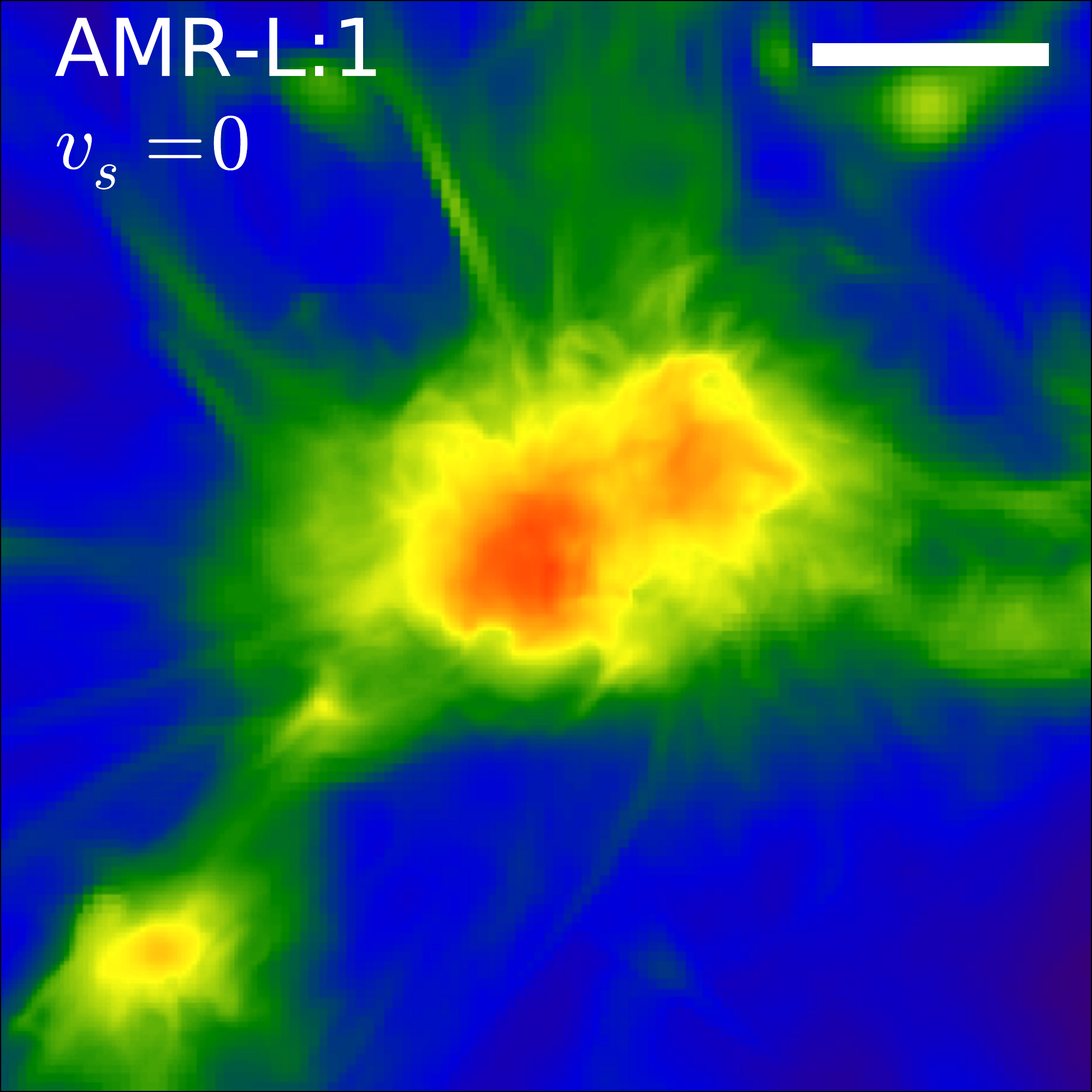} 
\includegraphics[scale=0.21]{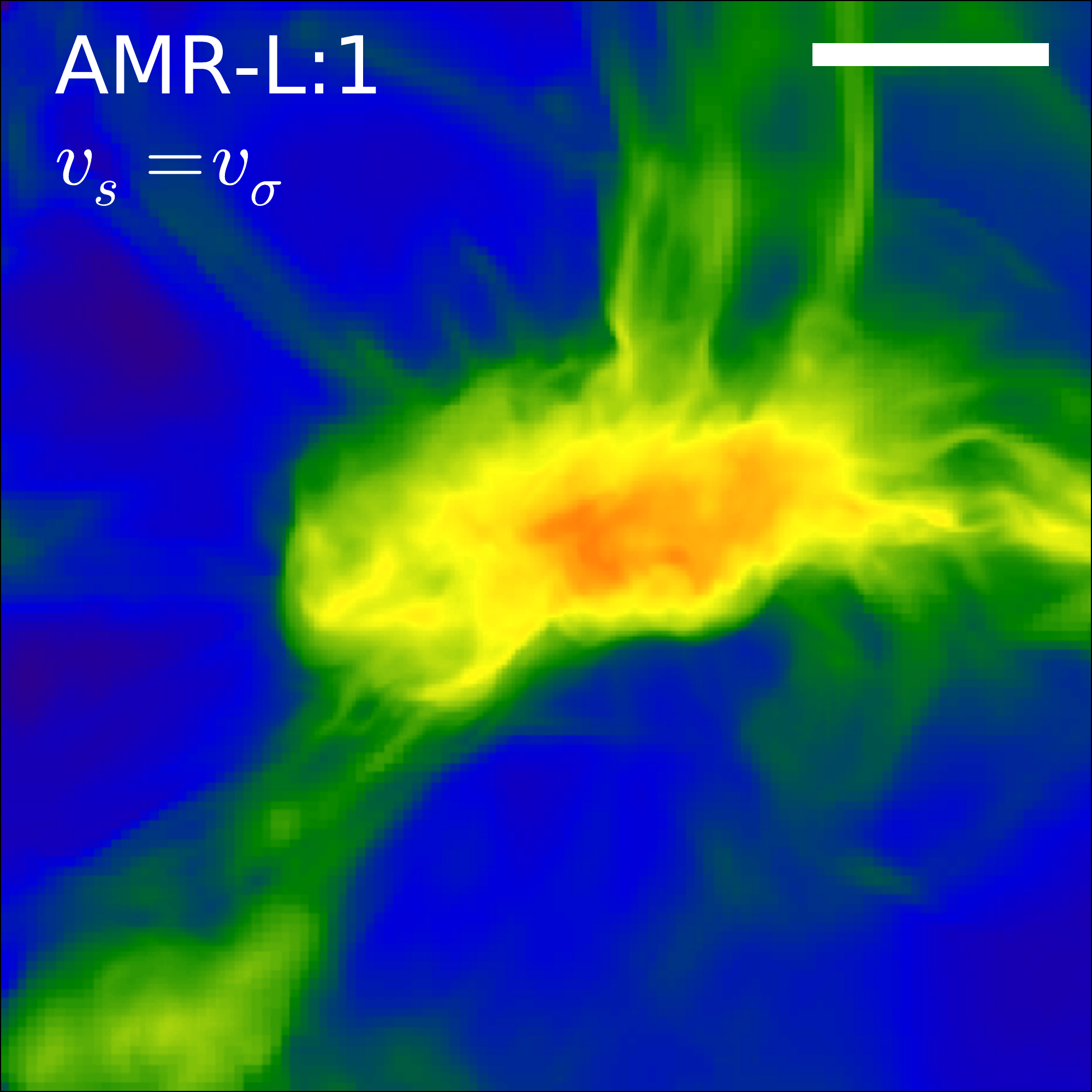} 
\includegraphics[scale=0.21]{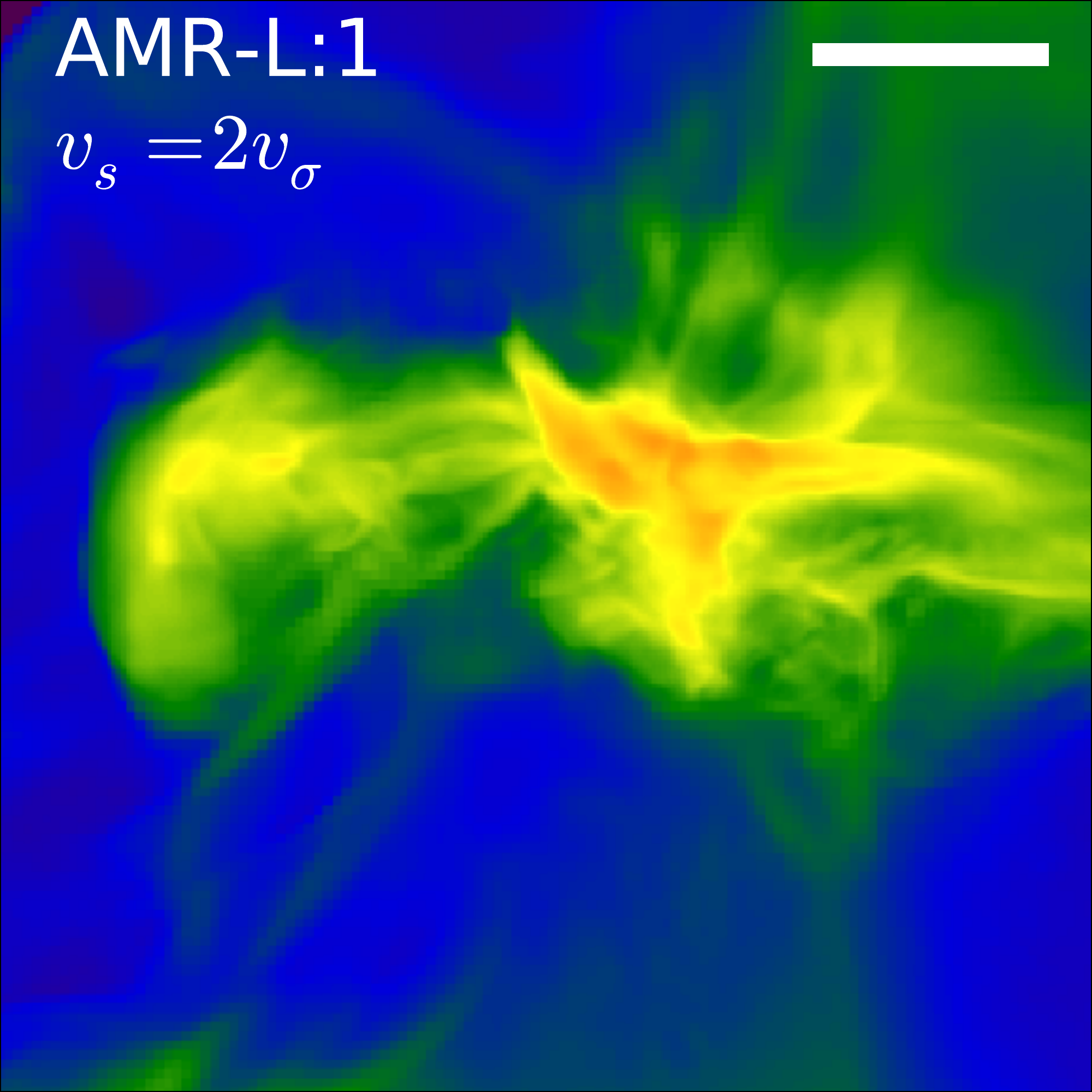} 
\includegraphics[scale=0.1565]{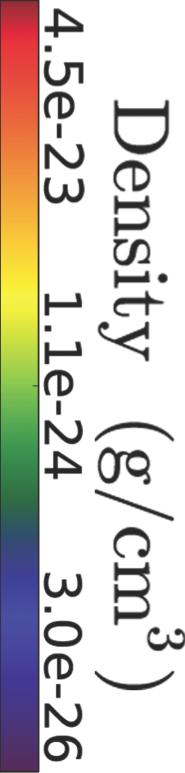} 
\includegraphics[scale=0.21]{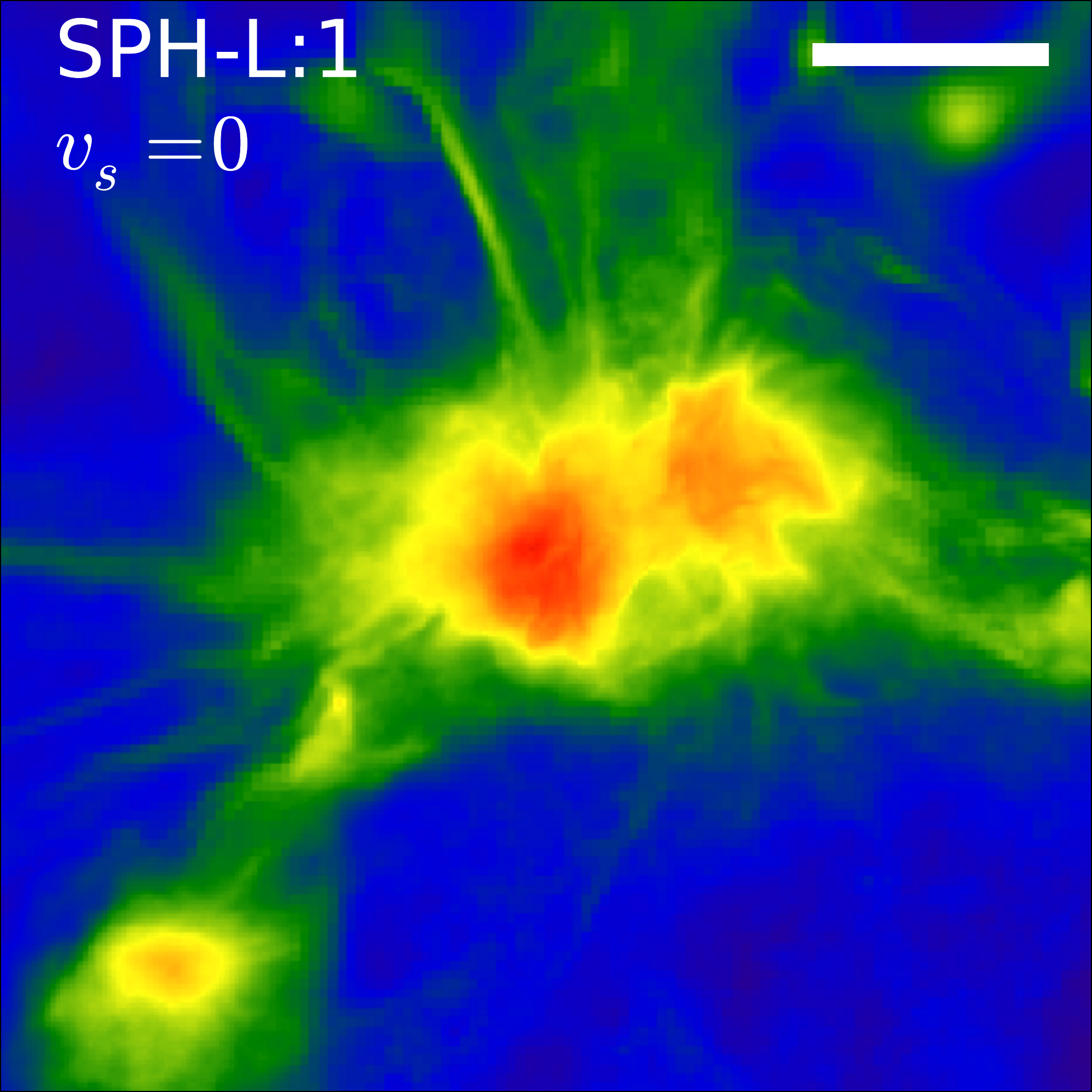} 
\includegraphics[scale=0.21]{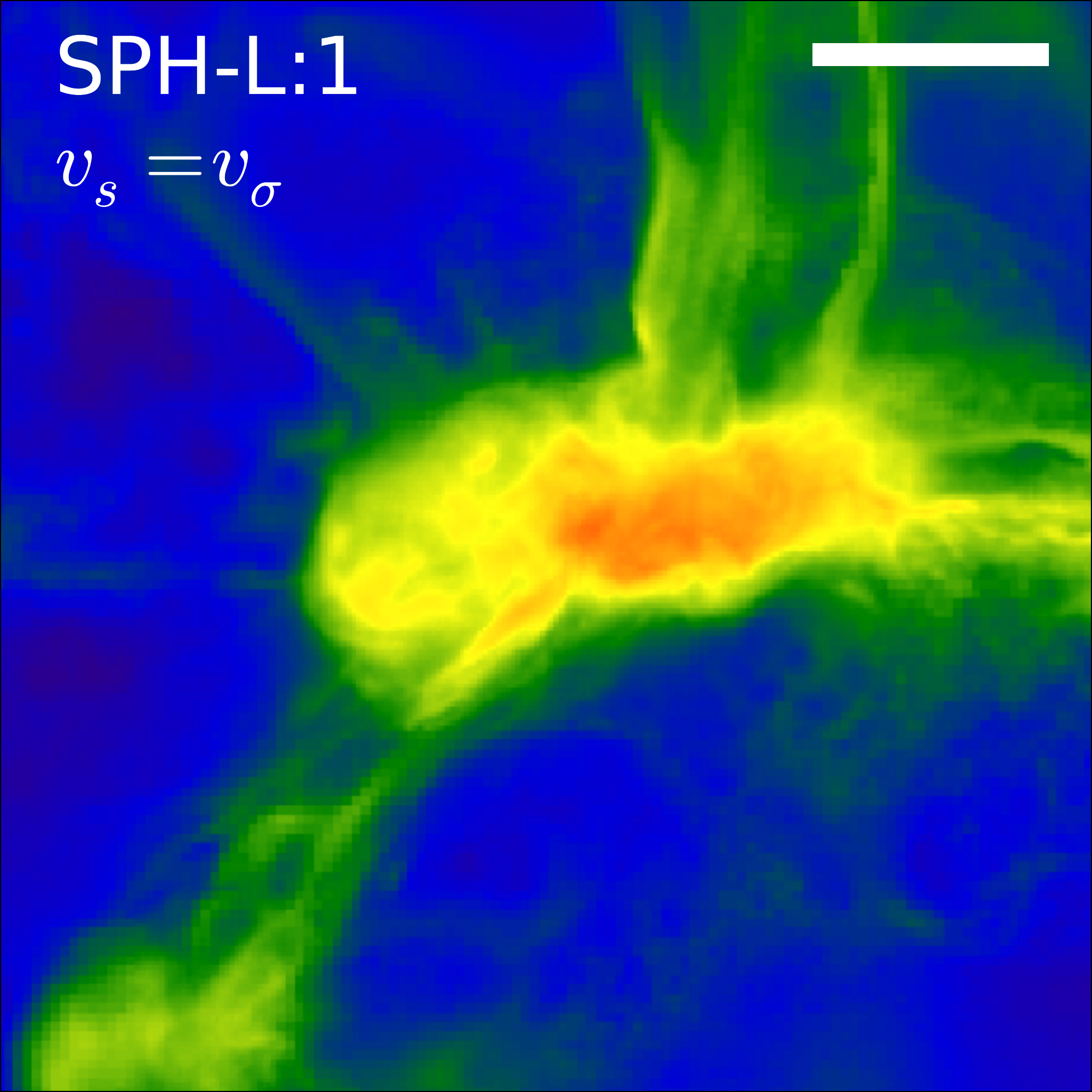} 
\includegraphics[scale=0.21]{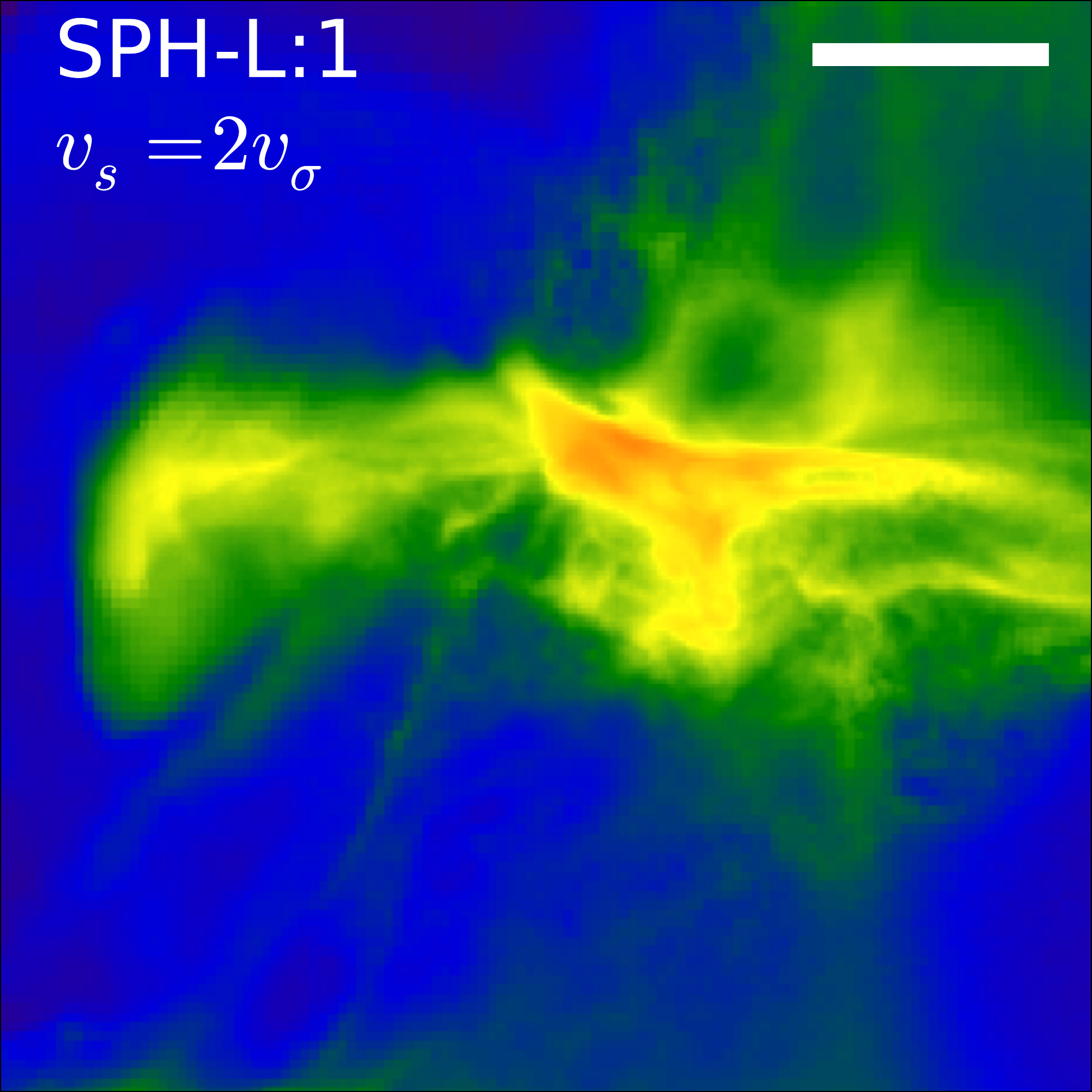} 
\includegraphics[scale=0.1565]{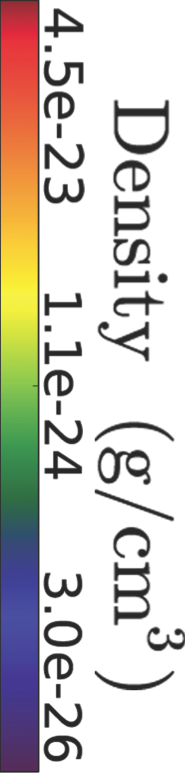} 
\includegraphics[scale=0.21]{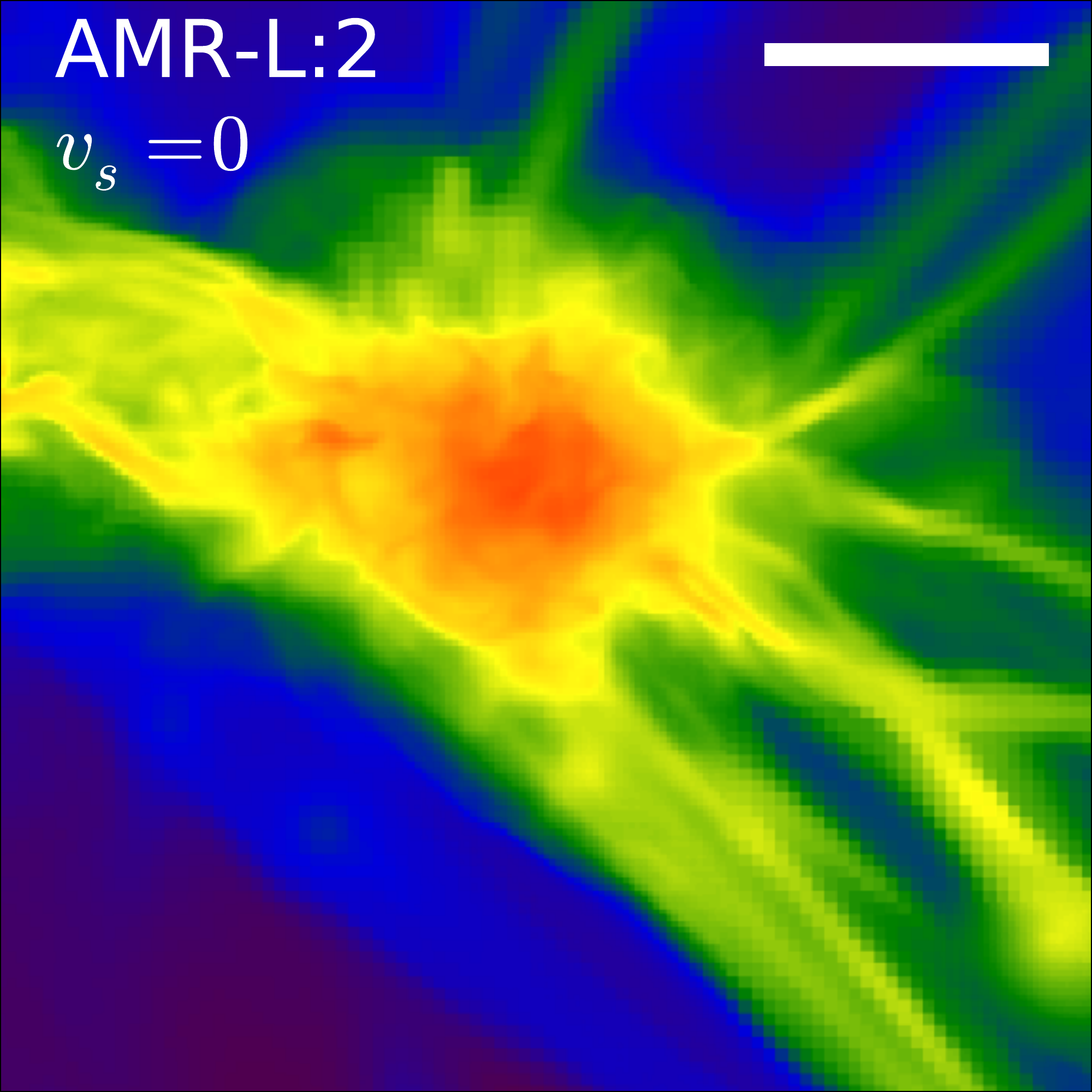} 
\includegraphics[scale=0.21]{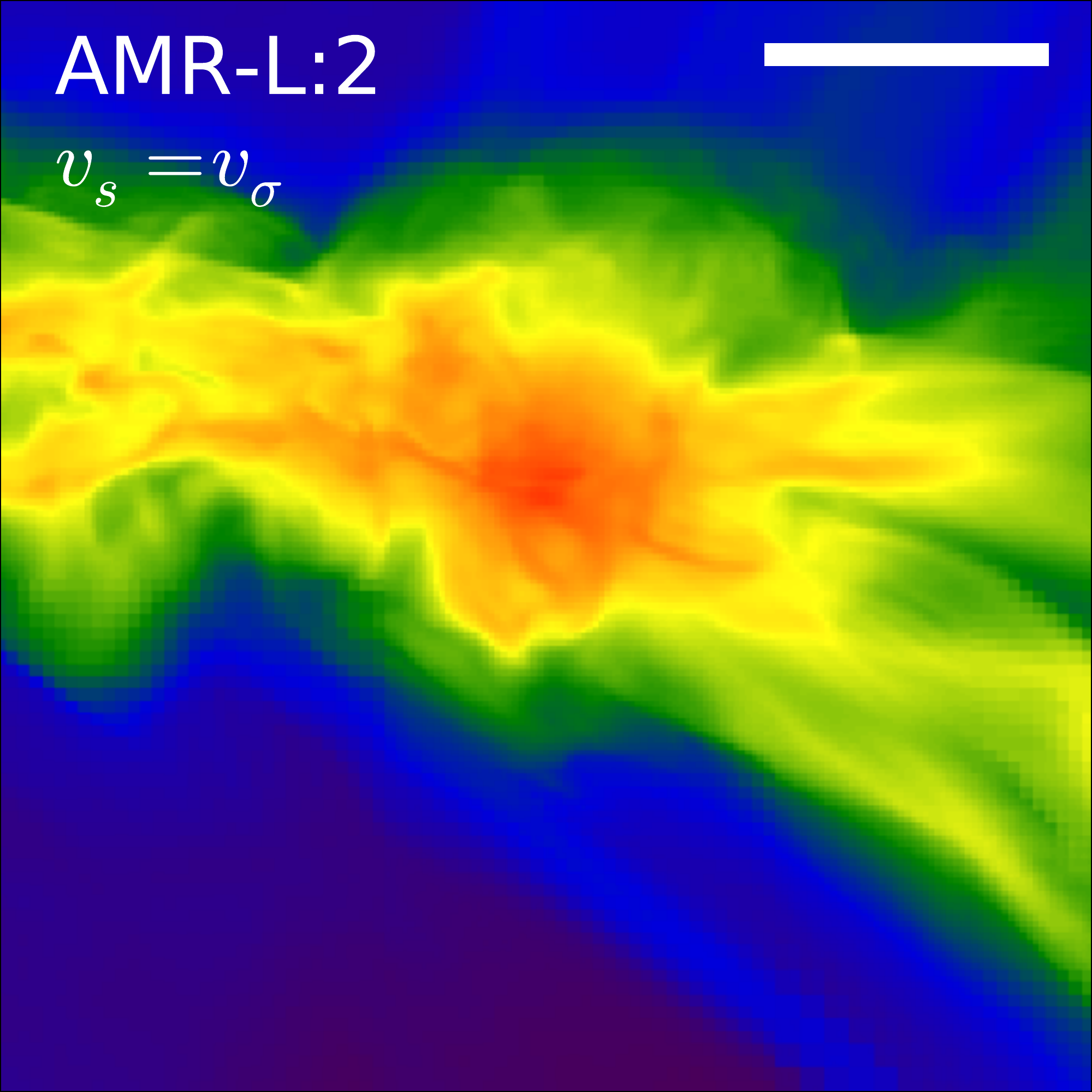} 
\includegraphics[scale=0.21]{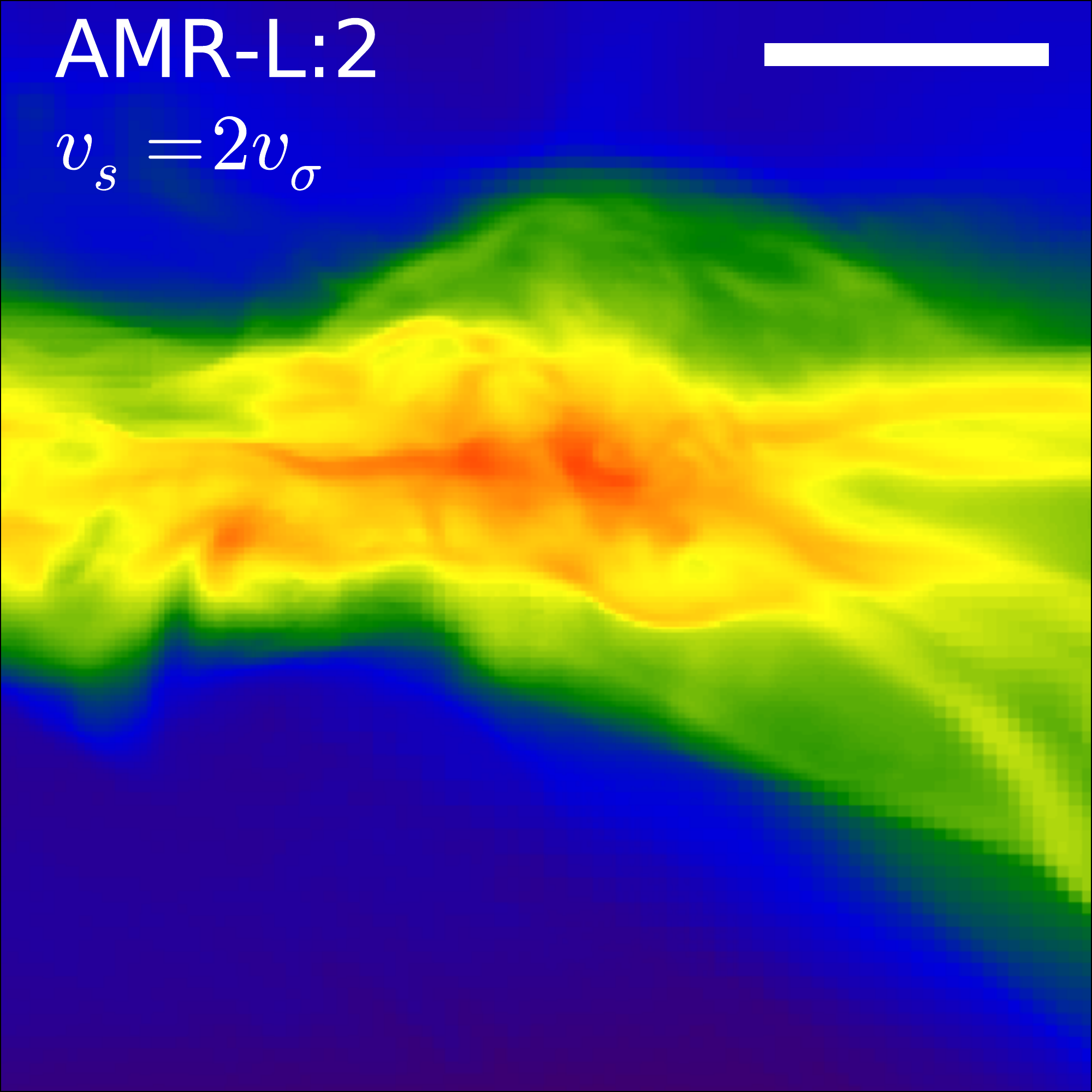} 
\includegraphics[scale=0.1565]{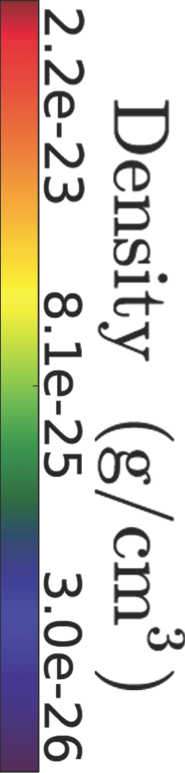} 
\includegraphics[scale=0.21]{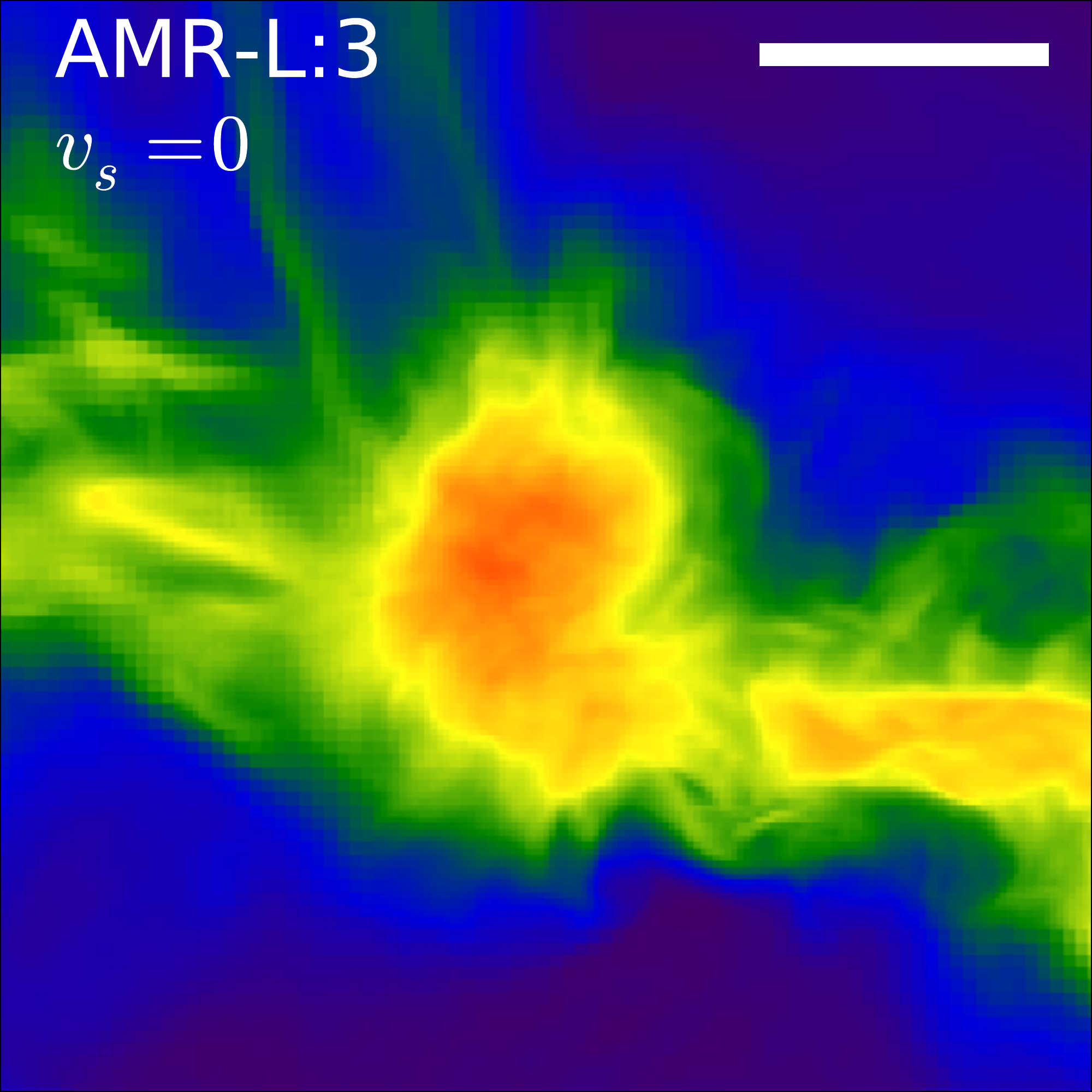}  
\includegraphics[scale=0.21]{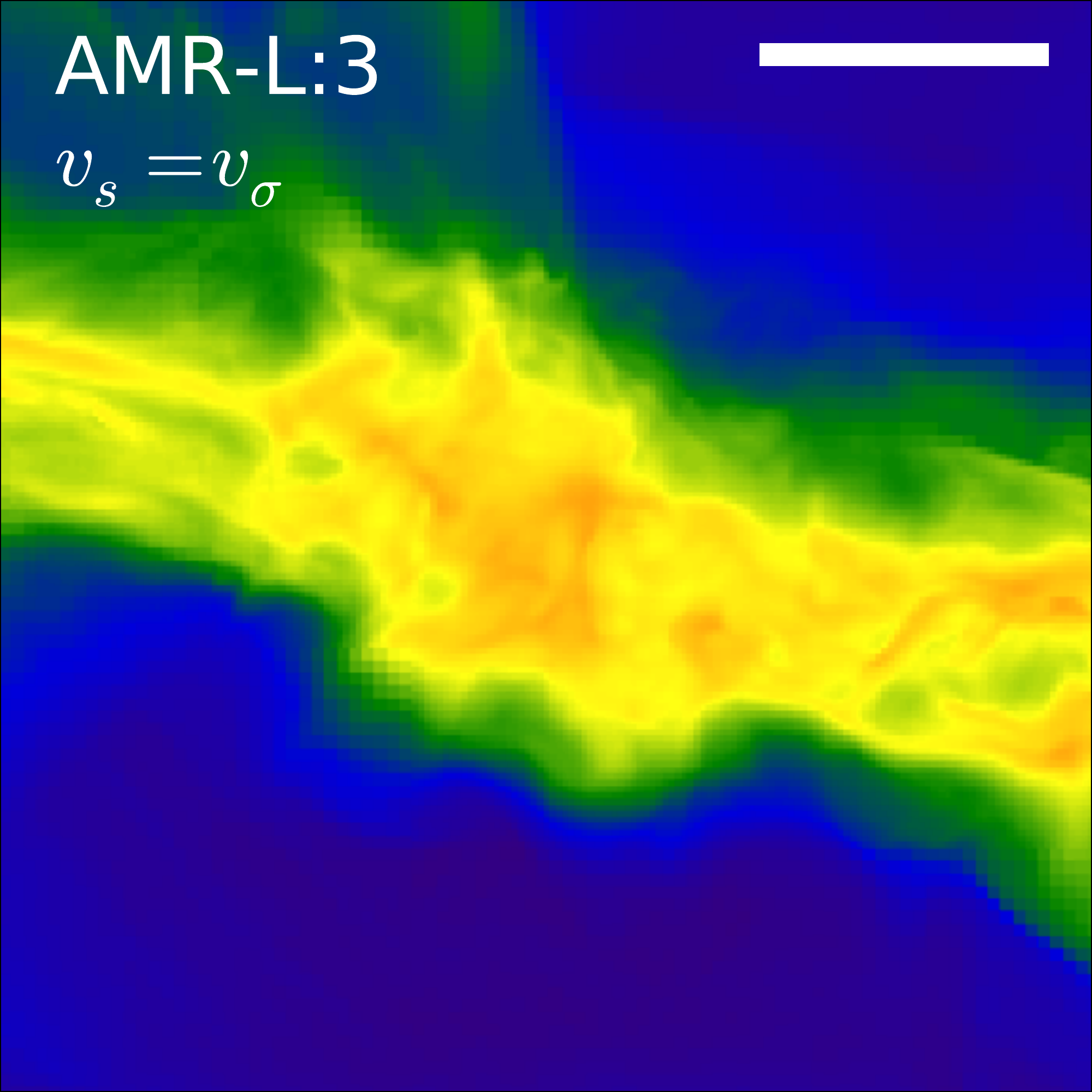}
\includegraphics[scale=0.21]{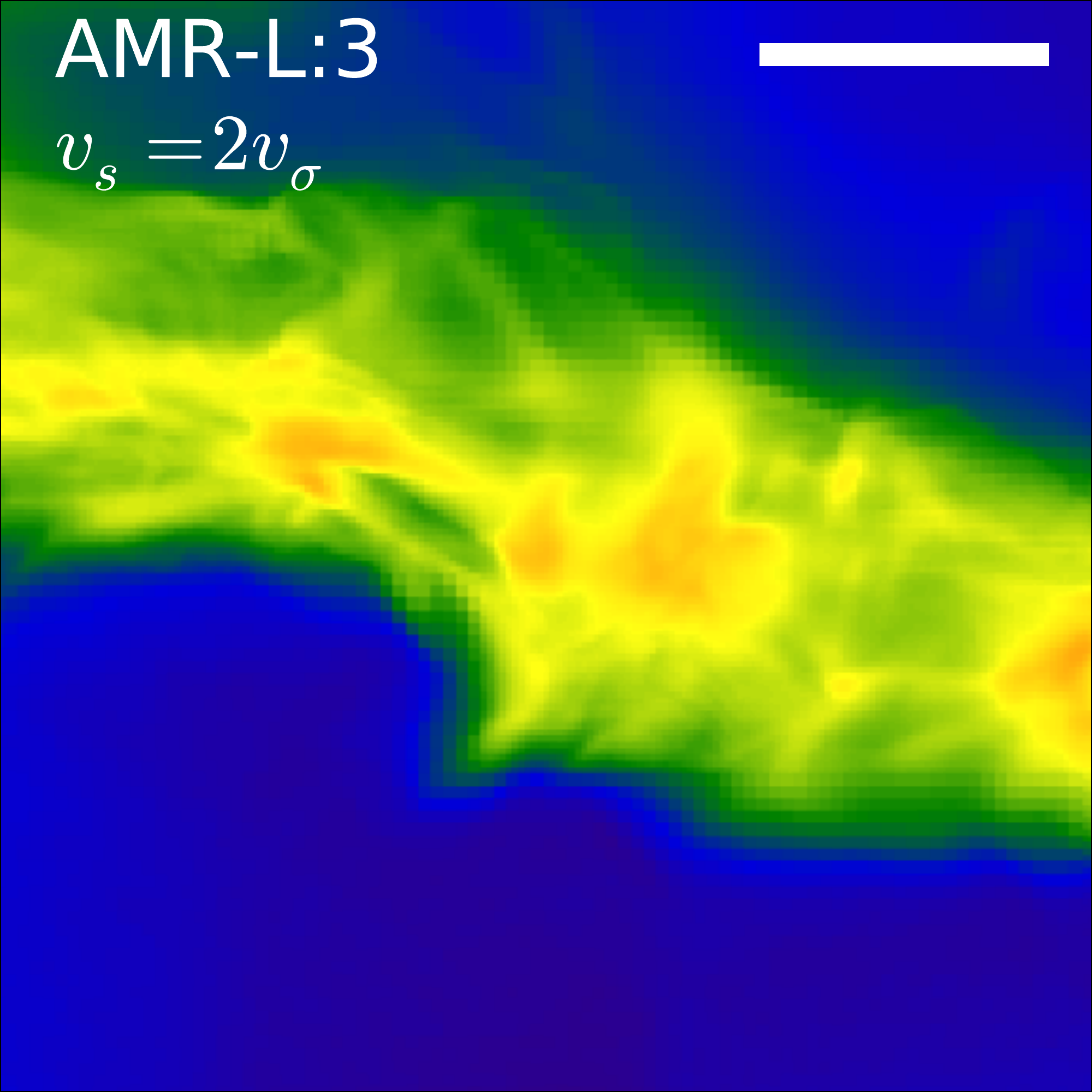}
\includegraphics[scale=0.1565]{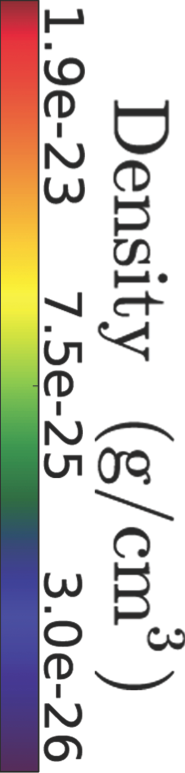} 
\caption{\footnotesize{Gas density projections of the three most massive halos in the $\mathcal{L}_{\rm i}$ AMR simulations, which in the $v_{\rm s}=0$ simulation have masses of $31 \times 10^5 \Msun$ (top row), $13 \times 10^5 \Msun$ (third row), and $8.9 \times 10^5 \Msun$ (fourth row), and the most massive halo in the SPH run (second row).
In all rows,  the left, middle and right columns correspond to  $v_{\rm s} /v_\sigma = 0,1,2$, respectively. In all panels, the white horizontal bar has a physical length of 100 pc, 
and the stream velocity is from the left to the right. This figure is generated using the yt toolkit (Turk et al. 2011, http://yt-project.org/).}}
\label{fig_threeHaloProj_L}
\end{figure*}

We begin by looking at the qualitative effects of a stream velocity on structures at $z=17.18$.  
In both of the $v_{\rm s} =0$ simulations, we identified the three largest halos, all of which had their masses in the most affected mass regime presented by
\citet{Tseliakhovich10}. We then identified these same halos in the $v_{\rm s} \ne 0$ simulations. 
In \fig{fig_threeHaloProj_L} we show gas density projections centered on the
three largest halos in the $\simLi$ simulations, labeled AMR-L:1-3, which in the $v_{\rm s} = 0$ simulation have dark matter masses of 31, 13 and 8.9 $\times 10^5 \Msun$, respectively.
The center of mass of each halo is at the center of each spherical projection, where the spherical region projected has a radius
 eight times larger than the average distance of the halo dark matter particles
from its center of mass. This allows us to examine the effect of
the stream velocity on a range of scales. 

Here we see that, in the presence of a stream velocity, structure is less evolved on all scales. 
Within the halo, the overall gas mass is reduced and the peak gas density is decreased.
Outside the halos,  the accretion lanes have lower densities and are displaced further downwind.
By comparing the structures around the two most massive halos, we can also see the dependence of streaming effects on orientation.
In the top row, where two accretion lanes are normal to the stream velocity, we see a dramatic reduction in structure as $v_s$ moves from 0 to 2$v_\sigma$, while in 
the middle row,  where the accretion lanes are parallel to the stream velocity, we see that structure reduction is much more moderate.

A comparison between  the first two rows in \fig{fig_threeHaloProj_L} illustrates that 
 differences between the SPH simulations and the AMR simulations are mostly minor, which
gives us confidence that our mapping tool reproduces the SPH simulation reasonably well in the AMR initial conditions.
Low density
gas in the AMR simulations appears slightly smoother, since it is better able to resolve these regions. In particular, in AMR schemes, low density accretion lanes look more defined
in the cases with non-zero stream velocity, showing that these structures require significant resolution to accurately model
their evolution. Furthermore, the SPH simulations appear slightly denser in the core of the halo. This may be due to over-efficient cooling in the SPH models, which we discuss further below.  Note that here and in all comparisons with the SPH runs, we have  mapped the SPH data to the AMR grid before working with it, so the analysis is identical. 

\begin{figure*}[t!] 
\centering
\includegraphics[scale=0.21]{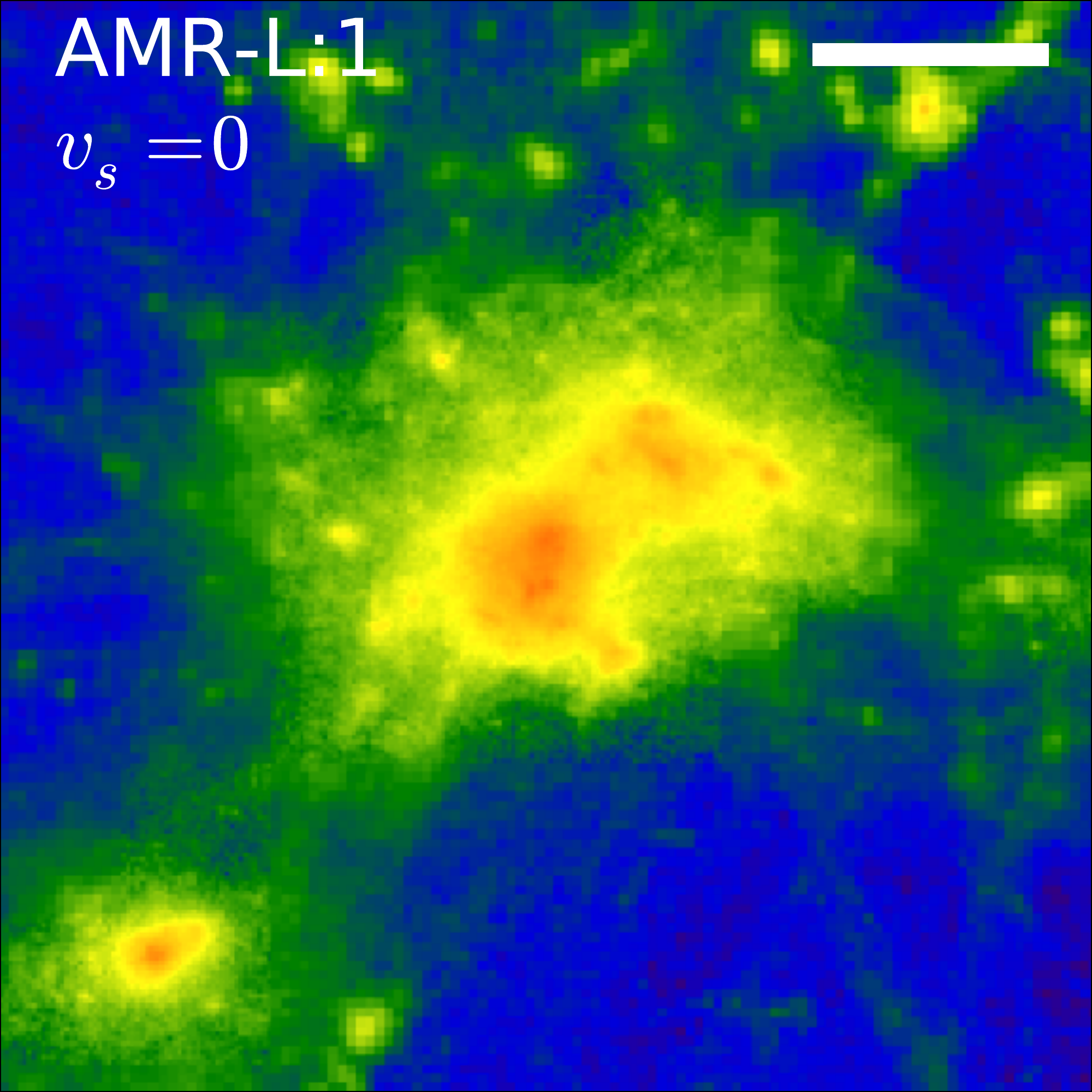} 
\includegraphics[scale=0.21]{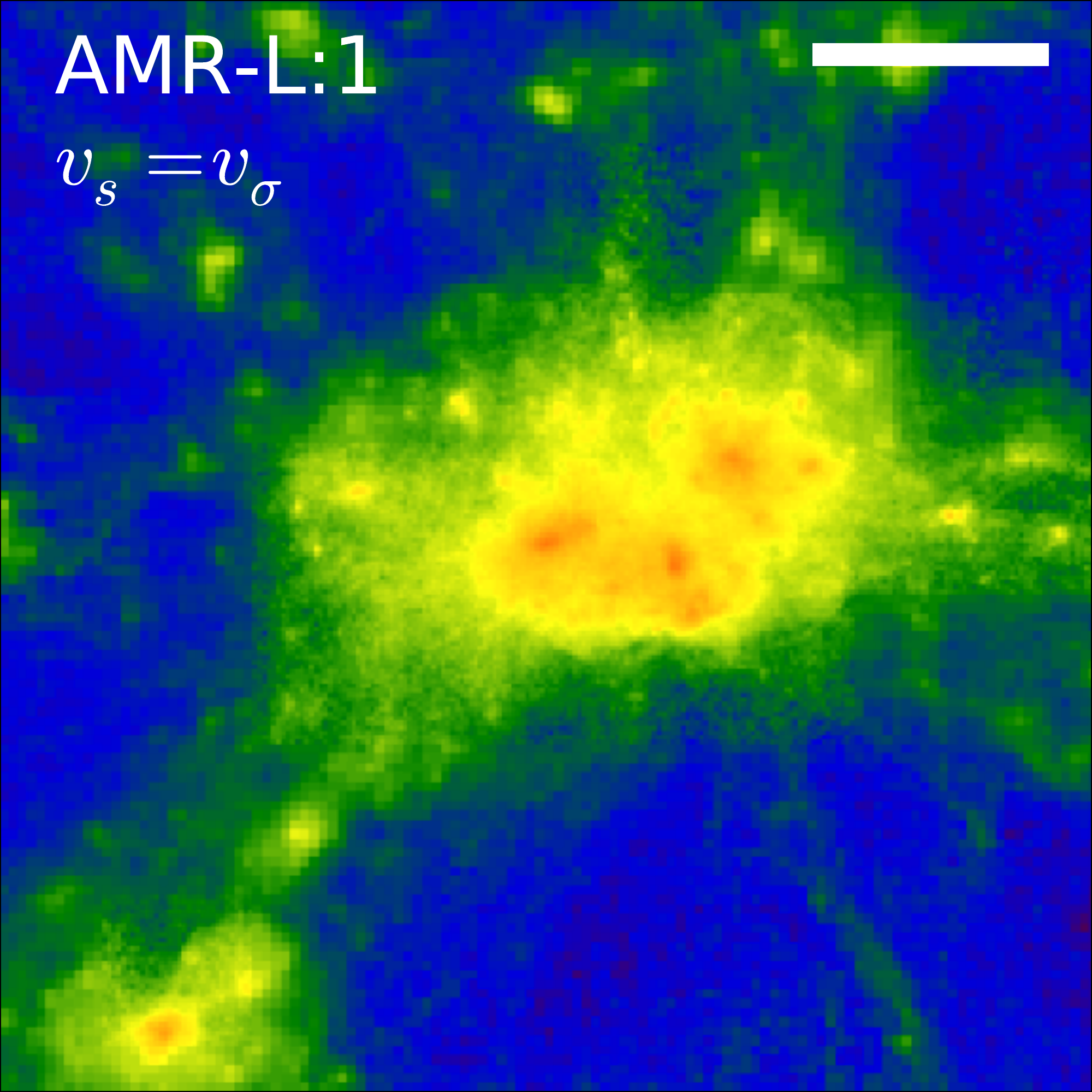} 
\includegraphics[scale=0.21]{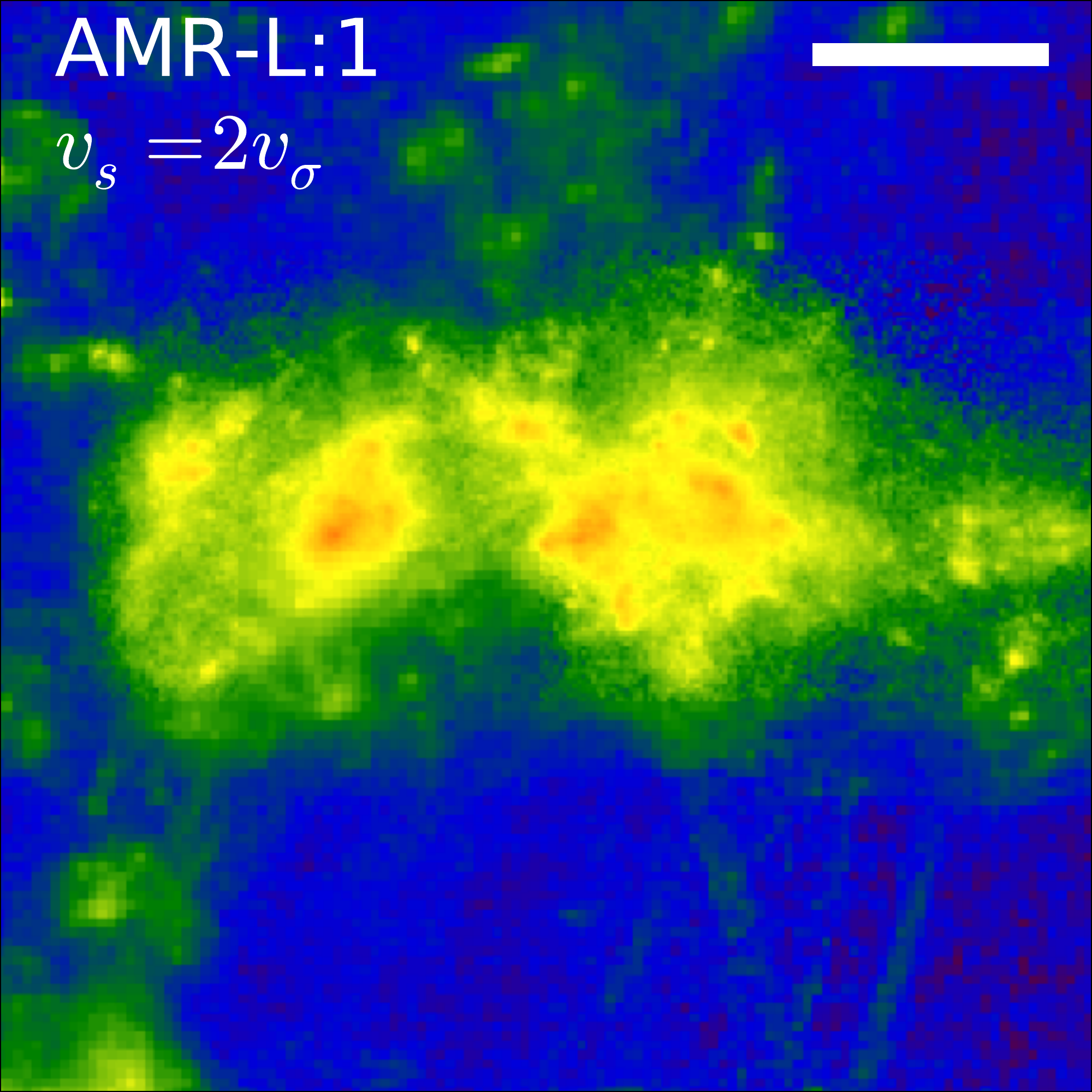} 
\includegraphics[scale=0.1565]{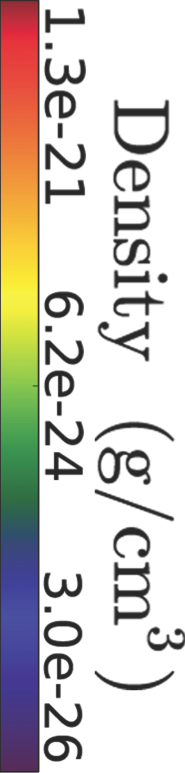} 
\includegraphics[scale=0.21]{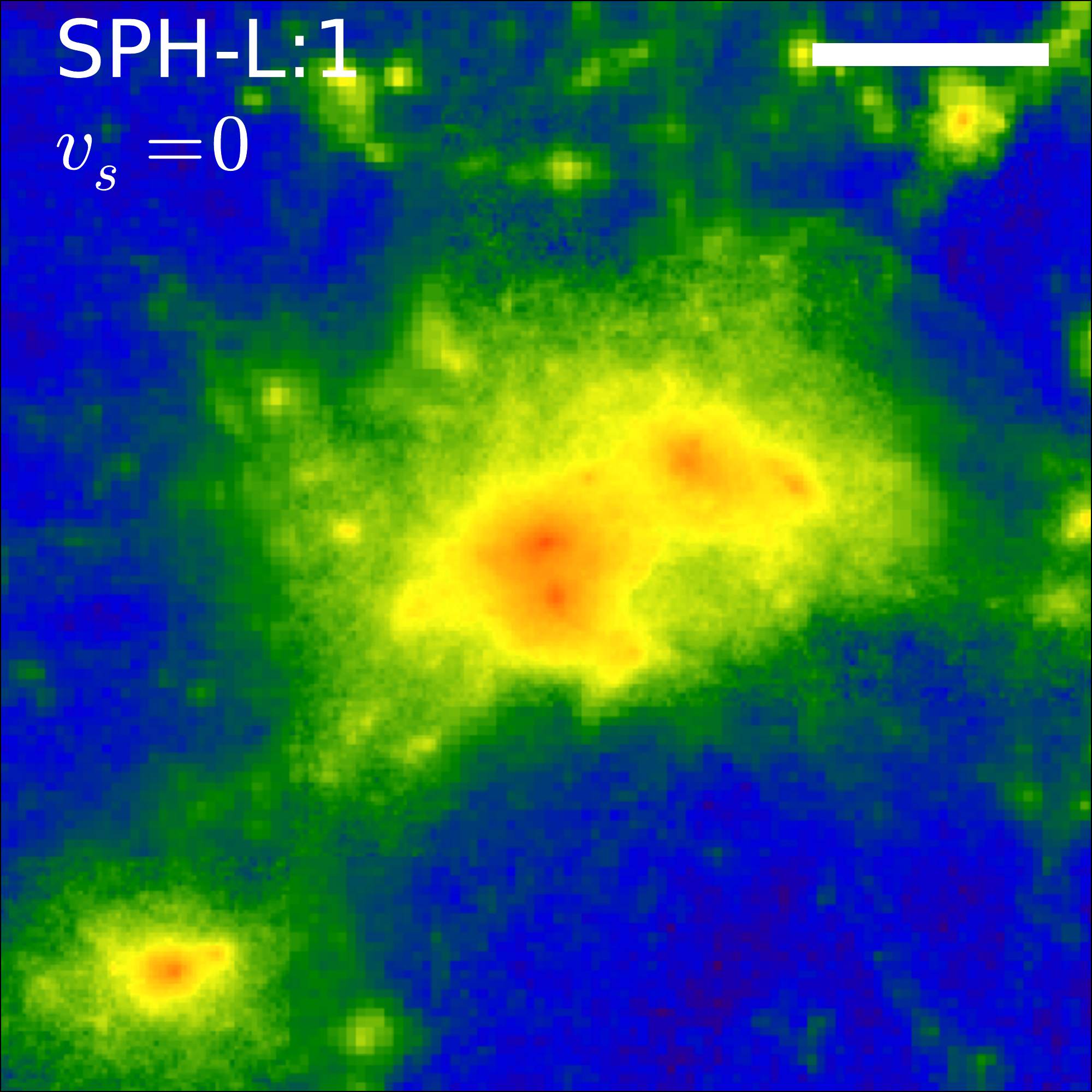} 
\includegraphics[scale=0.21]{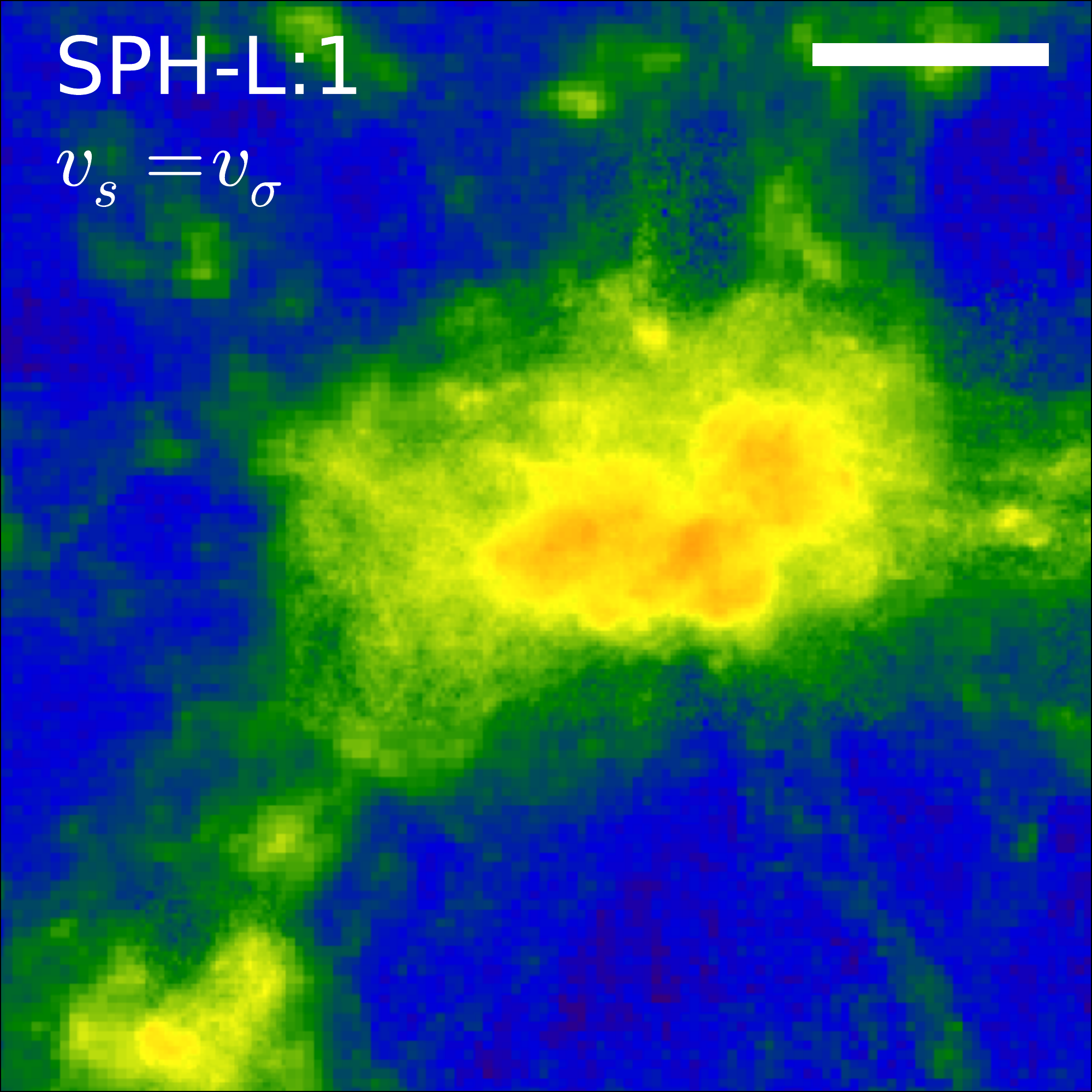} 
\includegraphics[scale=0.21]{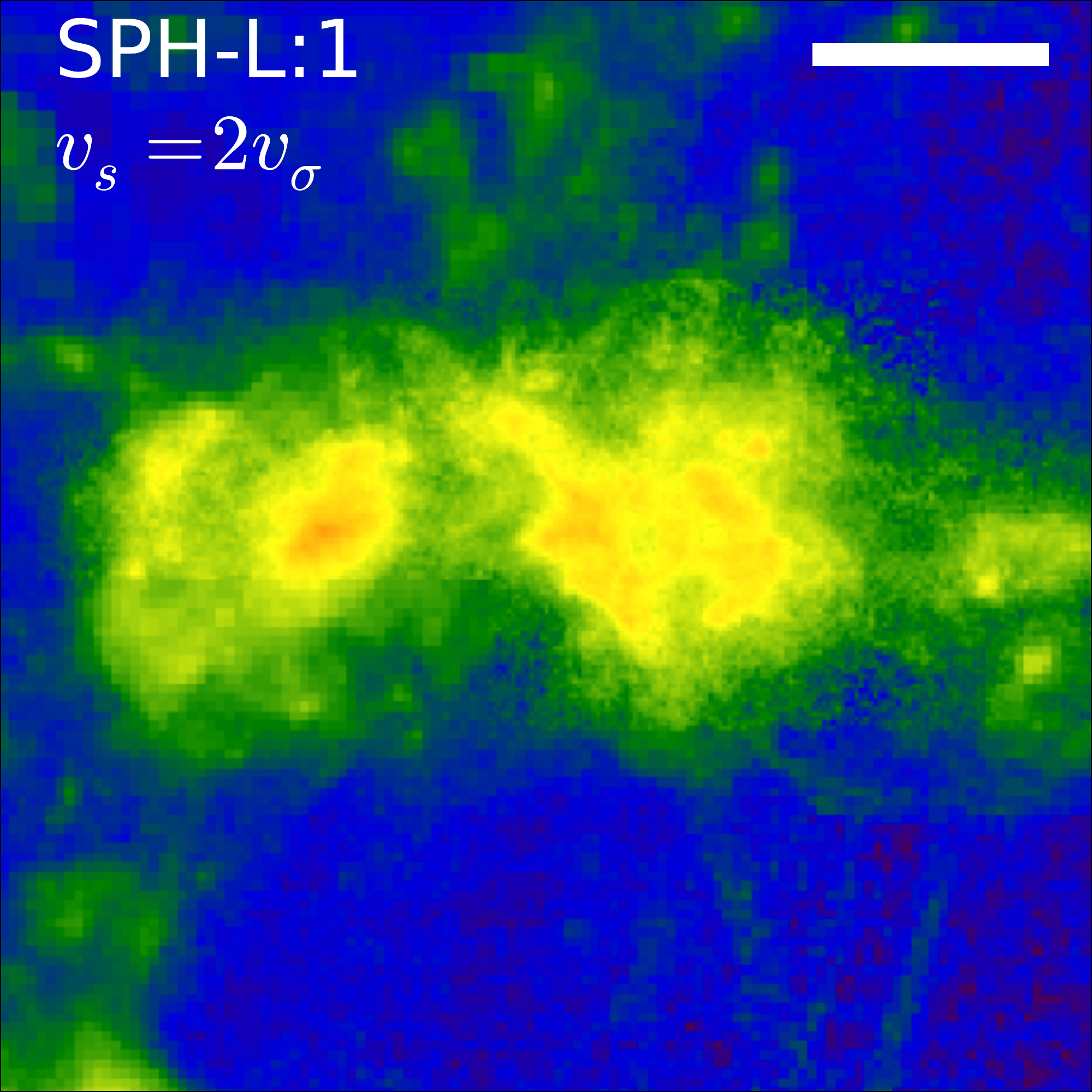} 
\includegraphics[scale=0.1565]{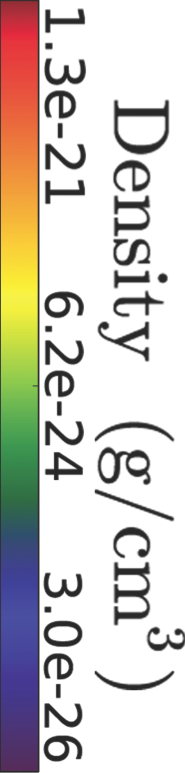} 
\includegraphics[scale=0.21]{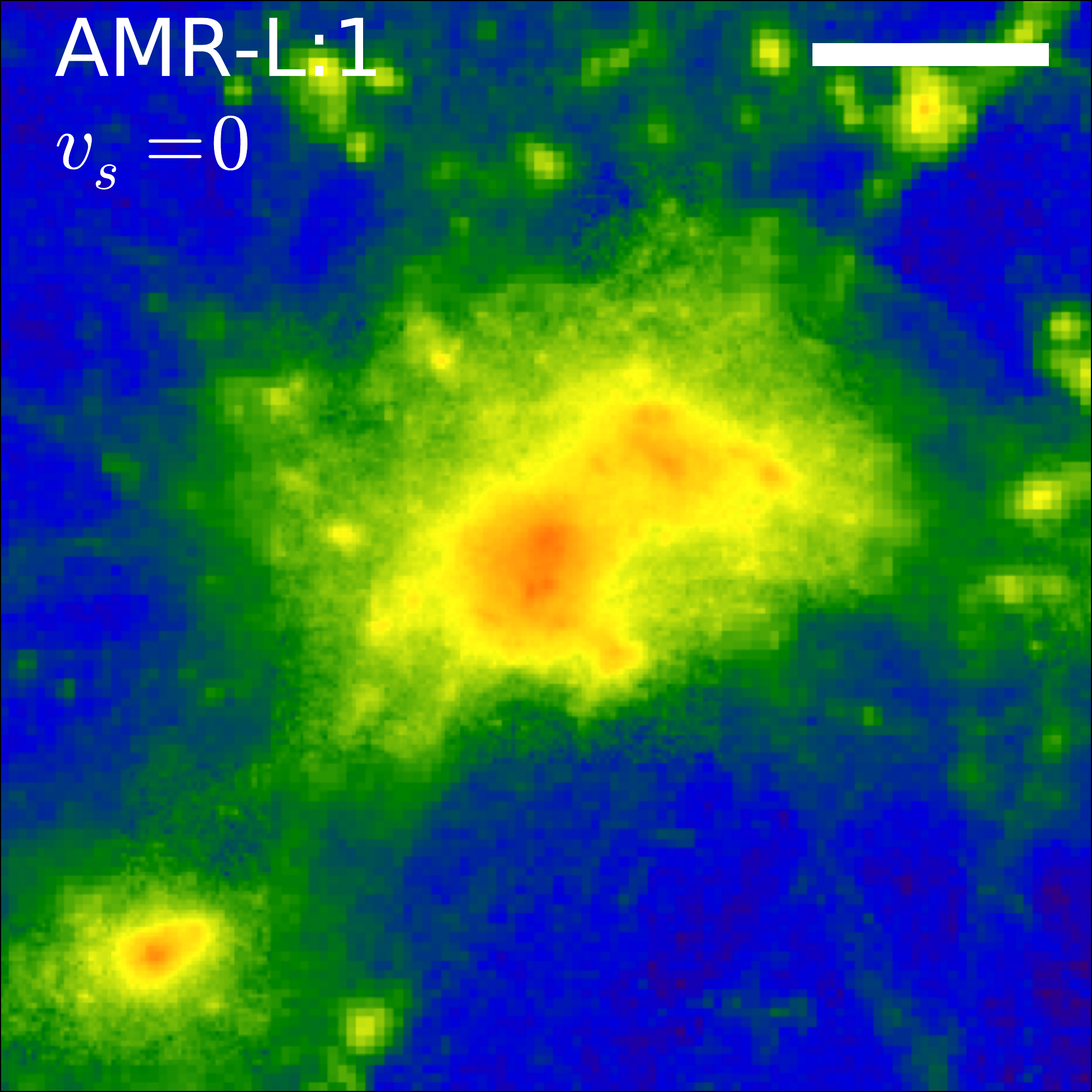} 
\includegraphics[scale=0.21]{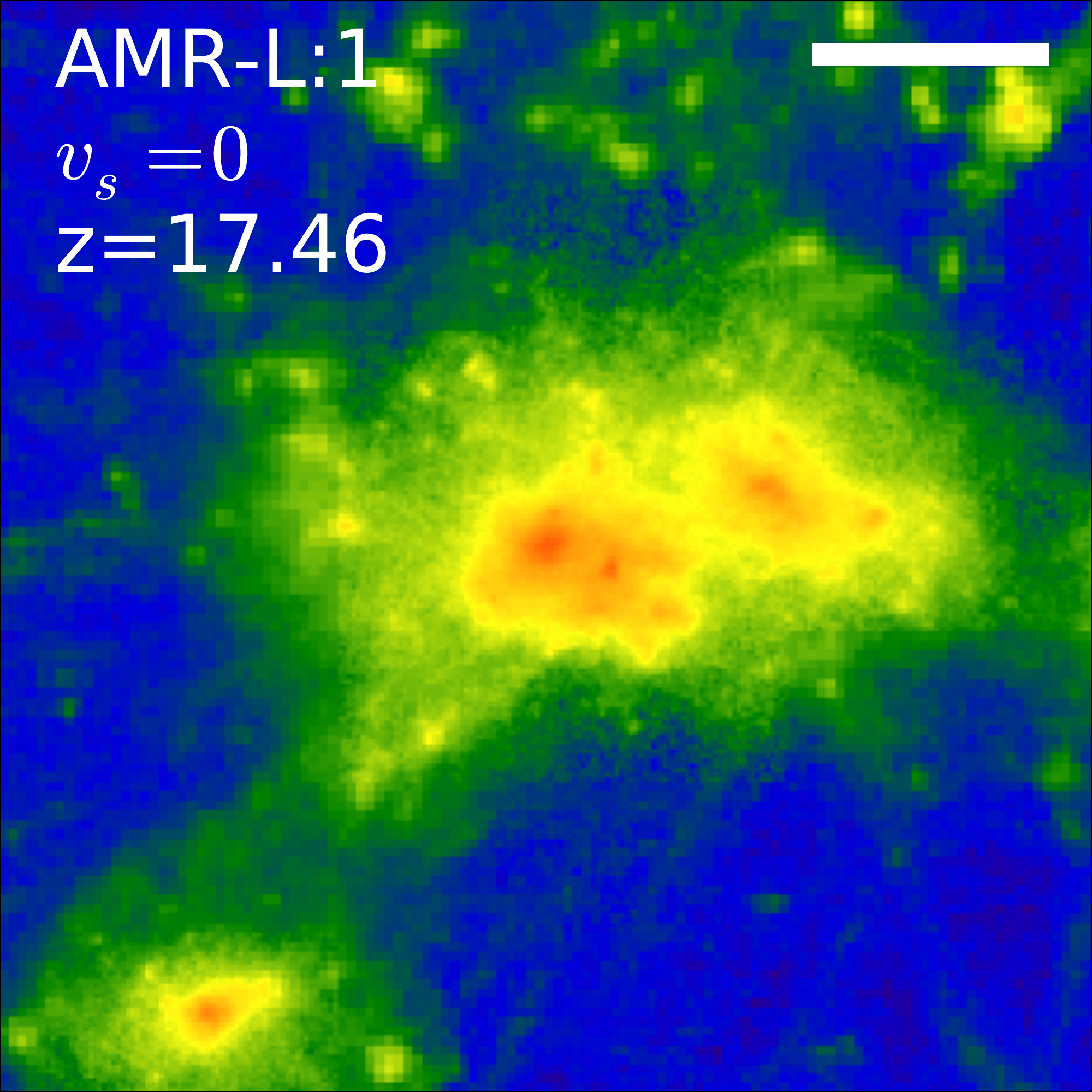} 
\includegraphics[scale=0.21]{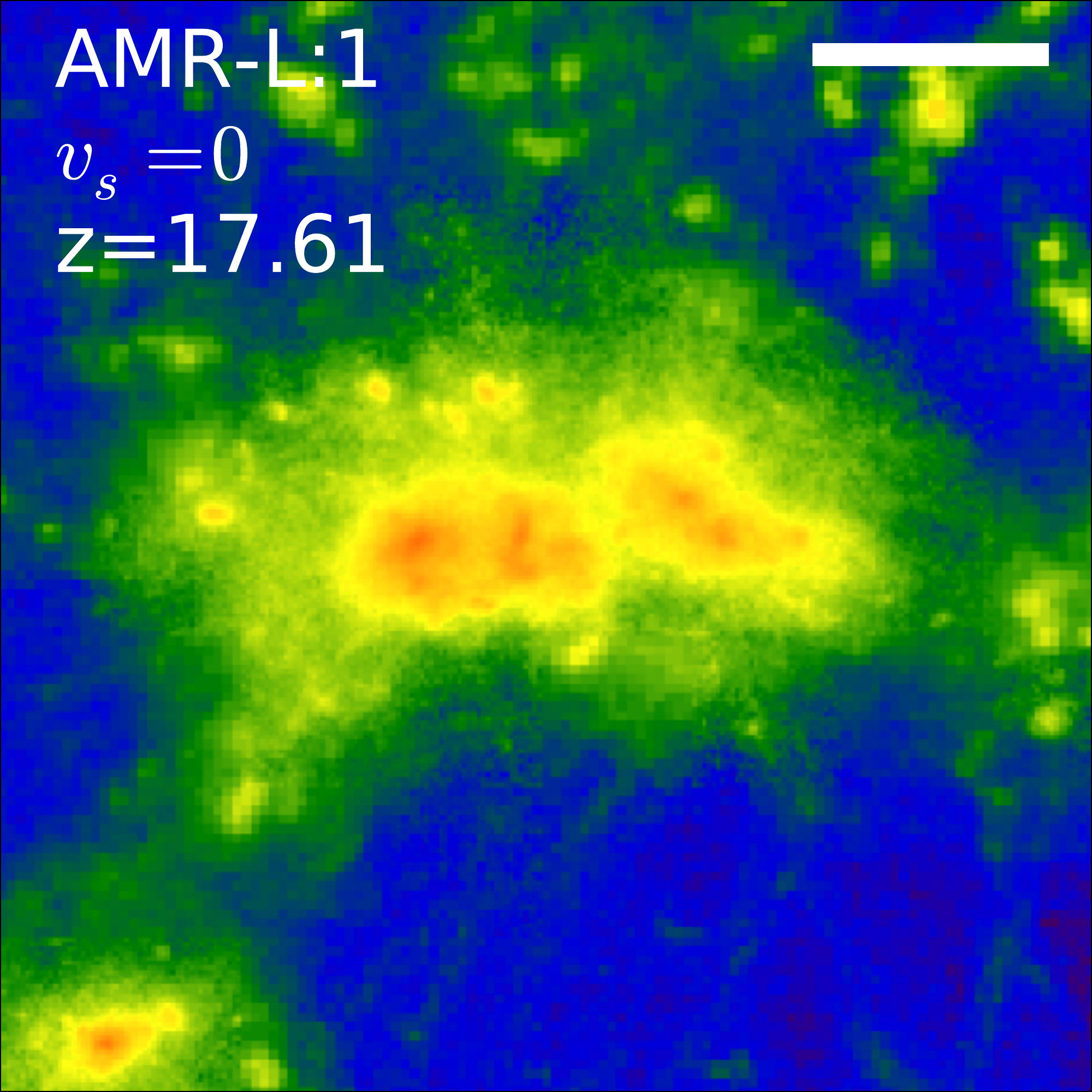} 
\includegraphics[scale=0.1565]{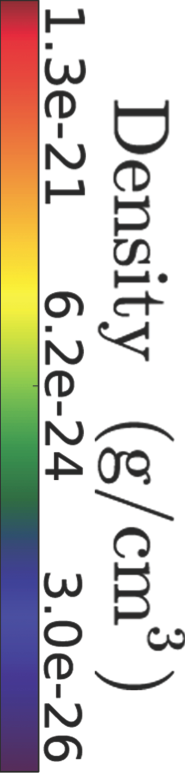} 
\caption{\footnotesize{Dark matter projections of the largest halo in the $\mathcal{L}_{\rm i}$ AMR simulations (top row) and SPH simulations (middle row), at $z=17.18$ with the left, middle and right image corresponding to  $v_{\rm s} /v_\sigma = 0,1,2$, respectively. The white horizontal bar has a physical length of 100 pc. The bottom row shows projections
of dark matter in the $\simLA$ run at  $z=17.18$, $z=17.45,$ and $17.60,$ for comparison with the $v_{\rm s} \ne 0$ runs.}}
\label{fig_HaloProj_L_p}
\end{figure*}

In \fig{fig_HaloProj_L_p}, we show projections of the largest halo's dark matter distribution for the $\simLi$ AMR and SPH simulations at $z=17.18$. The dark
matter is near identical between the two methods, and in both methods the presence of a stream velocity has only a minor effect on the dark matter, resulting in a slightly delayed evolution in the core of the halo. This is possibly caused by the reduced gravitational potential resulting from the missing gas. To show this, we plot earlier epochs from the 
AMR $\simLA$ simulation in the bottom row. We can see some similarities
between the core at $z=17.45$ and the $z=17.18$ $\simLB$ simulation, while the $z=17.18$ $\simLC$ simulation appears to be even less evolved than the $z=17.60$ projection. This suggests that the stream velocities can be thought of as a damping term, delaying the collapse of the densest structures.

\begin{figure*}[t!] 
\centering
\includegraphics[scale=0.21]{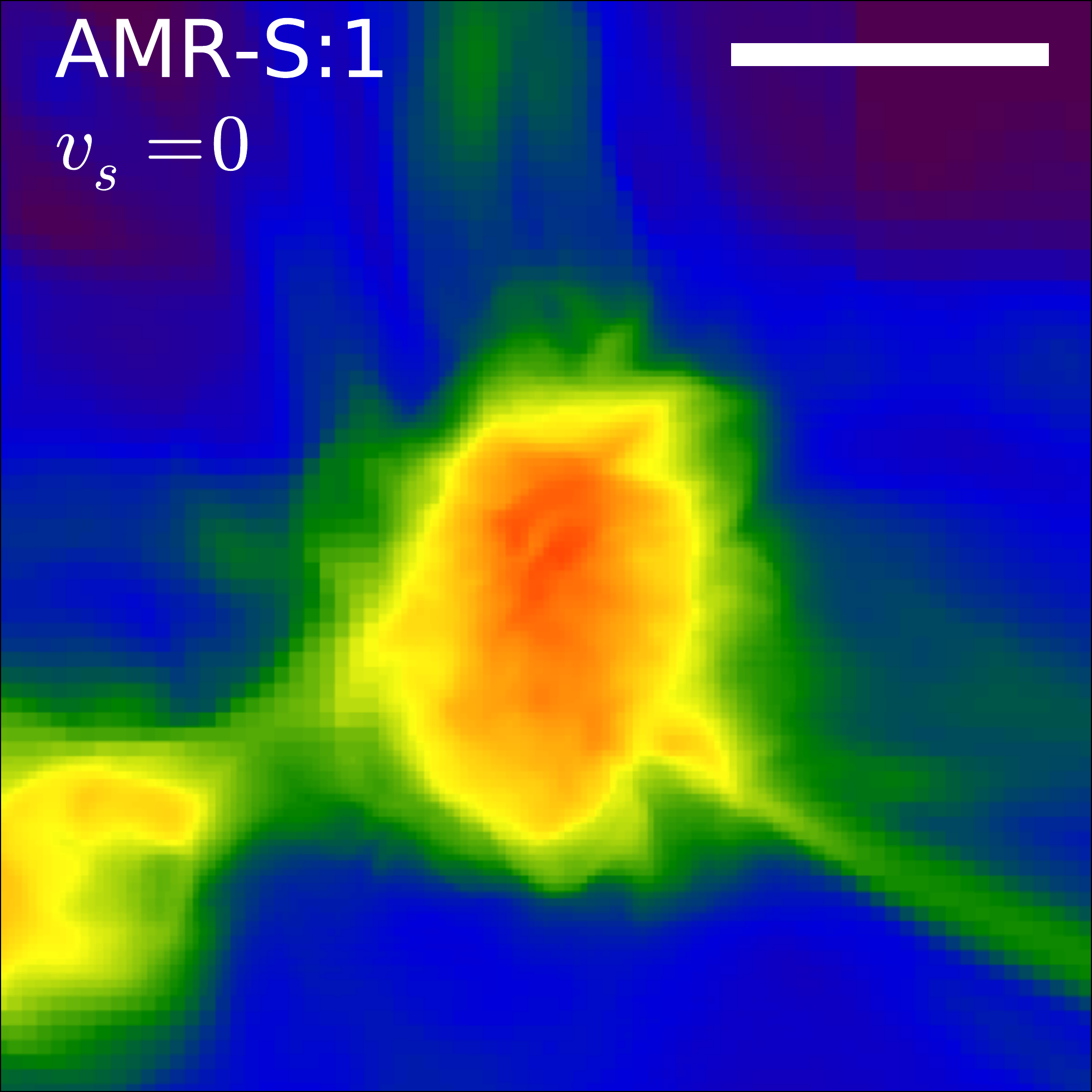} 
\includegraphics[scale=0.21]{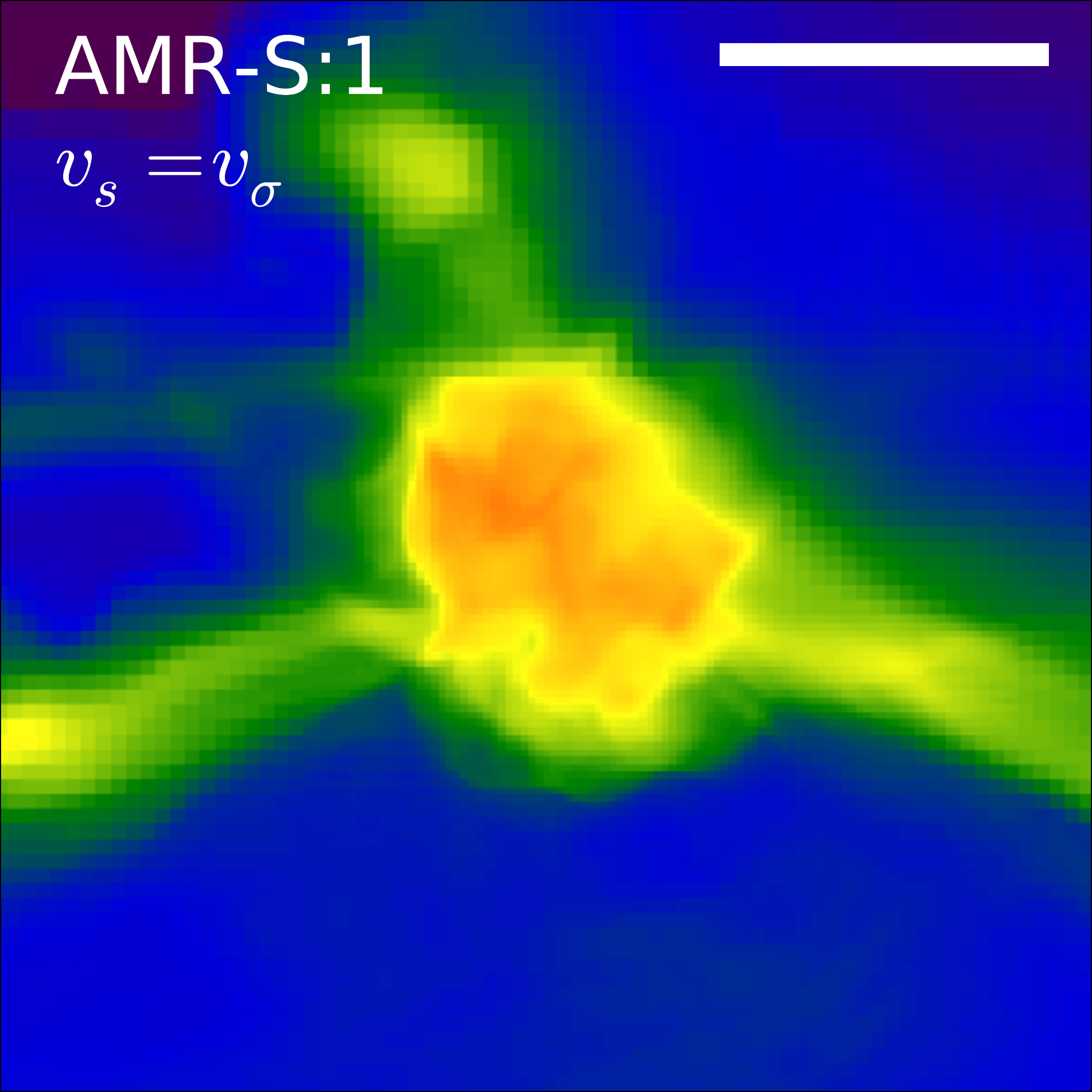} 
\includegraphics[scale=0.21]{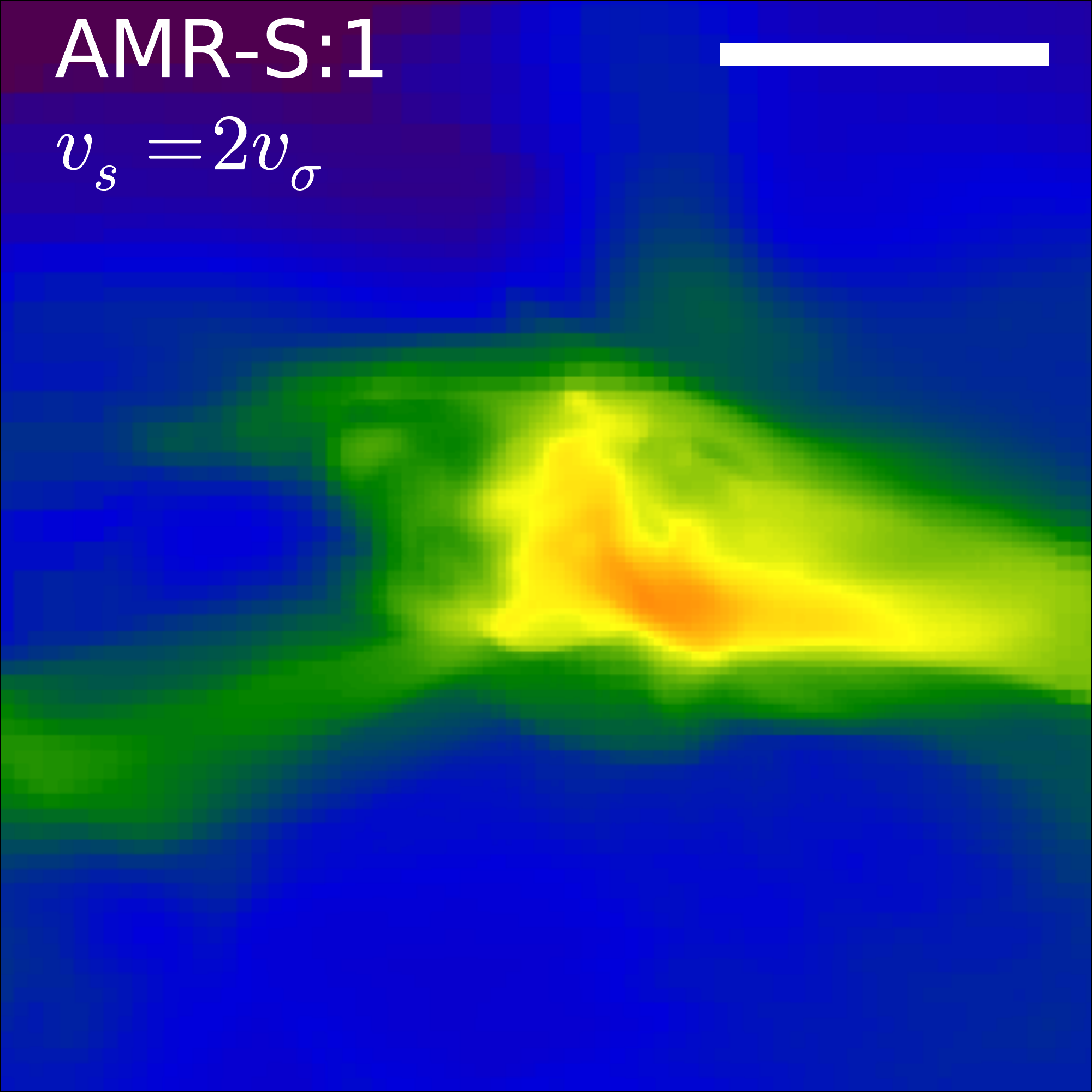} 
\includegraphics[scale=0.1565]{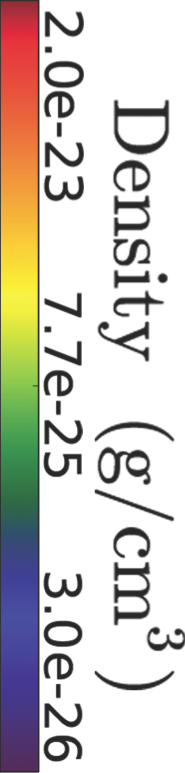} 
\includegraphics[scale=0.21]{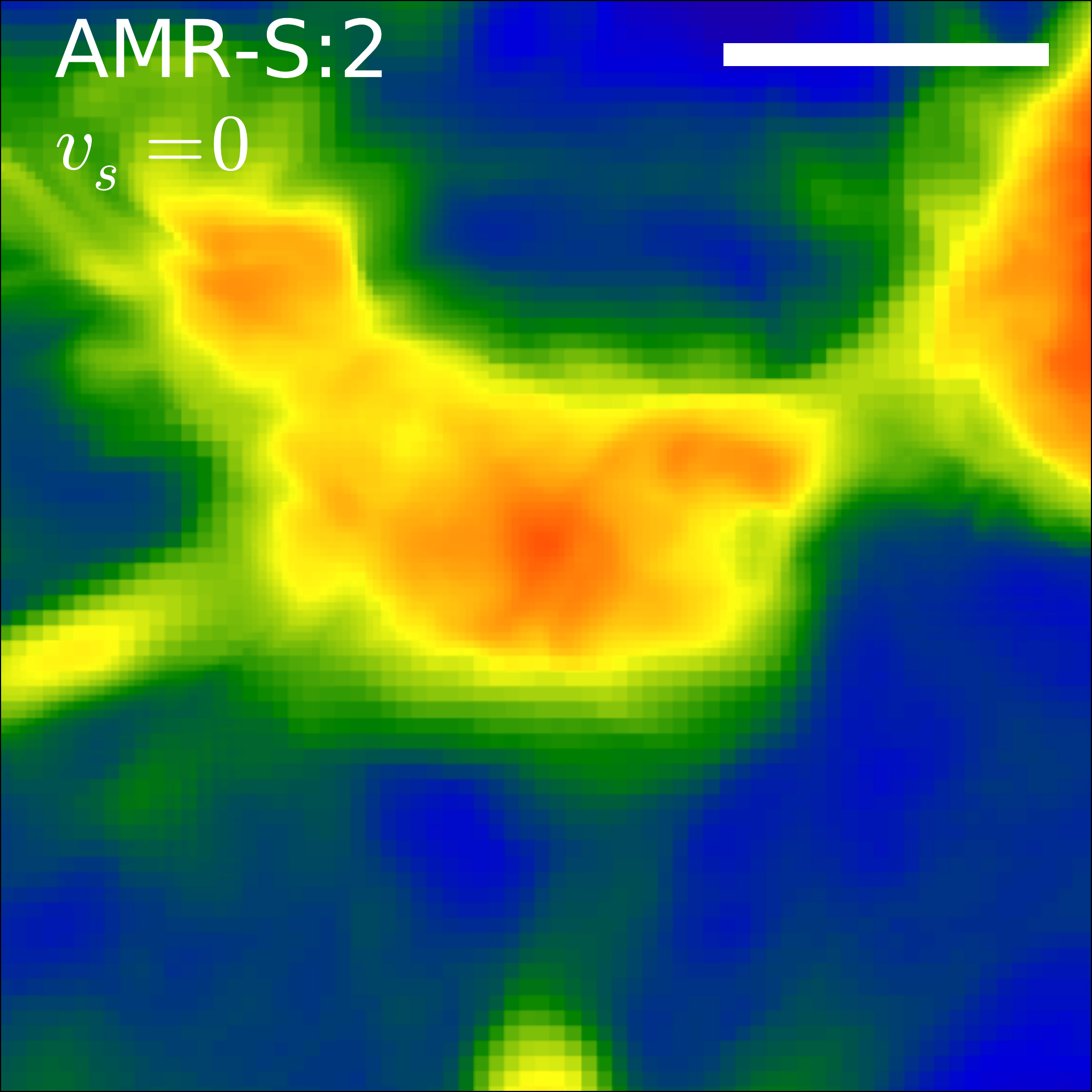} 
\includegraphics[scale=0.21]{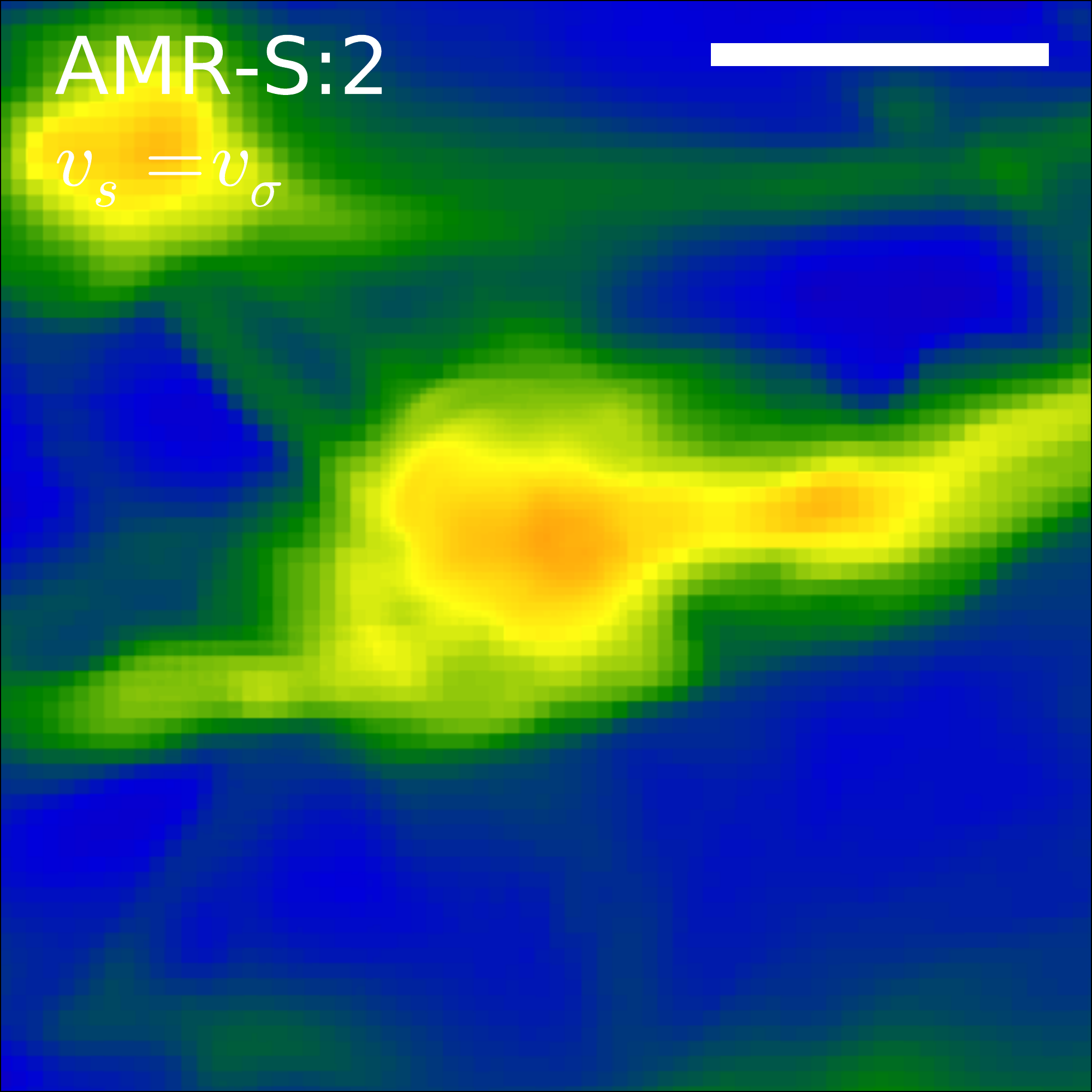} 
\includegraphics[scale=0.21]{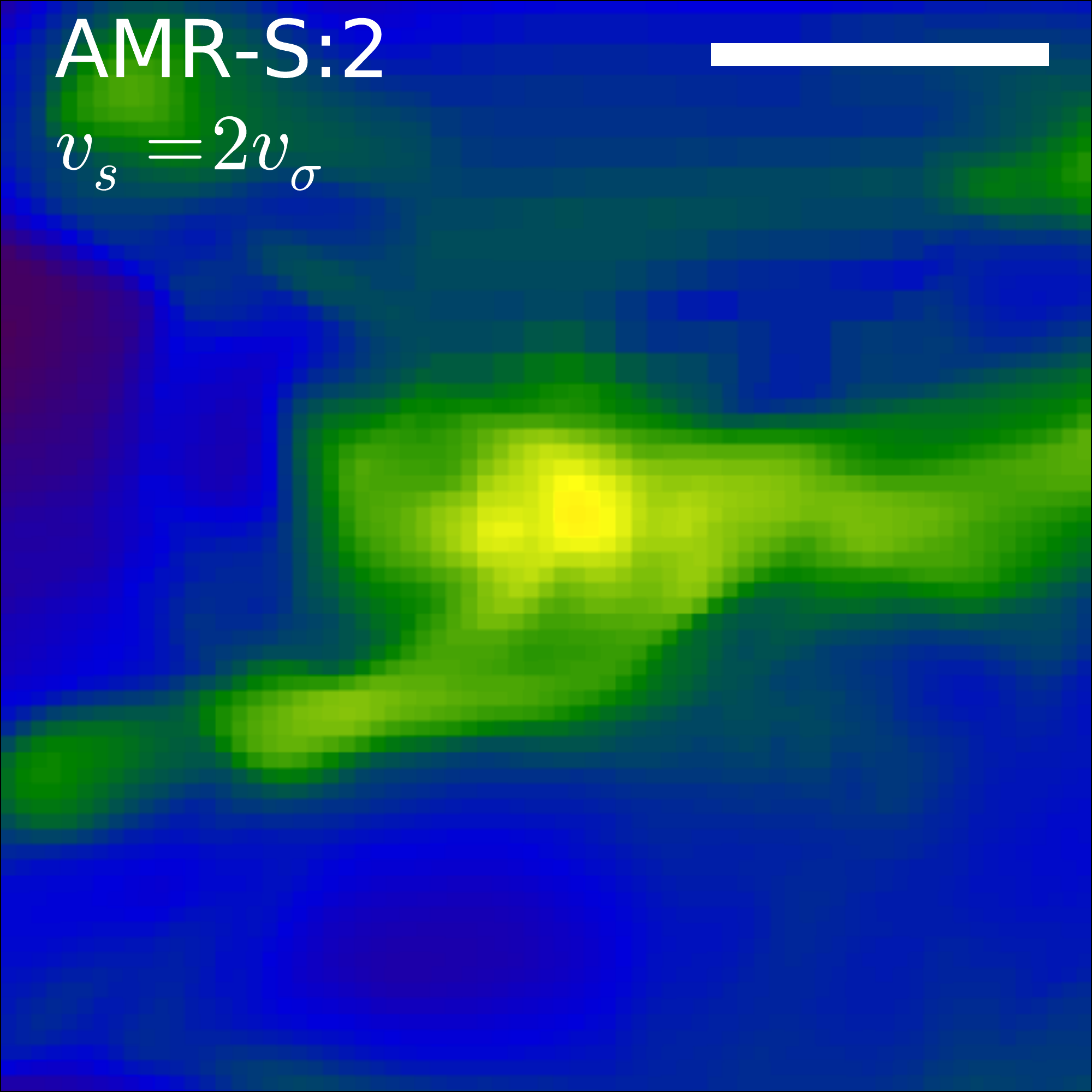} 
\includegraphics[scale=0.1565]{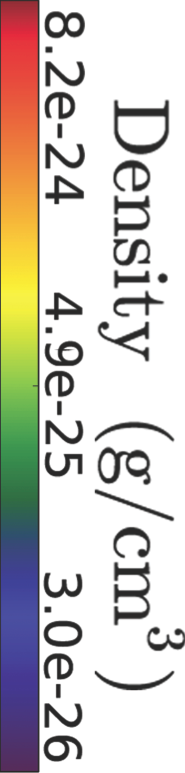} 
\includegraphics[scale=0.21]{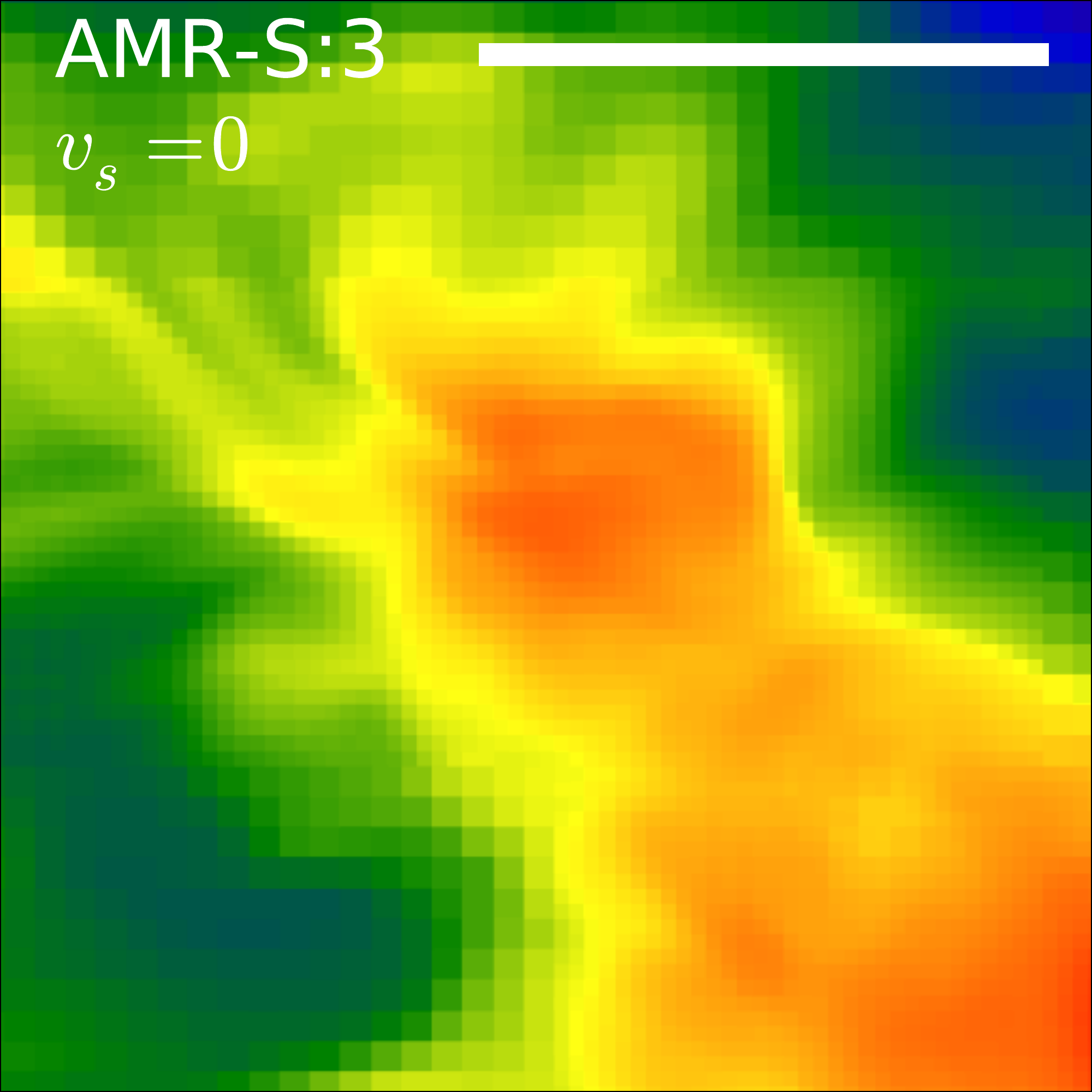}  
\includegraphics[scale=0.21]{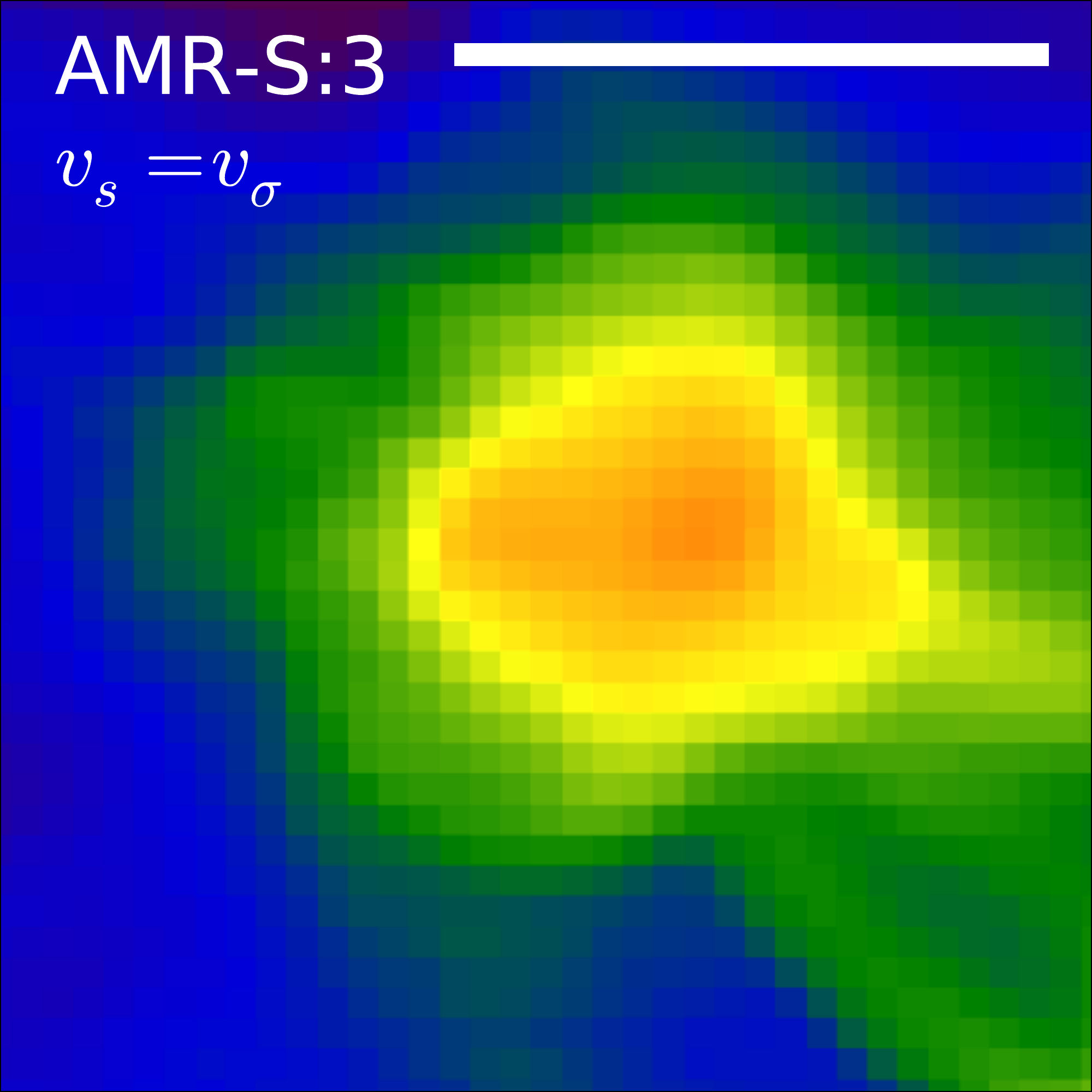}  
\includegraphics[scale=0.21]{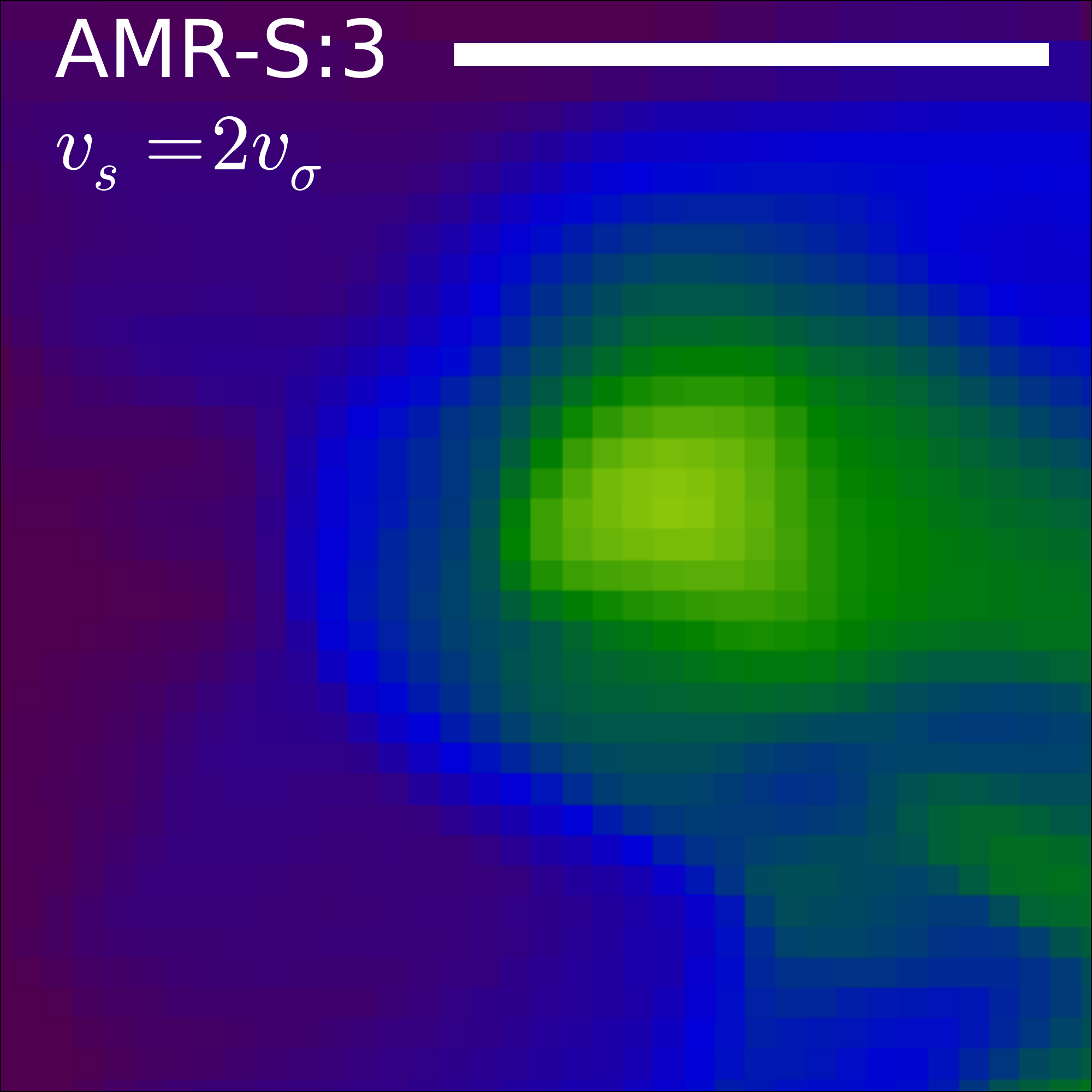}  
\includegraphics[scale=0.1565]{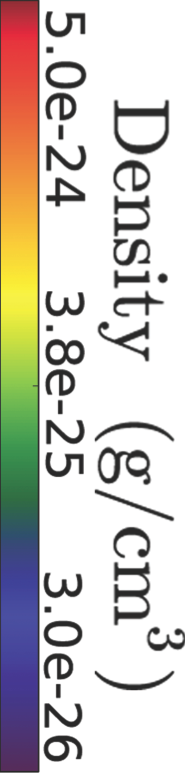} 
\caption{\footnotesize{Density projections of the three largest halos in the $\mathcal{S}_{\rm i}$ simulations, with the top, middle
and bottom rows corresponding to  the first, second, and third largest dark matter halos in the $v_{\rm s} =0$ simulation, respectively, and the left, middle, and right columns
corresponding to  $v_{\rm s} /v_\sigma = 0,1,2$, respectively.  The white horizontal bar has a physical length of 100 pc.   Note that the S:3 halo is a companion to halo S:2, and also appears in the upper-left corner of the plots in the second row.}}
\label{fig_threeHaloProj_S}
\end{figure*}

In \fig{fig_threeHaloProj_S}, we show gas density projections centered on the three largest halos in the $\mathcal{S}_{\rm i}$ simulations, labeled AMR-S:1-3. These halos have dark matter masses of 8.0, 2.5 and 0.88 $\times 10^5 \Msun$ in the $v_{\rm s} = 0$ case. Again we see a significant reduction in structure on all scales, and  given the weaker gravitational potentials involved,  this is even more dramatic than in the $\mathcal{L}_{\rm i}$ run.  Thus for example, the streaming velocity is able to  remove almost all of the gas from the 0.88 $\times 10^5 \Msun$ halo in the $v_{\rm s}  = 2 v_\sigma$ run, and the 
filament connecting S:2 and S:3, whose density is $\approx 10^{-24}$ g cm$^{-3}$ in the $v_s = 0$ run, is extremely tenuous in the $v_s = 1$ run, and absent almost completely in the $v_{\rm s}  = 2 v_\sigma$ run.  Finally, comparisons between the dark matter distributions between the runs, and comparisons between the AMR and SPH simulations reveal the same conclusions as for the $\simLi$ simulations, and thus we omit these plots for brevity.

To quantify the effects of the stream velocity, we followed each of these six halos through time and determined at what point their total matter density reached a threshold density of $\rho_{\rm thresh}=10^{-22}\gcc$, corresponding to roughly one order of magnitude below the densest structures that can form for $10^{11} \mbox{ erg g}^{-1}$ gas. We also find at what point their gas density reached a threshold density of $3\times10^{-24}\gcc$. The redshift at which these thresholds are reached is given in \tabl{tab_thresh}, except for the two smallest halos for non-zero stream velocities, which do not reach the threshold density. For these halos we include the peak density reached. 
Also included are the masses, virial radius as given by the HOP group finder \citep{Eisenstein98} included in the yt visualization and analysis toolkit (Turk et al. 2011, http://yt- project.org/),,
and the virial radius assuming a spherically uniform overdensity of 200. 
Smoothing lengths were determined for the dark matter particles by the HOP group finder. These lengths were given by the distance to the 49 nearest particles. 
\begin{deluxetable*}{lrrrcrll}[b!]
\tabletypesize{\scriptsize}
\tablewidth{0pc}
\tablecaption{Redshift of Halo Collapse$^*$
\label{tab_thresh}}
\startdata
Simulation: &  & $\mathcal{L}_{\rm i}$ &  &  \ \ &  & $\mathcal{S}_{\rm i}$ &  \\ \hline \hline
M$_{\rm DM}$ ($10^5\Msun$) & 31 & 13 & 8.9 & \ \ & 8.0 & 2.5 & 0.88 \\
$r_v$ (pc) & 240 & 190 & 200 & \ \ & 170 & 170 & 89 \\
$r_{v,{\rm sph}}$ (pc) & 270 & 200 & 180 & \ \ & 170 & 120 & 83 \\ \hline
$\rho_{\rm gas} >3\times 10^{-24} \gcc$ \\
$z(v_{\rm s} = 0)$ & 30.2 & 30.8 & 26.6 &  \ \ & 24.6 & 28.0 & 21.8 \\
$z(v_{\rm s} = v_\sigma) $ & 25.1 & 26.4 & 21.9 &  \ \ & 21.9 & 22.8 & $2.63^a$\\
$z(v_{\rm s} = 2v_\sigma) $ & 20.9 & 18.5 & 20.3 &  \ \ & 18.7 & $1.65^a$ & $0.65^a$ \\ \hline
$\rho_{\rm tot} >10^{-22} \gcc$ \\ 
$z(v_{\rm s} = 0)$ & 27.4 & 29.3 & 22.6 &  \ \ & 22.9 & 25.3 & 19.8 \\
$z(v_{\rm s} = v_\sigma) $ & 24.7 & 26.3 & 21.9 &  \ \ & 21.9 & 23.4 & $83^a$ \\
$z(v_{\rm s} = 2v_\sigma) $ & 23.7 & 22.9 & 21.4 &  \ \ & 20.2 & 20.7 & $78^a$
\enddata
\tablenotetext{a}{If threshold value not reached by $z=17.2$, peak density reached by that halo is given ($10^{-24}$ g cm$^{-3}$).}
\end{deluxetable*}

We see that the total density of larger halos typically reach $\rho_{\rm thresh}$ earlier, although this is a stochastic process resulting from major merging 
events. The inclusion of a stream velocity can delay the halo from reaching the threshold by $\Delta z \simeq 1-3$ for 
$v_{\rm s} = v_\sigma$ and $\Delta z \simeq 1$-7 for $v_{\rm s} = 2v_\sigma$. The delay is typically greater between the 
$v_{\rm s} = 0$ and 
$v_{\rm s} = v_\sigma $ simulations than it is between the $v_{\rm s} = v_\sigma$ to $v_{\rm s} = 2v_\sigma$ simulations.  This suggests that while the inclusion of a stream velocity will have a major effect on the formation of structure in this mass
range, the magnitude of this effect does not increase arbitrarily with the magnitude of the stream velocity.  The delay of collapse is even more dramatic when looking at the gas density, with $\Delta z \simeq 3$-5 for $v_{\rm s} = v_\sigma$ and
$\Delta z \simeq 6$-12 for $v_{\rm s} = 2v_\sigma$. To determine the expected value of delay for a given streaming magnitude would require a more thorough analysis of multiple halos' evolution with multiple stream velocity magnitudes, which is well beyond the scope of this work.

\subsection{Radial Profiles}\label{oneHalo}

\begin{figure*}[h!] 
\centering
\includegraphics[scale=0.45]{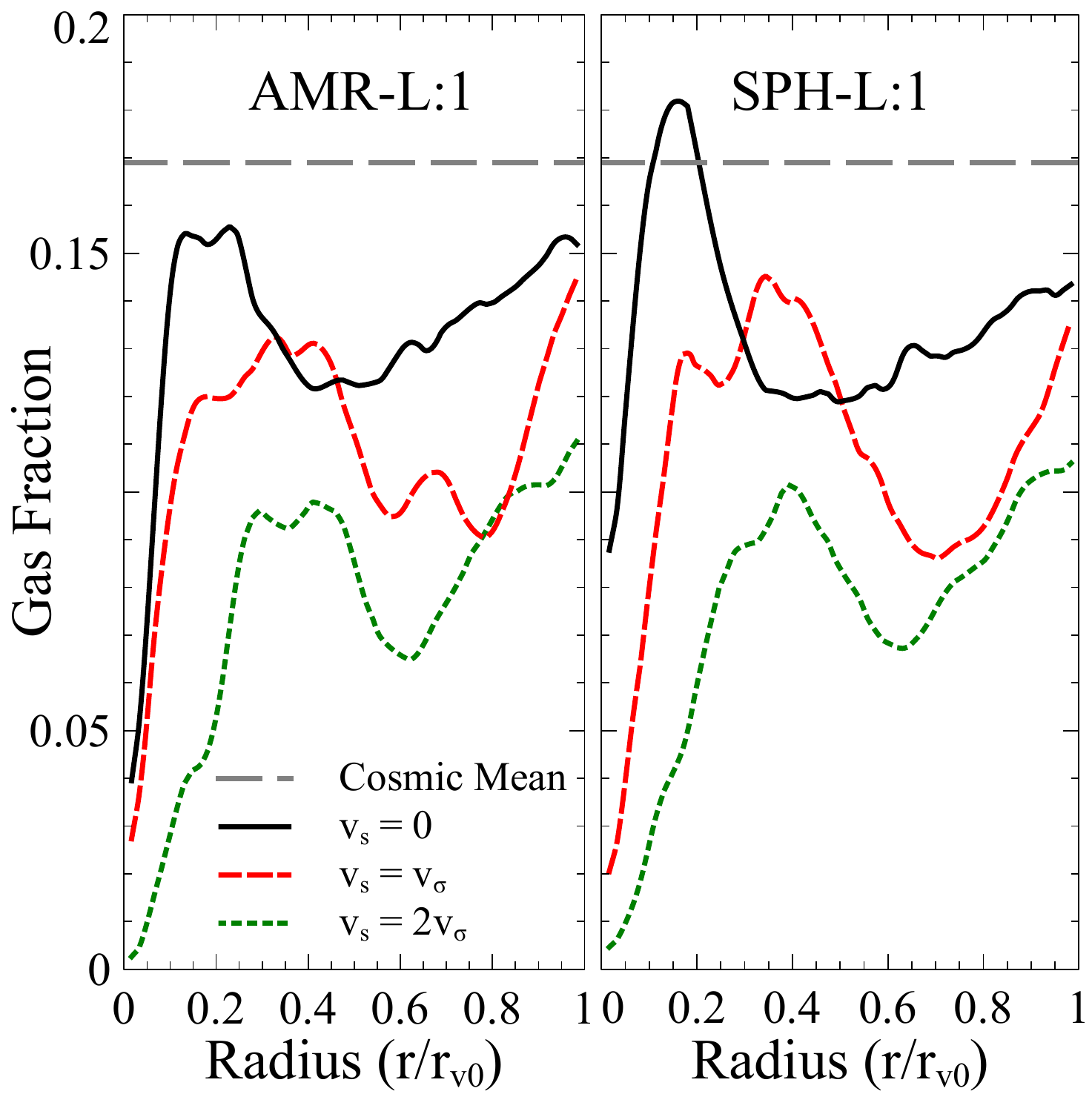}
\includegraphics[scale=0.45]{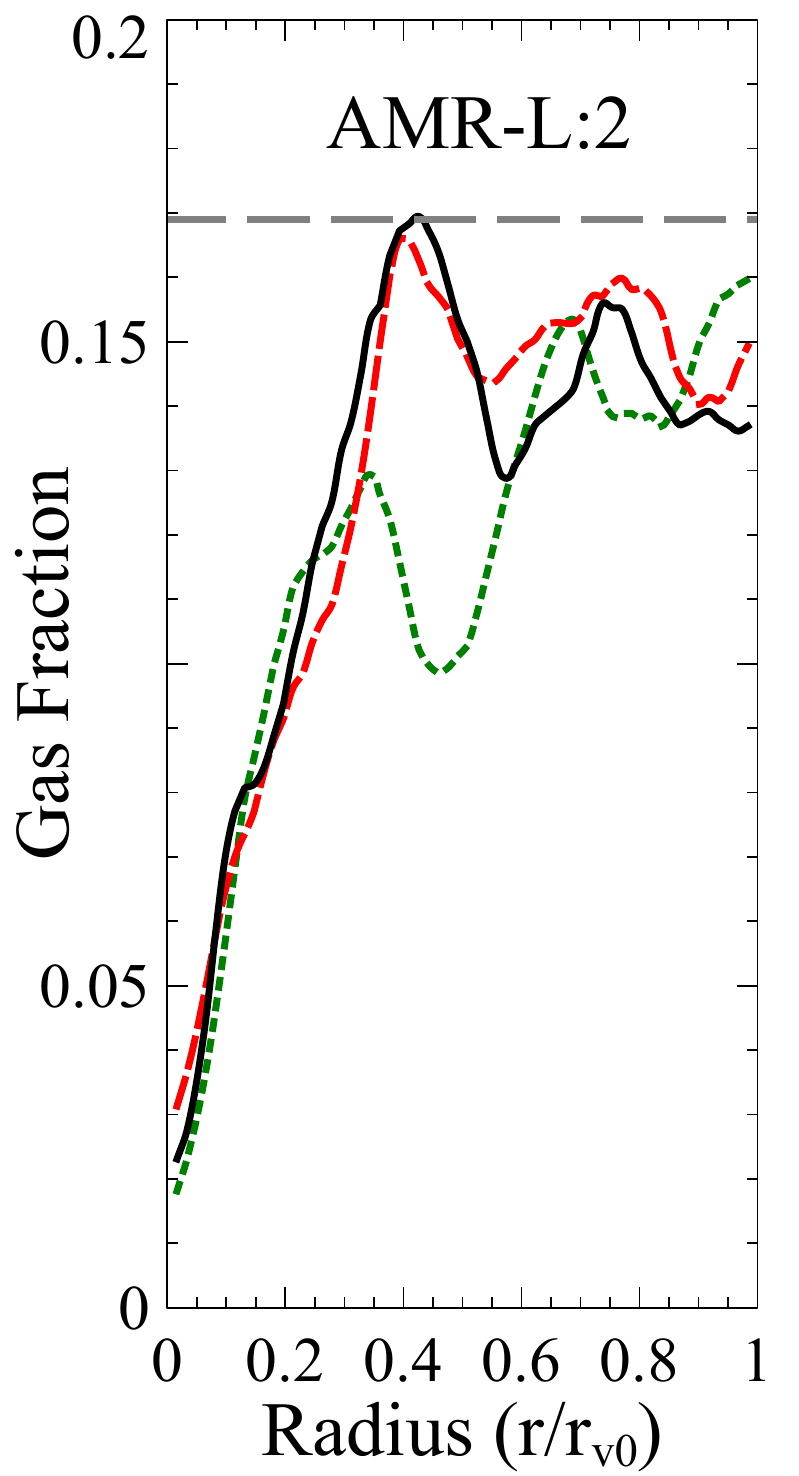}
\includegraphics[scale=0.45]{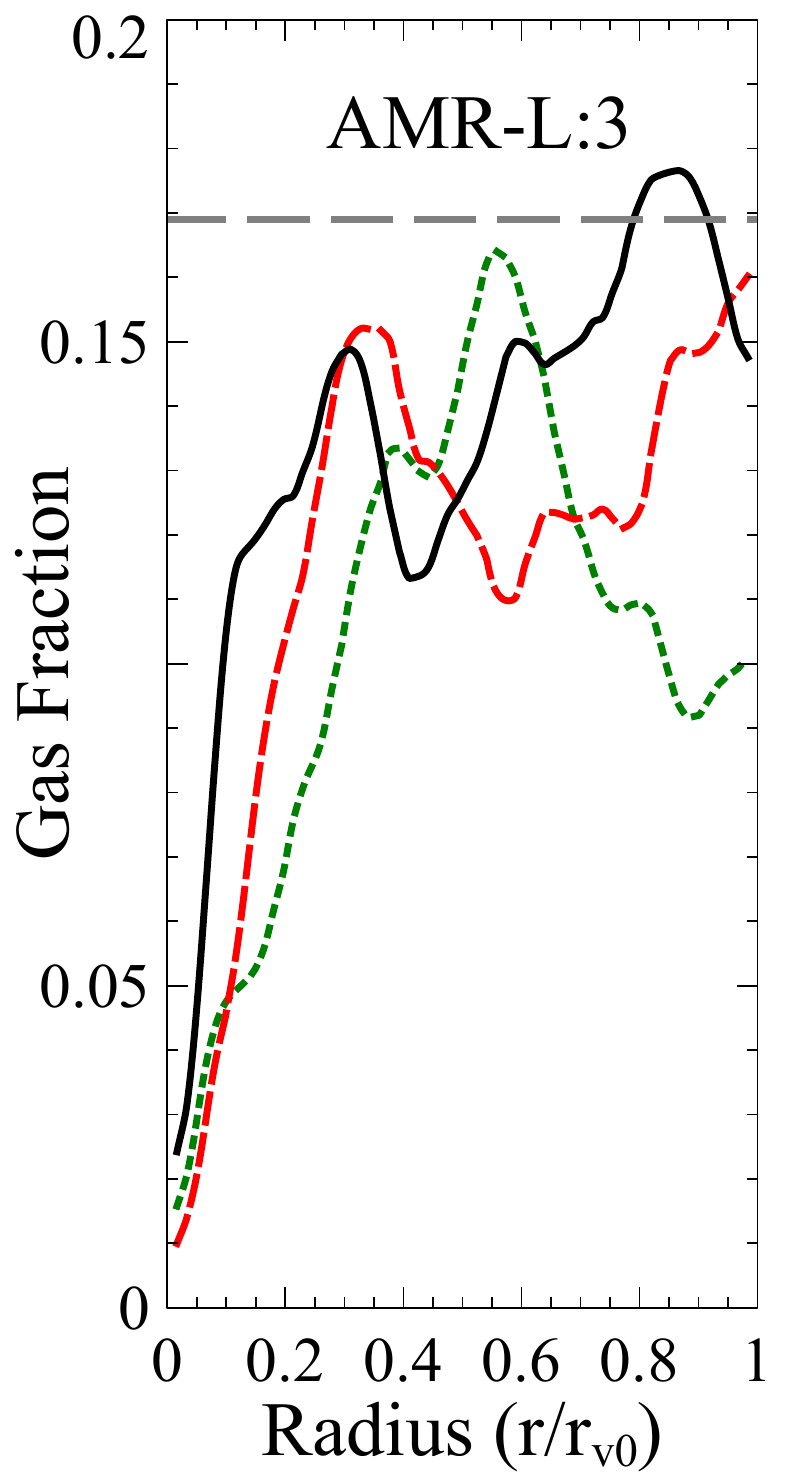}
\includegraphics[scale=0.45]{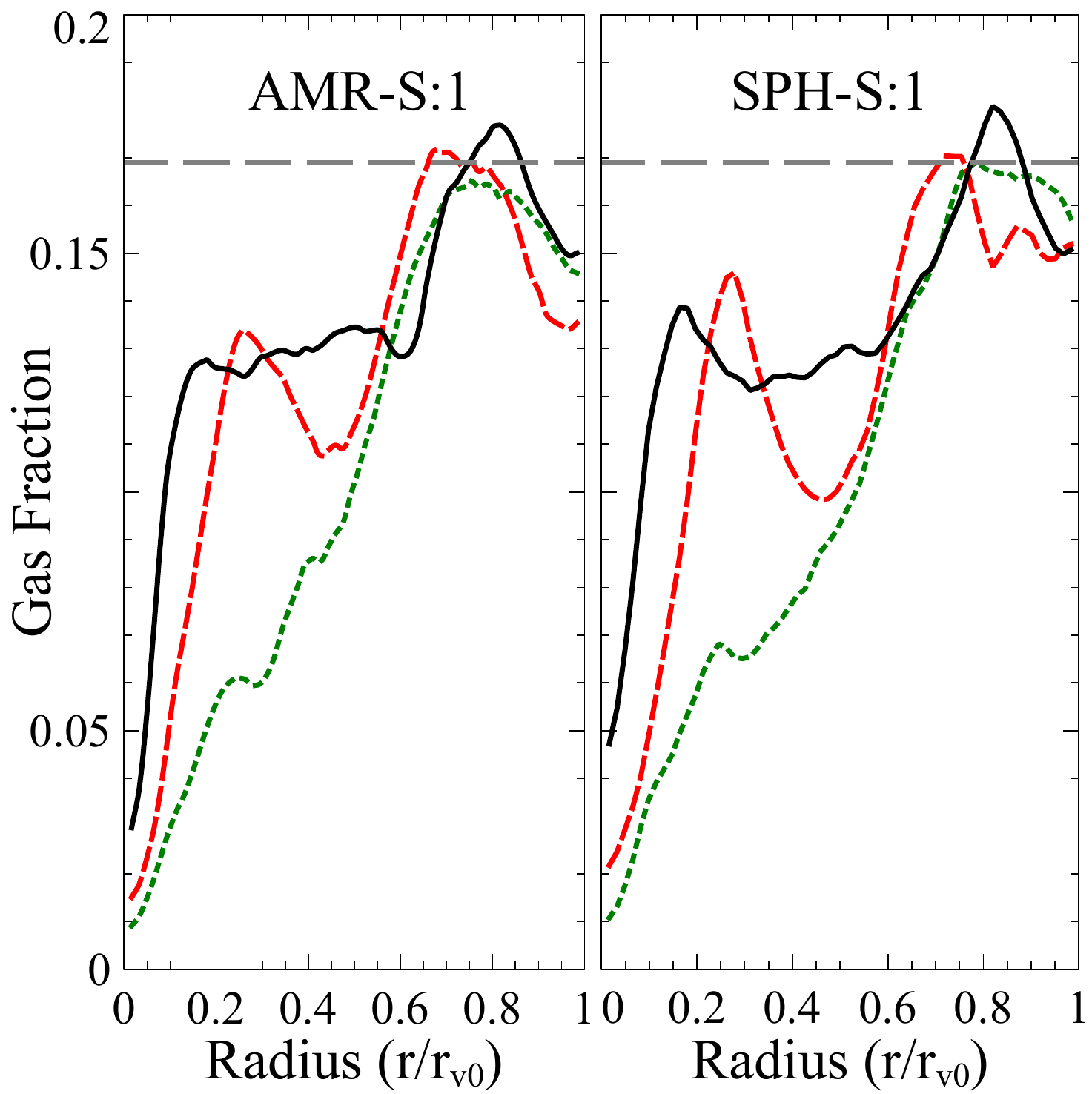}
\includegraphics[scale=0.45]{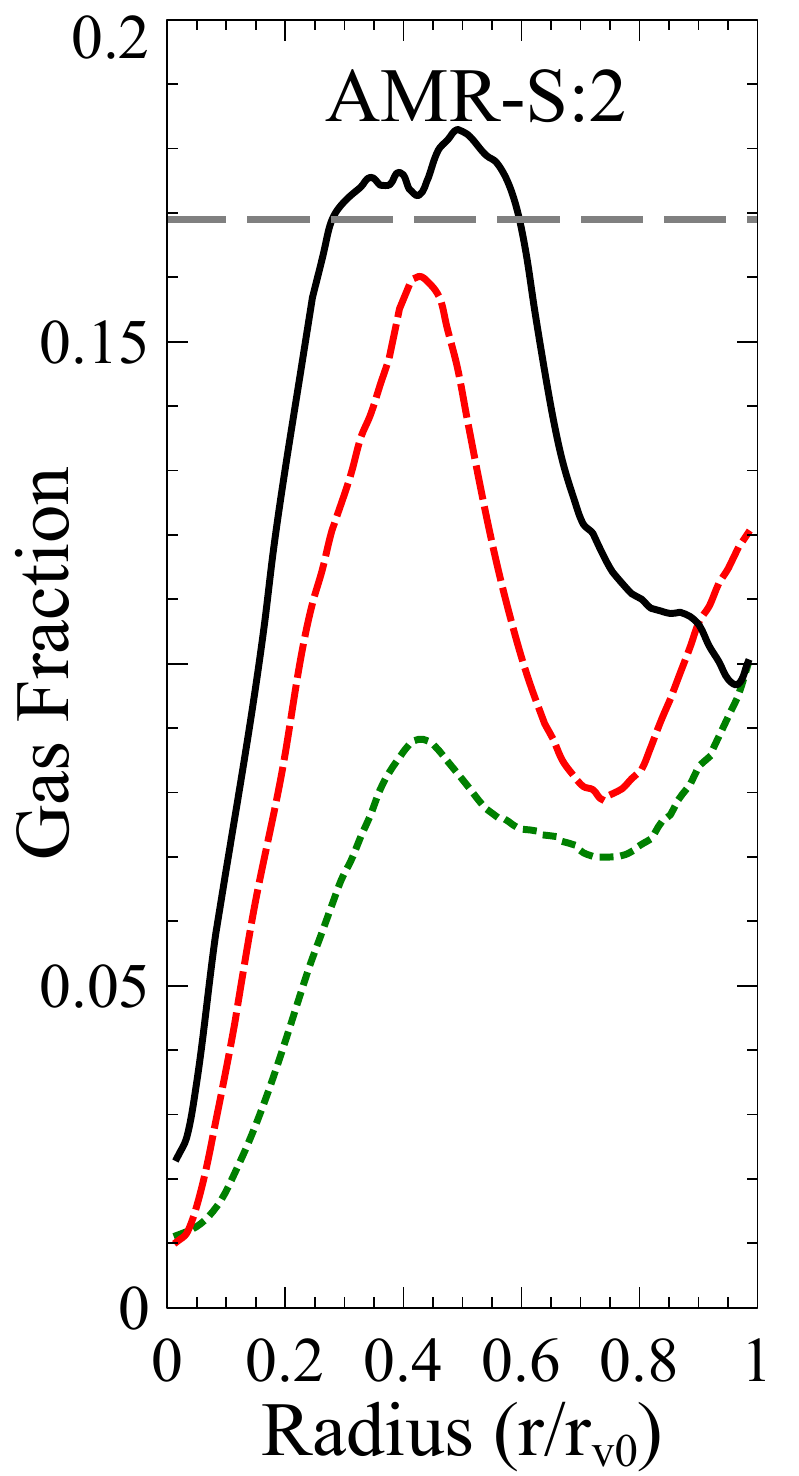}
\includegraphics[scale=0.45]{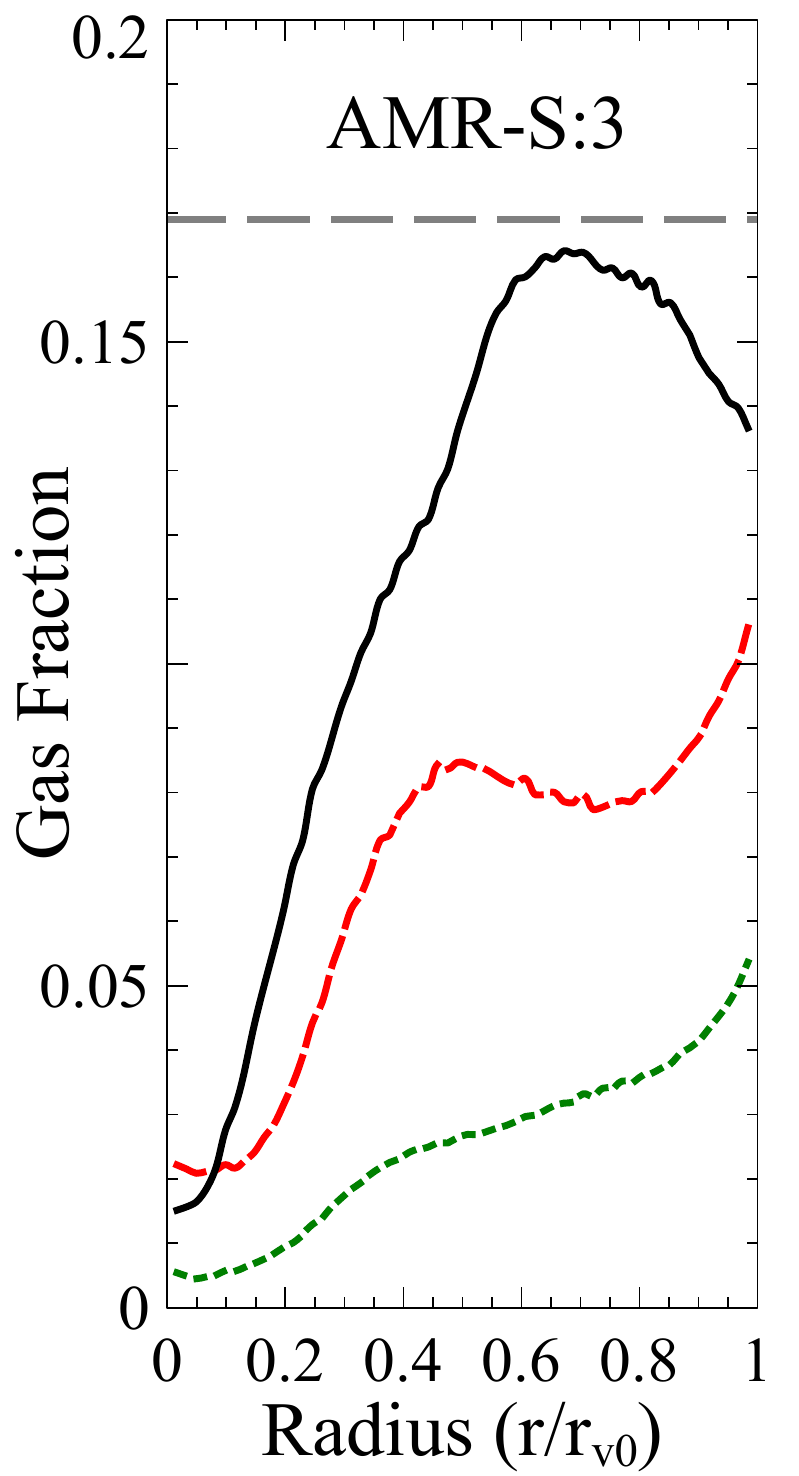}
\caption{\footnotesize{Radial profiles of the gas fraction for the three largest halos in the $\simLi$ (top) and $\simSi$ (bottom) runs, with the largest halo on the left and the smallest halo on the right. For the largest halo in $\simLi$ and $\simSi$, we compare the AMR result (left) with the SPH result (right). The
horizontal dotted line is the cosmic average. The solid, dashed, and dotted lines are for $v_{\rm s} /v_\sigma = 0, 1,$ and $2$, respectively. Radius is plotted in units
of HOP virial radius of the fiducial run (see \tabl{tab_thresh}). For the SPH simulations the gas and dark matter particles have gravitational softening lengths of roughly 1 pc. For the AMR simulations, the dark matter has a softening length equal to the grid resolution, 3 pc. The gravitational softening lengths are therefore quite small with respect to each halo's virial radius, with $r/r_{\rm v} < 0.05$ for all.  }}
\label{fig_gf}
\end{figure*}

Radial plots of the gas fractions in the three largest halos in the $\mathcal{L}_{\rm i}$ and $\mathcal{S}_{\rm i}$ simulations are given in  \fig{fig_gf}. Here, we have smoothed
the dark matter mass over a flat kernel with a radius of 2.25 times the width of the cell containing the dark matter particle. The central gas density profiles are very flat since they heat to the virial temperature but have
not had time to cool and collapse further, while the dark matter density profiles are cuspy (see \sect{oneHalo}). The result is a central dip in the gas fraction. The top 
row corresponds to the three largest halos in the AMR $\simLi$ simulations
with the largest on the left and smallest on the right, with the radial gas fraction of the largest halo in the SPH $\simLi$ simulations added for comparison. The second row corresponds to the three largest halos in the $\simSi$ simulations, with again the largest halo from the SPH simulations added for comparison. 
We see that the stream velocity typically causes a significant decrement in the gas fraction of the halos, with the effect most prominent in the least massive halos and the fastest 
stream velocities. Orientation
also appears to be important as AMR-L:2 has very little variation in gas fraction for different stream velocities, while Halo-L:1 has a significant reduction in gas fraction, even though it is the most massive halo.
We also see subtle variations in between the AMR and SPH simulation, but some of this may be due  to imperfect centering.

\begin{figure*}[h!] 
\centering
\includegraphics[scale=0.45]{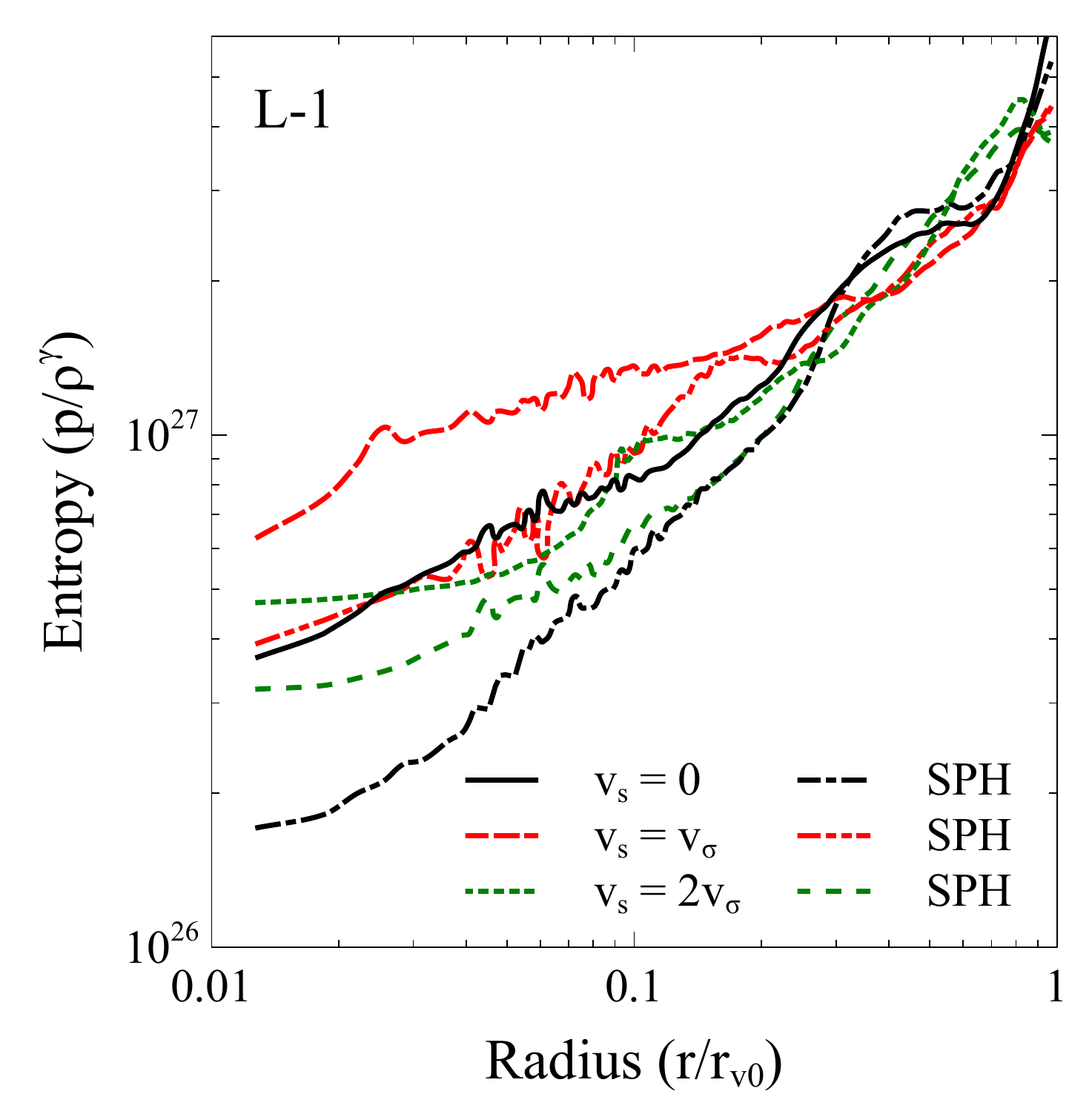}
\includegraphics[scale=0.45]{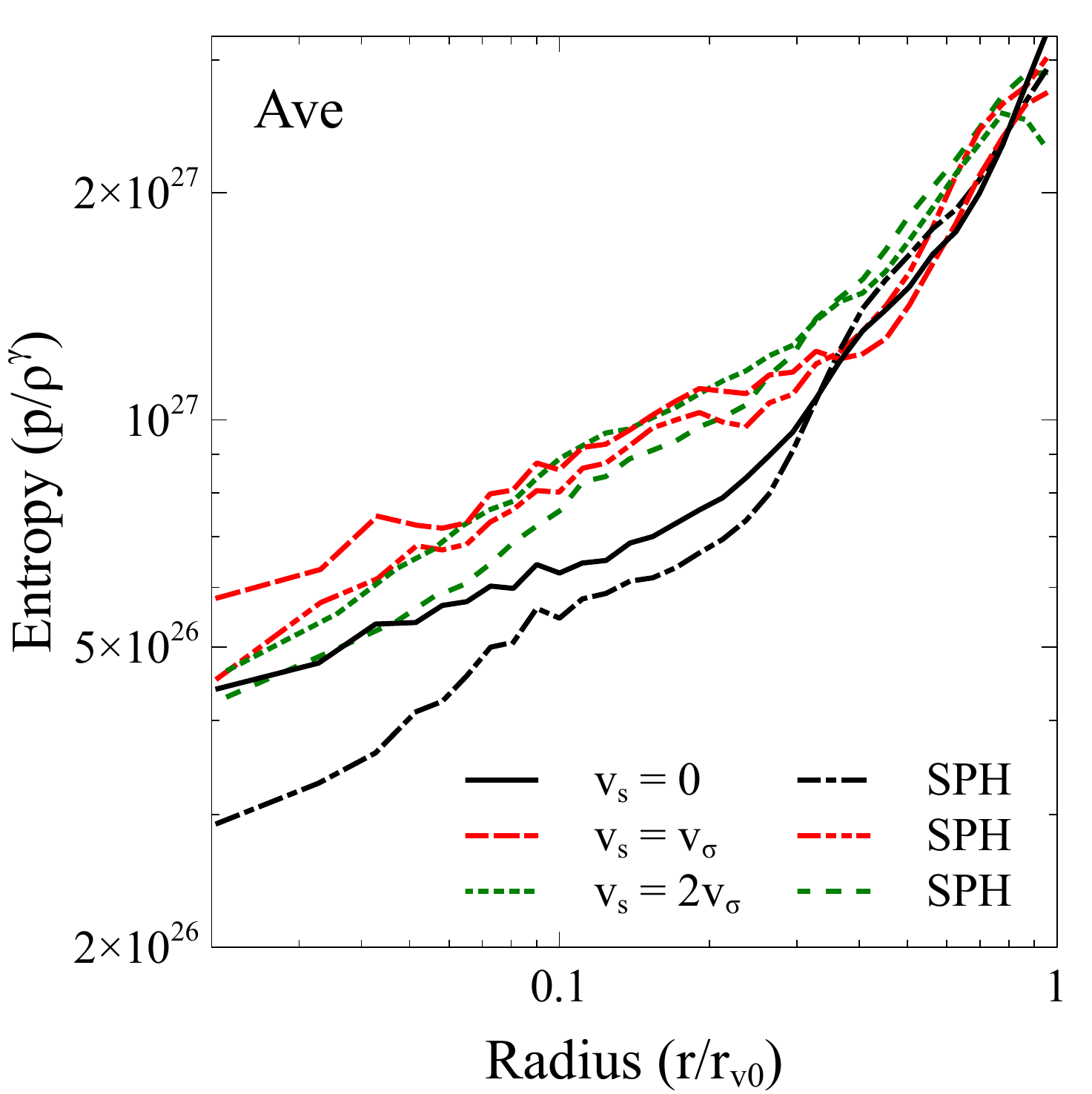}
\caption{\footnotesize{Radial profiles of the gas entropy, $S\propto p/\rho^{\gamma}$, for the largest halo in the $\simLi$ simulations (left) and averaged over all six largest halos in the $\simLi$ and $\simSi$ simulations (right). The solid, long-dashed, and dashed lines are for the AMR $v_{\rm s} /v_\sigma = 0, 1,$ and $2$, respectively, while
the dashed-dotted, dashed-double-dotted, and spaced dashed lines are their respective SPH results.}}
\label{figEntr}
\end{figure*}

To further compare the halos developed in the AMR and SPH simulations, and contrast their differences with other works in the literature, we plot their 
radial entropy ($S\propto p/\rho^{\gamma}$) profiles in \fig{figEntr}. We see, particularly in the $v_{\rm s} = 0$ case, that SPH has a lower entropy in the cores of halos as noted in many previous studies (e.g., see 
Mitchell \etal\ 2009; Sijacki \etal\ 2011), while at farther distances the discrepancy is reduced. Surprisingly, we also see that this discrepancy is reduced with increasing
stream velocity. \citet{Mitchell09} showed that the majority of the central entropy missing from SPH simulations is due to undermixing in these runs with respect to the
AMR simulations. We thus conclude that the inclusion of a stream velocity reduces this undermixing, perhaps by reducing gas structure on the smallest scales.
We also see that the inclusion of a stream velocity leads to a somewhat larger entropy profile, although this effect is not monotonic.

\begin{figure*}[h!] 
\centering
\includegraphics[scale=0.35]{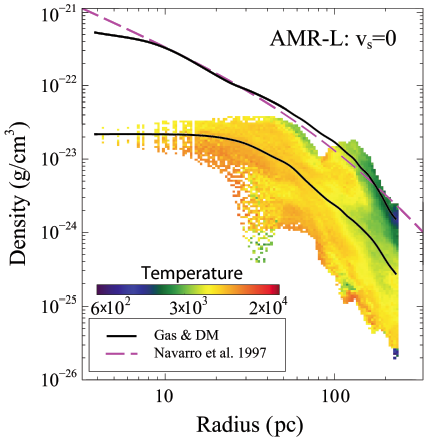}
\includegraphics[scale=0.35]{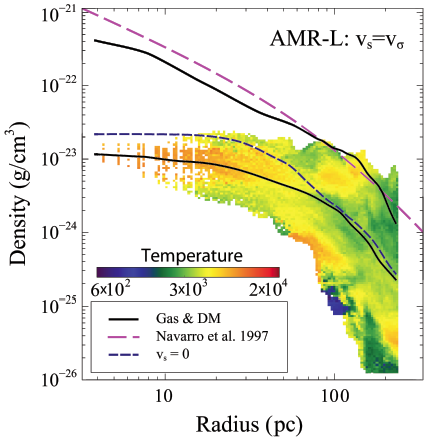}
\includegraphics[scale=0.35]{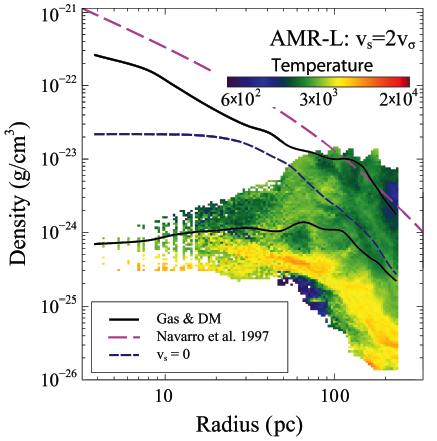}
\includegraphics[scale=0.35]{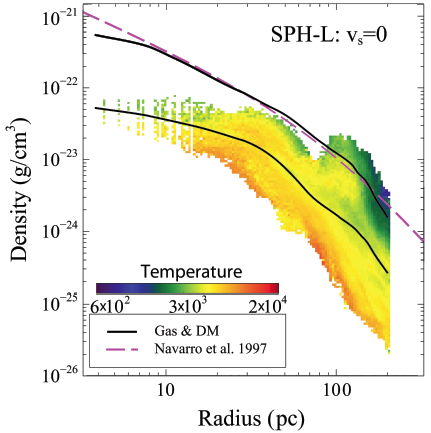}
\includegraphics[scale=0.35]{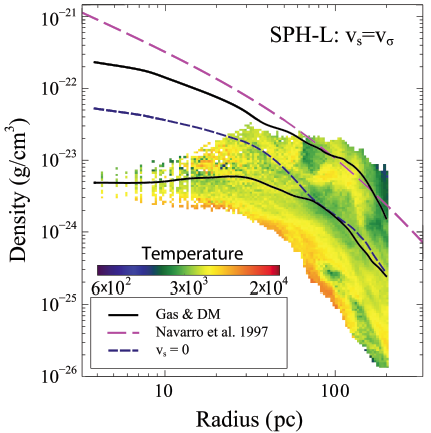}
\includegraphics[scale=0.35]{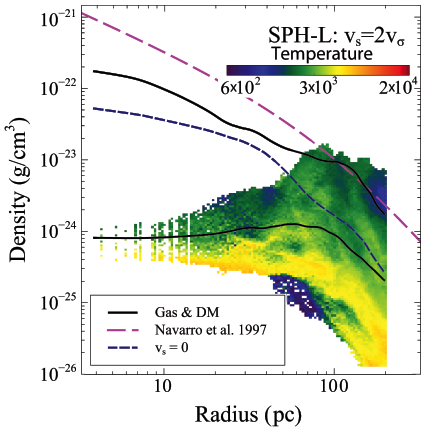}

\caption{\footnotesize{Average radial density profiles overlaid on phase diagrams for the largest halo in $\simLi$ for the AMR (top) and SPH (bottom) runs, with $v_{\rm s} = 0, v_\sigma,$ and $2v_\sigma$ for the left, center, and right columns, respectively.  In all panels,  the color corresponds to the average temperature as a function of radius and density,
 the solid black lines give the average gas and dark matter density profiles, and the dashed magenta line gives a \citet{Navarro97} profile, using a concentration parameter of 1.7, and an overdensity of 500. For comparison, in the $v_{\rm s} \ne 0$ plots, we also plot the $\simLA$ results as a blue dashed line. These are plotted out to the virial radius of the $\simLA$ case. For the SPH simulations the gas and dark matter particles have gravitational softening lengths of roughly 1 pc. For the AMR simulations the dark matter has a softening length equal to the grid resolution, 3 pc.  }}
\label{fig_phaseL}
\end{figure*}

For the largest halos in the six simulations, we investigated the effect of the stream velocity on the radial density and temperature profiles. In \fig{fig_phaseL} we plot 
a phase diagram of the largest halo in each of the $\simLi$ simulations with the AMR runs on the left and the SPH runs on the right, and overplot the average gas and dark matter density profile, and a \citet{Navarro97} profile as a magenta dashed
line, given by:
\begin{eqnarray}
\rho(r) &=& \frac{\rho_{\rm c}}{cx(1+cx)^2}\frac{c^3}{3F(c)} \mbox{ g cm}^{-3}, \\
\rho_{\rm c} &=& \Delta \Omega_{\rm m}  (1+z)^3 \rho_{\rm crit}, \\
F(t) &\equiv& \ln(1+t) - \frac{t}{1+t},
\end{eqnarray}
where $c$ is the concentration parameter, $\Delta$ is the overdensity, $\rho_{\rm crit}$ is the critical density, and $x$ is the dimensionless radius with $x = r/r_{\rm vir}$, the virial radius. In \fig{fig_phaseL}, this profile uses a concentration parameter of 1.7 and an overdensity of 500 (expected since the HOP virial radius is smaller than a spherical overdensity corresponding to 200). 
The overplotted dark matter density profiles are the result of slight smoothing, which can cause a slight decrease in the central density as it smooths into the outer cells, corresponding to $2.15\times10^{19}$ cm, where the dark matter density drops below the \citet{Navarro97} profile. For the $v_{\rm s} \ne 0$ plots, we also plot the $\simLA$ results as a blue dashed line for comparison. 

We can clearly see that although the addition of a stream velocity reduces the overall gas density and temperature, it has a much smaller effect on the dark matter
density profile. We can also see the hot dense virialized gas in the interior, and cold dense matter from an accretion lanes, which appear to 
penetrate the halo much more efficiently at large stream velocities. The SPH runs are very similar, and are cooler in the cores, suggesting that they either shock-heat to a lower virial temperature, or are able to cool more efficiently. This is a similar result as seen in \citet{Frenk99}, where the SPH simulations consistently had declining
radial temperature profiles in the core of their cluster, while grid-based methods had an increasing radial temperature profile. In general, the SPH central under-density ratio for the $v_{\rm s} \ne 0$ cases, $\rho(r$$=$$0;v_{\rm s})/\rho(r$$=$$0;0)$, are smaller than the AMR underdensity ratio for the same $v_{\rm s}$, and this discrepancy is more pronounced for larger $v_{\rm s}$.

\begin{figure*}[h!] 
\centering
\includegraphics[scale=0.35]{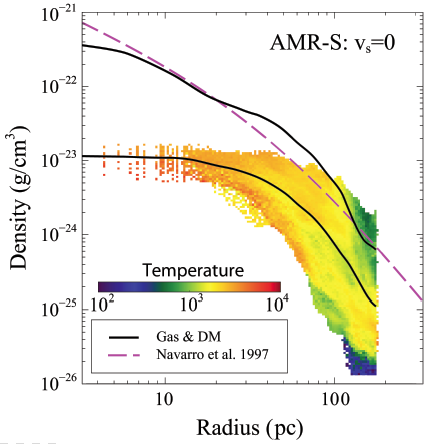}
\includegraphics[scale=0.35]{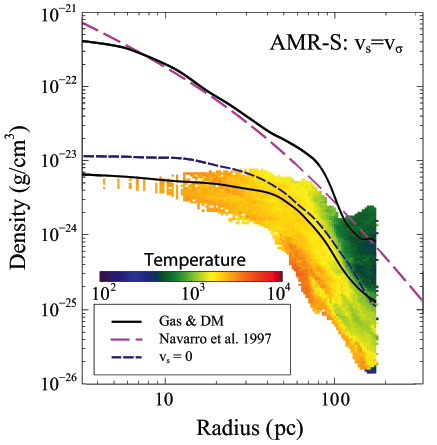}
\includegraphics[scale=0.35]{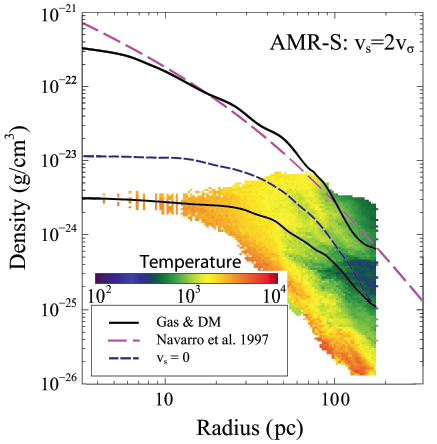}
\caption{\footnotesize{Same as \fig{fig_phaseL}, but with the largest halo from the $\simSi$ simulations. The \citet{Navarro97} fit uses a clumping factor
of 3.8.}}
\label{fig_phaseS}
\end{figure*}

In \fig{fig_phaseS} we show the same results for the largest halo in the $\simSi$ simulations, and see identical trends, where here the \citet{Navarro97} profile has a clumping factor of 3.8 and an overdensity of 200. Again, the gas density and temperature are reduced when
a stream velocity is added, while in this case, the dark matter profile is essentially the same across the runs.

\subsection{Halo Statistics}\label{HMF}

To understand quantitatively how the stream velocity affects halo
number density and gas fractions as a function of mass and streaming magnitude, we again
identified halos using the HOP method, rejecting objects that include low-resolution particles, have less than 100 particles, or do not have an 
overdensity of 160. HOP can form halos with severely triaxial dimensions, thus to properly account for gas mass 
in the halo we determined the ellipsoid based on the moment of inertia tensor:
\begin{equation}\label{MOI}
T_{i,j} = \sum_{n=1}^N (x_i^n - x_i^{\rm COM}) (x_j^n - x_j^{\rm COM}),
\end{equation}
where we sum over the $N$ particles in the halo, each particle has position $\mathbf{x^n}$ with components 
$x^n_i$, and the halo has a center of mass coordinate $\mathbf{x^{\rm COM}}$. We determine the 
eigenvalues of this tensor, which are the squares of the ellipsoids three principal axis dimensions, 
rescale these values to be consistent with the average distance of a particle in the halo from its center-of-mass, and finally
calculate the gas and dark matter mass inside this ellipsoid. 

\begin{figure*}[t!] 
\centering
\includegraphics[scale=0.54]{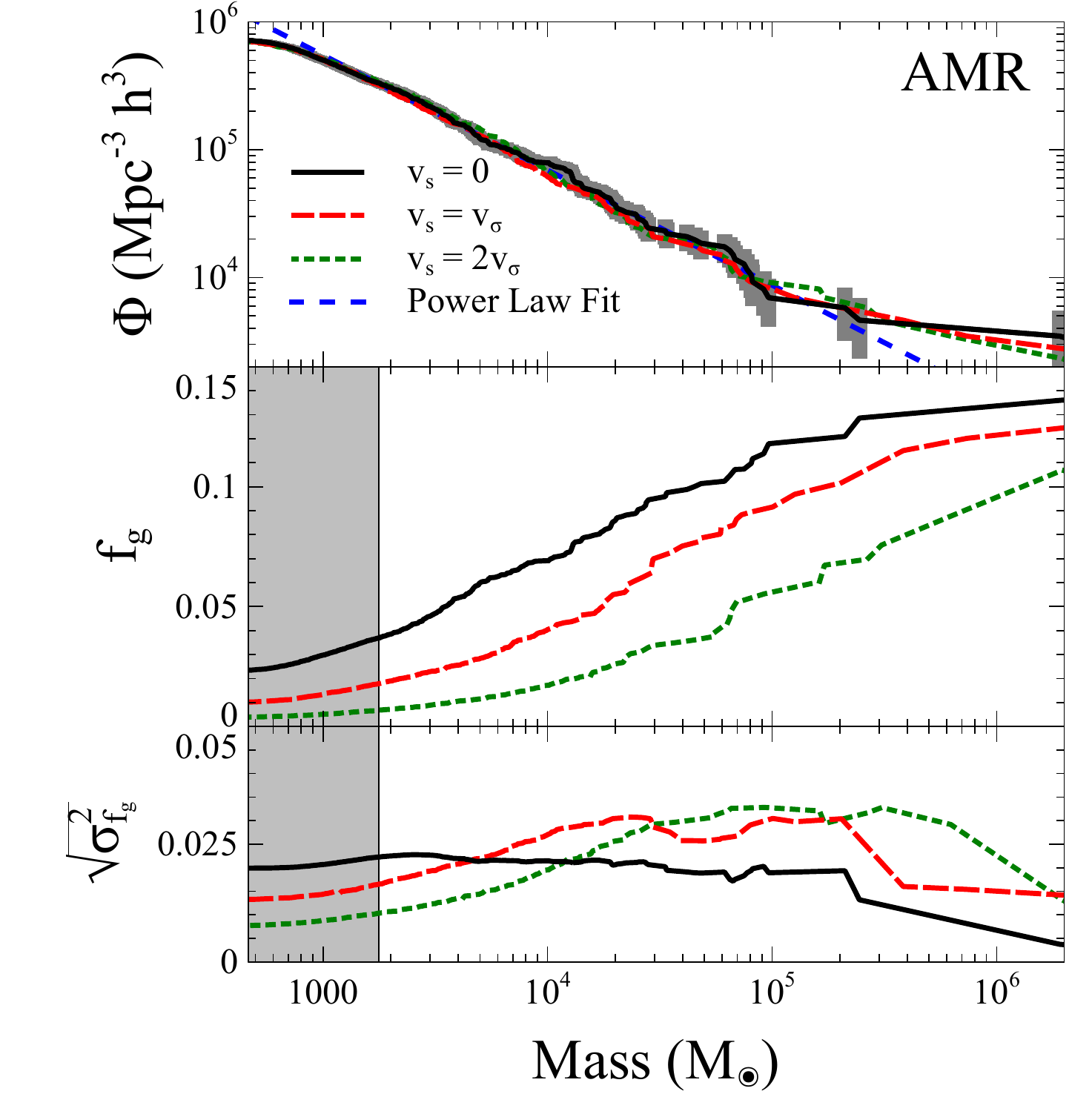}
\includegraphics[scale=0.54]{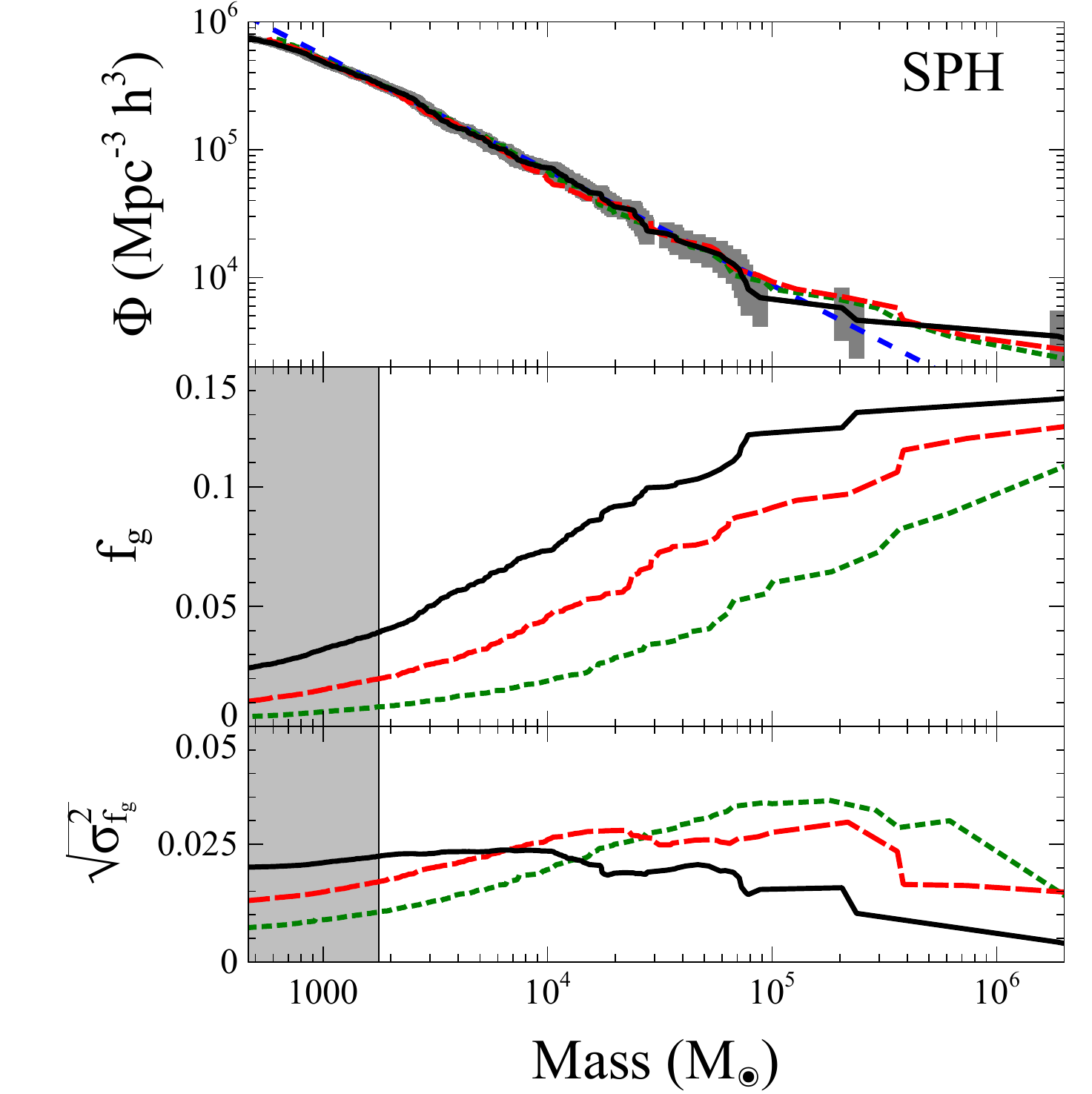}
\caption{\footnotesize{Halo statistics for our combined AMR (left) and SPH (right) simulations. Solid black, dashed red, and dotted green
lines corresponding to the $v=0,1,$ and $2v_\sigma$ simulations, respectively. Top: cumulative number density as a function of total mass, with gray boxes
illustrating the Poisson noise. Middle: average gas fraction of halos above a total mass. Bottom: square root of the variance of gas fraction of halos above a total mass. The gray hatch highlights the mass regime where some halos have less than 500 dark matter particles, the number necessary to resolve the gas fraction 
within 20\% \citep{Naoz09}.}}
\label{fig_hmf}
\end{figure*}

We determine the cumulative number density of 
these halos as a function of the total mass within the ellipsoid, shown in the top left of \fig{fig_hmf} for the 
$\mathcal{L}_{\rm i}$ and $\mathcal{S}_{\rm i}$ simulations combined. The simulations are shown in solid black, dashed red, 
and dotted green lines corresponding to the $v_{\rm s} /v_\sigma=0,1,$ and $2$ simulations, respectively. Included are also
our Poisson noise for the $v_{\rm s} = 0$ case, given by gray boxes.

We see that the stream velocity has little effect on the mass density function at lower masses, as seen previously in \citet{Naoz12}. This mass range is
well below the regime of peak suppression, $M=2\times10^6\Msun$ \citep{Tseliakhovich10}, which we omit here due to only having a small sample in such a small box. At higher masses near this regime we can
see a small reduction in the number density. If we assume a standard Press \& Schecter (1974) formalism, we should expect the cumulative number density to be given by:
\begin{equation}\label{PS}
\Phi(M)dM \sim M^{-P}dM \sim M^{-1 + (n_{\rm eff}+3)/6}dM,
\end{equation}
where $n_{\rm eff}$ is the effective spectral index, the slope of the power spectrum at a particular scale. Similar to \citet{Naoz12} who found $P\simeq 0.94$ for their $256^3$ and $512^3$ simulations, we find $P=0.9$, or $n_{\rm eff} = -2.4$. We would
expect $n_{\rm eff} \simeq -2.6$ for this mass range based on the CAMB power spectrum. The most significant cause of this difference is that we are looking at a rare piece of the
universe, where larger structure is prevalent, which artificial flattens the spectral index. The second reason is that we are discarding physically compact halos,
and are not resolving less massive halos, both of which would act to flatten the spectral slope. At very small mass the cumulative halo density leaves the power law as we are not resolving 
the halos well, with only about $10^2$ particles in a single halo, and are throwing out a majority of the halos as they are too compact to sample the gas.

For each halo we also determine the average gas fraction, $f_g$, where:
\begin{equation}
f_{\rm g} = \frac{M_{\rm g}}{M_{\rm DM} + M_{\rm g}}=\frac{M_{\rm g}}{M_{\rm tot}},
\end{equation}
where $M_{\rm g}$ is the gas mass, $M_{\rm DM}$ is the dark matter mass, and $M_{\rm tot}$ 
is the total mass within the halo's ellipsoid. These are then averaged across all halos above a given total mass, 
shown in the left middle panel of \fig{fig_hmf}, while the gas 
fraction variance of these halos, $\sqrt{\sigma^2_{f_g}}$, is shown in the bottom left panel. Note that we have very few large halos 
since we have such a small box, and our statistics are particularly noisy above $\approx 10^5 \Msun.$ At smaller masses we highlight the regime where some
halos have less than 500 dark matter particles, the number necessary to resolve the gas fraction within 20\% in previous SPH simulations \citep{Naoz09}. For consistency we also highlight the same mass 
range in the SPH simulations.

In the $v_{\rm s} = 0$ simulations, the gas fraction is quite small in the least massive halos, and approaches the cosmic mean in the largest halos, similar to that found in \citet{Naoz13}. We see that by including a stream velocity the gas fraction is significantly reduced, with the total average gas fraction dropping by almost $50\%$ as $v_s$ goes from 0 to $v_\sigma$ and by almost 50\% again as it increases to 2$v_\sigma$. This is to be 
expected, as gas is moving with respect to the dark matter gravitational potential, and is less able to be captured.
In the $v_{\rm s} = v_\sigma$ simulations the effect of streaming is roughly an absolute reduction in gas fraction by $\approx 0.01$ at all masses, while doubling the stream velocity roughly doubles this effect, with a slightly more dramatic reduction at $10^5 \Msun$. Clearly halos are unable to maintain their gas content which is translated out of their
potential. In the right side of \fig{fig_hmf}, we plot the same quantities  for the SPH results at $z = 17.18$. We see almost the exact same results as the AMR 
simulations, which gives us confidence in both the ability for SPH and AMR methods to give consistent statistical results in cosmological
simulations.

In both earlier work by \citet{Greif11}, and our work looking at specific massive halos below, we consider the
possibility that halos preferentially pointing along the stream velocity may be affected less severely. However, our analysis of the gas fraction as a function
of the $x$-component of the principal halo axis revealed no discernible dependence on orientation. This is a bit at odds with the individual halo results we discuss above. To better understand this in the future would require a larger volume, with a larger number of high-resolution halos.

\section{Summary and Conclusions}\label{Conc}
Due to the impact of radiation pressure, substantial relative stream velocities between the gas and dark matter persist after recombination, and these can have an important impact on the formation of the first cosmic structures.    To study the effect of such stream velocities in detail, we have run six high-resolution SPH cosmological simulations from $z=199$ to $z=39,$
converted these SPH datasets to an AMR data sets, and continued the simulations down to $z=17.2,$  the epoch at which the first nonlinear structures were able to accrete gas.
We have also continued the SPH simulations down to $z=17.2,$ to provide a check of our mapping routine and a way to highlight differences in studying this problem using two computational methods. Differences may arise from the contrasting methods of modeling physics in the two methods, such as the gravitational potential calculation, the ability to conserve angular momentum, and the Galilean (in)variance of advected flow, to name a few. In general we found that while the AMR method did provide better resolution in low density regions, this did not have a dramatic effect on the properties of the halos formed in our simulations, with the exception
of the central entropy profiles.

In each simulation set, we investigated the properties of the three most massive halos in detail.
Similar to \citet{Greif11} and \citet{Oleary12} , we find that the presence of a stream velocity suppresses structure formation both within and around these objects,
and this suppression is strongest at the smallest halo masses and for the largest stream velocities. 
Within the halos, the presence of a stream velocity reduces the core gas density, and lowers the overall gas content.
On larger scales, we see possible indications that gas accretion flows are particularly affected if they are perpendicular to the stream velocity. 
For halos in simulations including stream velocities, the dark matter projections were also somewhat less evolved, consistent with smaller gravitational potentials due to the reduction in gas content.

To quantify the delay of collapse produced by stream velocities, we determined when each massive halo achieved a 
total mass peak density of 
$10^{-22}\gcc$, and when the gas density reaches a peak of $3\times10^{-24}\gcc$.
We find a $\Delta z \simeq 1-3$ between the fiducial runs and the $v_{\rm s} = v_\sigma$ runs for the total mass density reaching its threshold, while the gas density can be delayed up to a $\Delta z = 5$. We find that in general there is a  
smaller delay between the $v_{\rm s} = v_\sigma$ runs and the $v_{\rm s} = 2v_\sigma$. This is roughly the same delay observed by \citet{Greif11},
although they were able to follow their gas to much higher densities. 
Radial gas fractions are severely reduced in the smallest of the halos, and  the loss of gas from the largest halos appears to depend on its orientation,
an effect similar to that seen in \citet{Greif11} and \citet{Oleary12}. Radial entropy profiles show that stream velocities act to increase the core entropy, although this effect is not 
monotonic with stream velocity. 

For the largest halo in each simulation set we plotted the radial gas and dark matter profiles along with a radial phase diagram
illustrating gas density and temperature. Our density profiles are well modeled by a \citet{Navarro97} profile with concentration parameters 
of 1.7 and 3.8, although the dark matter density is slightly reduced when we include a stream velocity. The gas is diminished with increasing stream velocity, and is cooler, with cold accretion flows penetrating deeper into the halo without showing signs of shock heating to the virial temperature. Thus the virial radii, mass and temperature of these halos are reduced by including a stream velocity. 

Looking at the statistical properties of the full halo population,
we find that including stream velocities has almost no effect on the cumulative halo mass density  over the mass range of $10^3$ - $10^6 \Msun$,
up to and including the peak expected suppression scale \citep{Tseliakhovich10}. We stress however that our simulations span two regions specifically selected for their rare overdensity. 
On the other hand, the average gas fraction of each halo mass bin is significantly reduced by incorporating a stream velocity, and the effect is greater for larger
velocities. The gas fraction is approximately reduced by a factor of $2^{-v_{\rm s} /v_\sigma}$ from the fiducial run. Both of these results are consistent with a number of previous studies, including the low resolution work by \citet{Maio11} and \citet{Naoz12}. This reduction in gas fraction will lead to delayed star formation in a range of galaxy mass, with the largest effect in regions with the largest stream velocity.

Finally, inspired by the orientation effects seen in the largest halos we have simulated, 
we fit all halos with ellipsoids and attempted to measure an increase in gas loss for halos oriented perpendicular to the stream velocity.
Unfortunately, these results were inconclusive given the statistics available from our limited halo population.
Such orientation-dependence would make an interesting subject for further study, as it would act to moderate the effects of stream velocities in a select fraction of high-redshift halos.

Although the six largest halos are only barely large enough at $z=17.2$ to produce star formation via molecular cooling \citep{Yoshida03}, we can estimate their mass growth from linear theory by assuming that $\sigma(M)/(1+z),$ and thus the bias, remains roughly constant. This gives that the virial temperature of these halos,  $T_v = 720 (M/10^6 \Msun)^{2/3} (1+z)/10$ K, will be greater than $10^4$ K by $z \approx 13.$  This will permit atomic cooling, leading to even more efficient star formation. The increased entropy of these halos, along with their delayed collapse, suggests that the gas component of halos that include stream velocities will not grow as quickly, will take longer before atomic cooling is possible, and will have a delayed episode
of first star formation. Such a delay in star formation was observed in \citet{Maio11}. The epoch of reionization, driven by the ionizing flux from the earliest stars,
is likely to be delayed as well.

Meanwhile, it has also been suggested that gas-rich minihalos act to delay the evolution of reionization by shielding the intergalactic medium (IGM; e.g., Shapiro \& Giroux 1987; Barkana \& Loeb 
2002; Iliev et al. 2005; Ciardi et al. 2006; McQuinn et al. 2007). Since the gas fractions in our minihalos are reduced when stream velocities are included, their ability to shield the IGM is reduced. This suggests that while stream velocities act 
to delay the onset of reionization, they can also accelerate its evolution. Further understanding these competing effects will require high-resolution large-scale simulations that include ionization sources and radiative transfer. 
Recent work (e.g., McQuinn \& OÕLeary 2012; Visbal et al. 2012) has begun this investigation, finding to first order that stream velocity between baryons and dark matter leave an imprint on reionizationÕs effect on the 21cm background signal.The measurements of the onset and extent of reionization (e.g., Pritchard \& Loeb 2012; Bowman et al. 2008; Liu \etal 2013) from future 21 cm experiments, such as the Precision Array for Probing the Epoch of Reionization (Parsons et al. 2010), the Murchison Wide-field Array (Tingay et al. 2012; Bowman et al. 2013), and the Low Frequency Array (Rottgering et al. 2006), will be necessary to directly constrain these models.

This is the first study of its kind using the AMR method with initial conditions deirved from SPH datasets, and the consistencies between the SPH and AMR are an excellent demonstration
of the effectiveness of the two methods, as well as our ability to translate between them.  However, we do see that mixing in the low density regions, a likely occurrence with stream velocities included, may be slightly dampened in SPH methods, as expected from \citet{Agertz07}, while cool low-density accretion flows are more
tenuous in SPH methods. Also, radial entropy profiles show the typical result that the SPH methods result in lower central entropy values than their 
AMR counterparts \citep{Mitchell09}, mostly due to undermixing in the SPH cores. This discrepancy is less pronounced with increasing stream velocity, perhaps
due to an overall reduction of structure at the smallest scales. Future work refining the details of star formation in the first structures should keep these issues in mind,  as one may need to consider grid methods to understand the essential accretion flows  and their role in providing cool gas, which can be affected by
  the presence of stream velocities.

M. L. A. R. was supported by NSF grant AST11-03608 and the National 
Science and Engineering Research Council of Canada. E. S. was also supported by the National Science Foundation under grant AST11-03608 and NASA theory grant NNX09AD106. R. J. T. is supported by a Discovery Grant from NSERC, the Canada Foundation for Innovation, the Nova Scotia Research and Innovation Trust and the Canada Research Chairs Program. Simulations were conducted with Arizona State
University Advanced Computing Center compute resources and the CFI-NSRIT funded {\em St Mary's Computational Astrophysics Laboratory}. We thank Rennan Barkana,  William Gray, Daniel Jacobs, and Reid Landeen for discussions that greatly improved this study.

\end{document}